\documentclass{comnet}%%%%where comnet is the template name
\bibpunct{[}{]}{,}{n}{,}{;}
\usepackage{amsmath}
\usepackage{lipsum}
\usepackage{tikz}
\usepackage{lipsum}
\usepackage{pgfplots}
\usepackage{pgfkeys}
\pgfplotsset{compat=1.14}
\usepackage{pgfkeys}
\usepackage{appendix}
\usepackage{url}
\newenvironment{customlegend}[1][]{%
        \begingroup
        \csname pgfplots@init@cleared@structures\endcsname
        \pgfplotsset{#1}%
    }{%
        \csname pgfplots@createlegend\endcsname
        \endgroup
    }%

    \def\addlegendimage{\csname pgfplots@addlegendimage\endcsname}
\pgfplotsset{
cycle list={%
{draw=black,mark=star,solid},
{draw=black, mark=square,solid},%densely dashed}, 
{draw=black,mark=+,solid},%dashdotted}, %every mark/.append style={rotate=90},
{black,mark=o},}}
\date*{}
\begin{document}

\title{Generating dynamic contact graphs with indirect links}

\shorttitle{Modelling co-location networks} %%%for recto running head
\shortauthorlist{M Shahzamal} %%% for verso running headhttps://www.overleaf.com/project/5c821347bbdc4666016db985

\author{%%%% First author details
\name{$^*$Md Shahzamal$^{1,2}$, Raja Jurdak$^{2,3}$, Bernard Mans$^1$, Frank de Hoog $^2$ and Dean Paini$^2$}
\address{$^1$Macquarie University, Sydney, Australia}
\address{$^2$ CSIRO, Australia}
\address{$^3$Queensland University of Technology, Brisbane, Australia}
\address{\email{$^*$Corresponding author: md.shahzamal@mq.edu.au}}
%%%%%%% Second author details
% \name{Raja Jurdak}
% \address{CSIRO, Australia}
%%%%%%%
% \and
%%%%%%% Third author details
% \name{Bernard Mans}
% \address{Macquarie University, Australia}}
}

\maketitle

\begin{abstract}
{Graph models are widely used to study diffusion processes in contact networks.
% Individuals in graph models are represented as nodes and links are created between two interacting nodes. 
Recent data-driven research has highlighted the significance of indirect links, where interactions are possible when two nodes visit the same place at different times (SPDT), in determining network structure and diffusion dynamics. However, how to generate dynamic graphs with indirect links for modeling diffusion remains an unsolved challenge. 
Here, we present a dynamic contact graph model for generating contact networks with direct and indirect links. Our model introduces the concept of multiple concurrently active copies of a node for capturing indirect transmission links. The SPDT graph model builds on activity driven time-varying network modelling for generating dynamic contact networks using simple statistical distributions. This model is fitted with a large city-scale empirical dataset using maximum likelihood estimation methods. Finally, the performance of the model is evaluated by analysing the capability of capturing the network properties observed in empirical graphs constructed using the location updates of a social networking app and simulating SPDT diffusion processes. Our results show that, in comparison to current graph models that only include direct links, our graph model's indirect links match empirical network properties and diffusion dynamics much more closely.   
}
{Social networks, network modelling, diffusion processes, epidemiology}
%%%% If classification number provided then
\\

\end{abstract}
\section{Introduction}
Graph modelling is an efficient method to understand and explore behaviours of diffusion processes in networked systems of individuals. In addition, modelling contact graphs is an attractive research topic due to its wide applications, from disease modelling to viral marketing~\cite{keeling2008modeling,moon2019spatio,nagatani2019epidemic,kovacs2019network, sekamatte2019individual}. Graph models represent individuals as nodes with interactions among individuals as links or edges~\cite{porter2016network,rushton2019transmission, xu2017synthetic, zhang2016modelling,starnini2013modeling}. A key function of graph models is to generate contact networks capable of characterising diffusion behaviours. In diffusion processes, infectious items  appear at a node or a group of nodes and then spread further through inter-node interactions (edges). The inter-node interactions create local (node level) transmission links of infectious items and a series of such local transmissions spread the infectious items throughout the network. This paper focuses on dynamic time varying graphs~\cite{shahzamal2017airborne,shahzamaROS,jacquet2010information}, which are more realistic though more challenging for capturing the temporal features such as contact frequencies or inter-contact times. 

Most current graph models assume that both infected and susceptible individuals are present together in a physical or virtual space to make an individual level transmission of infectious items~\cite{jacquet2010information,shahzamal2016smartphones}. These graph models implicitly assume diffusion processes where the only underlying transmission contact networks are created due to the direct interactions between individuals, i.e. presence at the same physical or virtual locations. However, there are many diffusion scenarios where susceptible individuals can receive infectious items in the absence of an infected individual due to delayed interactions with infectious items~\cite{han2014risk, sze2010review,shahzamal2017airborne,richardson2015beyond}. 
We refer to diffusion processes with indirect transmissions as SPDT (same place different time) diffusion and to diffusion processes with the direct transmission links as SPST (same place same time) diffusion. The enhancement in diffusion dynamics by the SPDT model over SPST models has been reported in our earlier work~\cite{shahzamal2017airborne,shahzamal2018impact,shahzamaROS}. One of the key properties of SPDT diffusion is that nodes in this model can create transmission links at multiple locations simultaneously: a direct transmission link at the current location and indirect transmission links at previous locations. Current graph models only consider that nodes can create links with other nodes at a single location. The other key properties of SPDT diffusion is that the time lag between the arrival of susceptible individuals and departure of infected individuals along with the duration of a link is crucial, as these temporal features determine the likelihood of transmission. Thus, a graph model that can explicitly generate these temporal features associated with indirect links is essential for modelling SPDT diffusion. 

In this paper, we propose  a generative SPDT graph model that supports both direct and indirect links, and that explicitly captures the temporal features of these links for more realistic diffusion modelling. The SPDT graph model adopts the principal of activity driven time varying networks (ADN) modelling to generate contact networks with both direct and indirect links~\cite{perra2012activity,sun2015contrasting,ogura2019optimal}. In the basic ADN model, a node is activated with a probability defined by the activation potential at each time step and create $m$ links with other nodes regardless of their activation status. The activation potential of the  node is often assigned according to a power-law distribution. At each time step, links among nodes are generated independently. In our graph model, the activation of a node means arrival of an individual at a location where at least one other individual is present. To model realistic scenarios where invididuals can visit a location of a period of time, we introduce the concept of active period that is defined as the consecutive range of time steps during which the individual is capable of spreading the infectious items to others.  In order to represent both direct and indirect transmission opportunities, an active copy of a node is created for each active period of the node. In the proposed model, links are created between the active copy of the host node and neighbour nodes. The active copy survives for the active period, when the host is present at the interaction location, in addition to an indirect transmission period, when the host leaves the location yet the contagious items persist at the location. In the SPDT graph, an active copy creates links at a time step within the active period and indirect transmission period. Thus, the SPDT graph evolves according to temporal changes of link and node status. Interestingly, the SPDT graph model supports multiple simultaneous copies of a node in the network that are active, where one copy is aligned the current location of the node, and any other copies represent the ongoing indirect transmission periods from recently visited locations of the same node. 

Our key contribution in this paper is the design of a graph generation method based on the definition of the SPDT graph model in which we define six co-location interaction parameters (CIP), capturing temporal dynamics and contact properties, to generate the transmission contact networks by the SPDT graph model. We use simple statistical distributions such as the geometric distribution to generate co-location interaction parameters. Various algorithms are developed using maximum likelihood estimation methods to fit model parameters with that of the networks constructed using the location updates of location based social networking app Momo. The model is validated by analysing network properties of the generated contact networks and by the capability of simulating SPDT diffusion processes. The diffusion dynamics are compared with that of the ADN model and SPST models. Our paper is organised as follows:

\begin{itemize}
\item We introduce a generative SPDT  graph model to include indirect transmission links with the direct transmission links and defined contact network generation methods in Section 2.

\item We develop data-driven algorithms to fit the proposed graph model with a  real dynamic contact graph and the estimated model parameters using a Momo data set in Section 3.

\item We analyse how the proposed model with active copies of nodes capture network properties of real world SPDT graph and simulated SPDT diffusion dynamics realistically, and we conduct a sensitivity analysis of the model's parameters in Section 4.
\end{itemize}

%%%%%%%%% SECTION-2 %%%%%%%%%%%%%%%%%%%%%%%%%
\section{SPDT graph model}

The SPDT graph model is developed using  the definition of SPDT diffusion. Before introducing the graph model, we recall briefly this diffusion model as introduced in~\cite{shahzamaROS}. In the SPDT diffusion model, an SPDT link is created through location and time, and may have two components: a direct transmission link and/or an indirect transmission link. SPDT links require the visits of an infected individual to the locations where at least one susceptible individual is present. Visits of an infected individual to locations where no susceptible individuals are present do not lead to transmission of disease. We thus first explain the link creation procedures in the SPDT diffusion processes. We then define the SPDT graph model by following contact network generation methods under this graph definition.

\subsection{Modelling Scenarios}
\begin{figure}[h!]
    \centering
    \includegraphics[width=0.80\linewidth, height=8.0 cm]{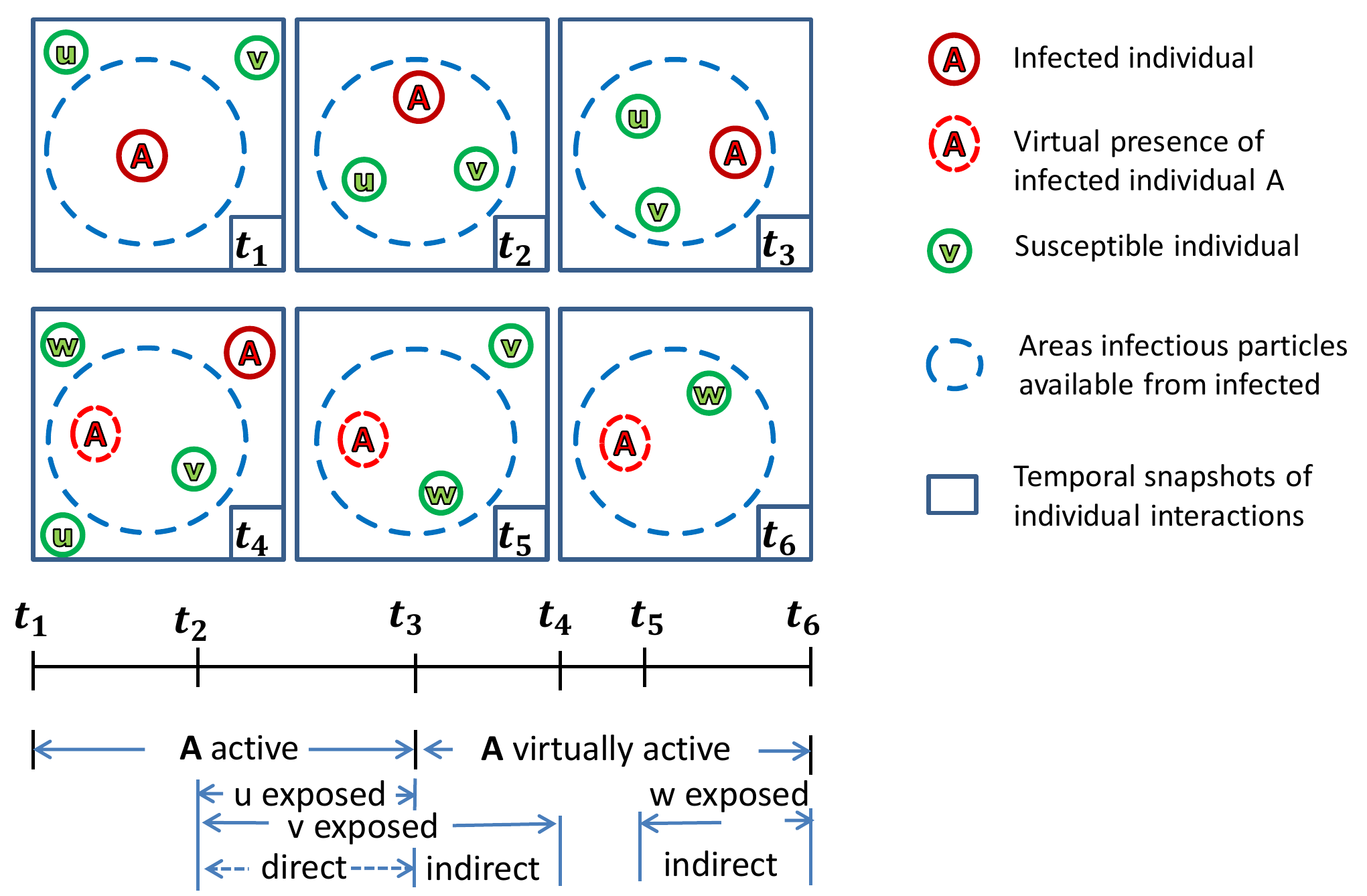}
    \caption{\cite{shahzamaROS} Disease transmission links creation for co-located interactions among individuals in SPDT model. The upper part shows the six snapshots of interactions over time and the lower part shows the periods of exposure through direct and indirect interactions. Susceptible individuals are linked with the infected individual if they enter the blue dashed circle areas within which infectious particles are available to cause infection}
    \label{fig:spdtp}
\end{figure}

The link creation procedure for the SPDT model can be explained by airborne disease spreading phenomena as shown in Figure~\ref{fig:spdtp}. In this particular scenario, an infected individual A (host individual) arrives at location L at time $t_{1}$ followed by the arrival of susceptible individuals u and v at time $t_{2}$. The appearance of v at L creates a directed link for transmitting infectious particles from A to v and lasts until time $t_{4}$ making direct contact during $[t_{2},t_{3}]$ and indirect contact during $[t_{3},t_{4}]$ (see Figure~\ref{fig:spdtp}). The indirect contact is created as the impact of A still persists (as the virtual presence of A is shown by the dashed circle surrounding A) after it left L at time $t_3$, due to the survival of the airborne infectious particles in the air at L. The appearance of u has only created direct links from A to u during $[t_{2},t_{3}]$. Another susceptible individual w arrives at location L at time $t_{5}$ and a link is created from A to w through the indirect contact due to A's infectious particles still being active at L. However, the time difference between $t_5$ (arrival time of w) and $t_4$ (departure time of A) should at most $\delta$ seconds, defined the maximum time that infectious particles can infect another person after A left at $t_4$.

\subsection{Graph definition}
\begin{figure}[h!]
\centering
\includegraphics[width=0.67\linewidth, height=3.5cm]{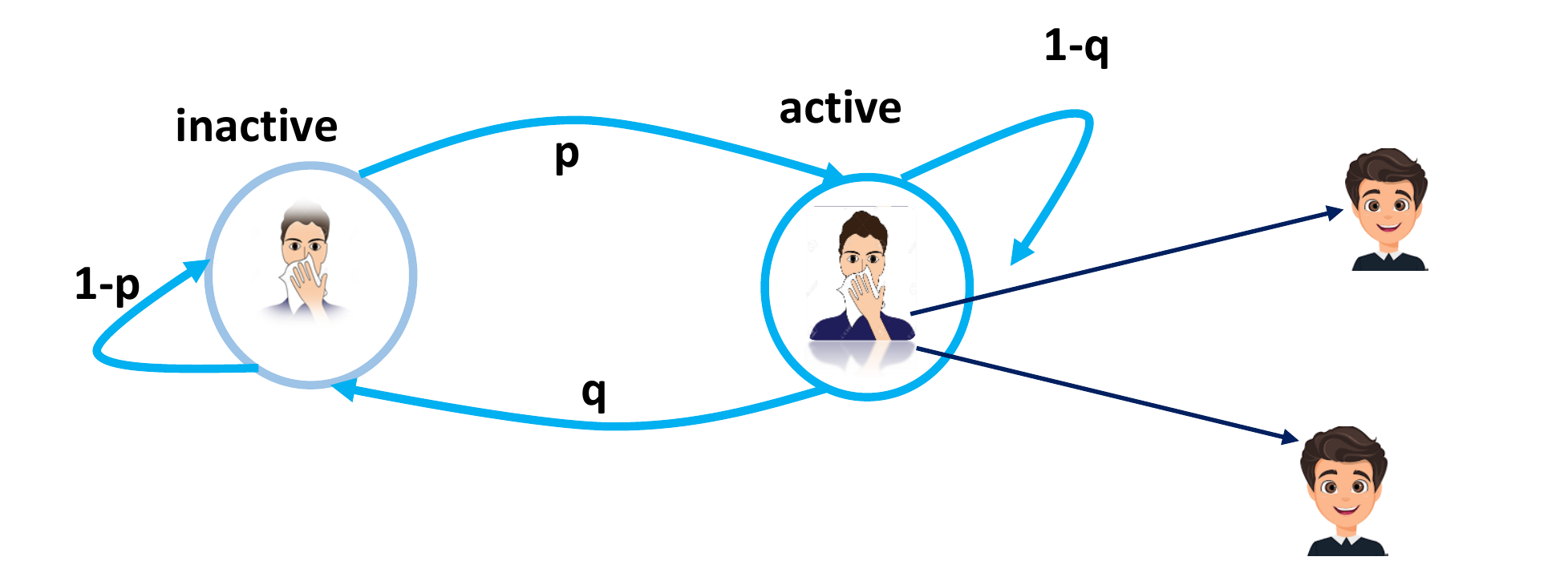}
\caption{Node's states switch with the transitional probabilities p and q. When a node is at inactive state, it switches to active state with a probability p while the probability of remaining in the inactive state is 1-p. Similarly, a node switches from active to inactive state with a probability q and remain in the active state with a probability 1-q}
\label{fig:ndswitch}

\vspace{1 em}
\includegraphics[width=0.9\linewidth, height=2.5cm]{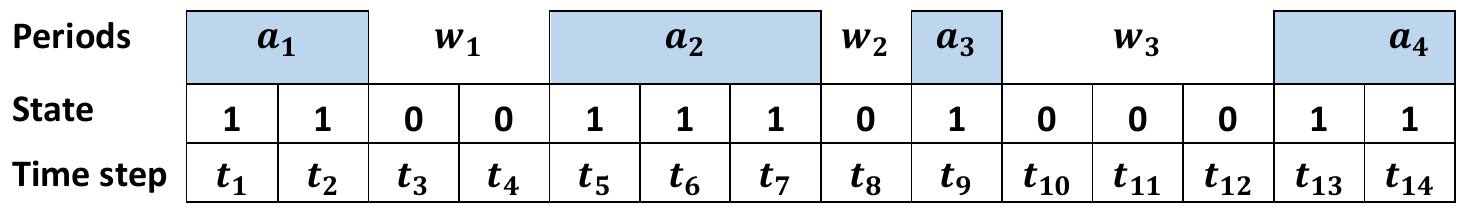}
\caption{State transitions of a node over time and how active and inactive periods are formed with the underlying states}
\label{fig:states}
\vspace{1 em}
\includegraphics[width=0.90\linewidth, height=2.5cm]{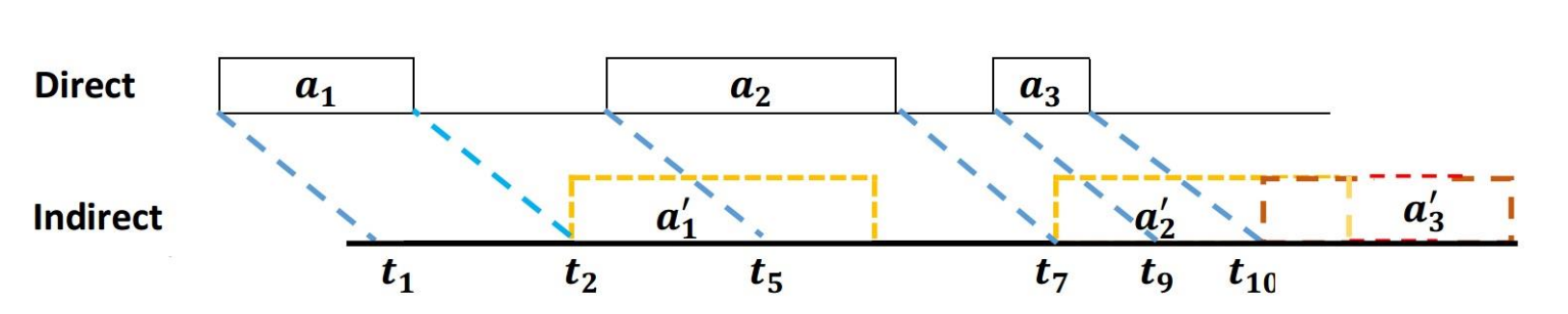}
\caption{Direct and indirect transmission periods of a node over time. Indirect transmission periods are formed following a direct period and can be concurrent with other direct and indirect periods}
\label{fig:timing}

\vspace{2 em}
\includegraphics[width=0.80\linewidth, height=3.2cm]{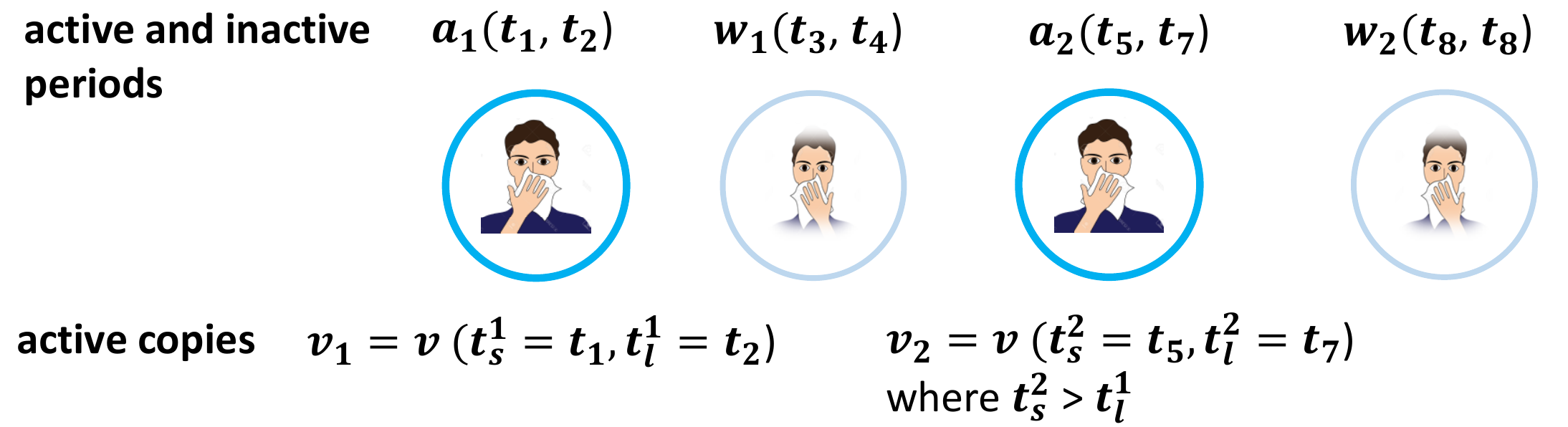}
\caption{Creation of active node copies corresponding to each active period. The active copies of a node are identified by the active period time and survive for active period duration plus $\delta$ time that is the duration when indirect links are created with active copy}
\label{fig:ndcopy}

\end{figure}

The main goal of the SPDT graph model is to support  indirect transmission links along with direct links for co-located interactions among individuals. The SPDT graph model focuses on link creations in the time domain to ensure scalability. Temporal modelling is sufficient to identify the nodes participating in possible disease transmission links. Accordingly, the link creation events in the proposed scenario can be represented as a process where an infected node activates for a period of time (staying at a location where susceptible nodes are present) and creates SPDT links. Then, the infected nodes become inactive for a period during which it does not create SPDT links. Inactive periods represent the waiting time between two active periods. Thus, the co-located interaction status of an infected node during an observation period can be summarised by a set $\{a_1,w_1,a_2,w_2,..\}$ where $a_i$ is active period and $w_i$ is inactive period (see Fig.~\ref{fig:states}).

The proposed SPDT graph is defined as $G_T=(Z, A, L, T)$ to represent all possible disease transmission links among nodes, where $Z$ is the set of nodes. The number of nodes in the graph is constant; however, nodes may have one or more active copies in the graph which captures their ability to spread diseases both at locations at which they are present and at locations from which they recently departed. The set of active copies of nodes is represented by $A$ and is dynamically updated as new copies are created and old ones expired. $L$ is the set of links in the graph. The graph is represented over a discrete time set T=$\{t_1,t_2,\ldots t_z \}$. Each node in the graph has a set of active and inactive periods $\{a_1,w_1,a_2,w_2 \ldots\}$. An active copy $v_{i}=v(t_s^{i},t_l^{i})$ is defined for an active period $a_i$ of node $v$, where $a_i$ starts at time step $t_s$ and finishes at $t_l$ (see Fig.\ref{fig:timing}). Thus, each node will have several such temporal copies for the observation period. For an active copy $v_{i+1}$ of a node $v$, $t_{s}^{i+1}$ should be greater than $t_{l}^{i}$ of $v_i$ to capture the requirement that a node should have left the first location before arriving in another location. In this graph, a link $e_{vu}\in L$ is defined between an active copy $v_i$ of host node $v$ and neighbour node $u$ (node $u$ visits the current or recent location of node $v$) as $e_{vu}=(v_i,u,t_{s}^{\prime},t_{l}^{\prime})$ where $t_s^{\prime}$ is the joining time and $t_l^{\prime}$ is departure time of $u$ from the interacting location. The value of $t_s^{\prime}$ should be within $t_s^{i}$ and $t_l^{i}+\delta$ , as an active copy $v_i$ of a node v expires after $t_l^{i}+\delta$ seconds. During the indirect transmission period $\delta$, node $v$ can start another active period at another location (see Fig.\ref{fig:timing}). However, if the infected node $v$ leaves and returns to the location of $u$ within a time period $\delta$, then there will be two active copies of $v$, each with a link to the susceptible node $u$. The first copy is due to the persistent of particles from $v$'s last visit, while the second copy is due to $v$'s current visit. 

The evolution of the proposed graph is governed by two dynamic processes: 1) switching of nodes between active and inactive states, as in Figures~\ref{fig:ndswitch} and~\ref{fig:states}; and 2) link creation and deletion for active copies of nodes. As stay times at locations are not fixed, a transition probability $\rho$ is defined to determine switching from active state to inactive state (modelling stay and departure events of a node at a location). This induces variable-length active periods. Similarly, another transition probability $q$ is defined to determine when a node switches from an inactive to an active state (modelling arrival of a node at location). A similar approach is taken to define link update dynamics in the graph. An active copy of a node creates a link to a newly arriving neighbour node with probability $p_{c}$ at each time step until the active copy expires. During an activation period, multiple neighbours can arrive at interaction locations. Thus, an activation degree probability $P(d)$ is defined to model the arrival of multiple new neighbours for an active copy. The created links break (neighbour node leaves the interaction area) with probability $p_{b}$ at each time step. 

\subsection{Network generation}
Based on the above SPDT graph model definition, dynamic contact networks among a set of nodes can be generated in various ways ranging from homogeneous to complex heterogeneous scenarios. The simplest homogeneous version of a model often provides useful insights about the studied processes, but cannot capture higher resolution interactions  of the actual system. On the other hand, the heterogeneous system is more representative at the cost of added complexity. Our proposed network generation process combines both approaches. To generate the SPDT contact network, we first use geometric distributions (because of their computational simplicity), and second a power-law distribution is added to allow heterogeneity in the generated networks. The SPDT graph contains a fixed number $N$ of nodes interacting with each other over the observation period $[0,T]$. The graph is implemented over discrete time with time step $\Delta t$. The link generation procedure in the SPDT graph model is implemented through five co-location interaction parameters (CIP) namely: active periods($t_a$), waiting periods($t_w$), activation degree(d), link creation delay ($t_c$) and link duration($t_d$). The generation procedures are as follows:

\subsubsection{Node activation}
Active copies of nodes are the key building blocks for constructing an SPDT graph. The active node copies are created over time according to the active periods. Thus, it is first required to generate active periods and intervening inactive periods. In the proposed model, determining whether a node will stay in the current state or transit into the other state at the next time step resembles a Bernoulli process with two outcomes. Thus, the number of time steps a node stays in a state can be obtained from a geometric distribution. With the transitional probability $\rho$ of switching from active to inactive state, the active period duration $t_a$ can be drawn from the following distribution as:
\begin{equation}\label{aprds}
Pr(t_{a}=t)=\rho (1-\rho)^{t-1}
\end{equation}
where $t=\{1,2, \ldots \}$ are the number of time steps. Similarly, the inactive period duration, $t_w$, with the transition probability $q$ can be drawn from the following distribution as:
\begin{equation}\label{iaprds}
Pr(t_{w}=t)=q (1-q)^{t-1}
\end{equation}
where $t=\{1,2, \ldots \}$ are the number of time steps. 

Knowing the active and inactive period duration, the graph generation process requires the initial states of nodes to generate the active copies of nodes. The model's state dynamics can be described with a two state Markov-process with transition matrix
\begin{equation*}
P=\begin{bmatrix}
q & 1-q\\
\rho & 1-\rho
\end{bmatrix}
\end{equation*}
for which the equilibrium probabilities that the node is in inactive state and active state are $\pi_0$ and $\pi_1$ respectively, where
\begin{equation}\label{son}
\pi_{0}=\frac{\rho }{q+\rho}
%\end{equation}
\mbox{\ \ \ \ \  and \ \ \ \ \ }
%\begin{equation}\label{soff}
\pi_{1}=\frac{q}{q+\rho}
\end{equation}
If the initial state of node $v$ is active, the first active copy $v_1$ is created for the time interval $(t_s^{1}=0,t_l^{1}=t_a)$. Otherwise, $v_1$ will be created for the interval $(t_s^{1}=t_w,t_l^{1}=t_w+t_a)$. Active copy creation continues over the observation period and the corresponding interval $(t_s,t_l)$ is defined according to the drawn $t_a$ and $t_w$. Active copies are generated for each node independently. The values of $\rho$ and $q$ are fitted with real data.

\subsubsection{Activation degree}
The next step in the graph generation process is to define interactions of neighbour nodes with an active copy. Multiple neighbour nodes can contact with an active copy. The number of neighbour nodes interacting with an active copy is noted activation degree $d$. The value of $d$ depends on the spatio-temporal dynamics of the graph and are drawn from a geometric distribution (Eq.~\ref{eq:actdgr}) instead of finding the arrival times of neighbour nodes separately.  
\begin{equation} \label{eq:actdgr}
Pr\left (d=k\right )=\left(1-\lambda\right)\lambda^{k-1}
\end{equation}
where $k=\{1,2,\ldots \}$ are the number of neighbour nodes of an active copy and $\lambda$ is a scaling parameter. If the same parameter $\lambda $ is assigned to all nodes in the network, it is a homogeneous contact network. However, individuals in the real world contact scenarios have heterogeneous propensity to interact with other individuals. This heterogeneity is incorporated into the model by selecting $\lambda$ from a power law distribution of Equation~\ref{accesb}:
\begin{equation} \label{accesb}
f\left( \lambda_{i}=x\right)= \frac{\alpha x^{-(\alpha+1)}}{\xi^{-\alpha}-\psi^{-\alpha}}
\end{equation}
where $\alpha$ is the scaling parameter, $\xi$ is the lower limit of $\lambda_i$ and $\psi$ is the upper limit which is approximately 1. The value of $\lambda_{i}$ defines the range of variations of $d$ for active copies of a node $i$ and Equation~\ref{eq:actdgr} ensures wide ranges for large values of $\lambda$. Combining geometric and power law distributions is known to generate more realistic degree distribution~\cite{chattopadhyay2014fitting} as is demonstrated in the model fitting section. Given the degree distribution, a link creation delay $t_c$, time gap between arrivals of host node and neighbour node $(t_s-t_{s}^{\prime})$, and a link duration $t_d$, the stay time of neighbour at the interacted location $(t_s^{\prime}-t_{l}^{\prime})$, are assigned to each link. 

\subsubsection{Link creation}
With the activation neighbour set, the graph generation process defines the arrival and departure dynamics of neighbour nodes for each link created with an active copy. A similar approach to the definition of active and inactive periods creation with transition probabilities is adapted. Here the assumption is that each link is created with probability $p_{c}$ at each time step during the life period $(t_s,t_{l}+\delta)$ of an active copy and is broken with probability $p_{b}$ after creation. For the link creation delay $t_c$, the time gap between arrivals of host node and neighbour node $(t_s-t_{s}^{\prime})$, the truncated geometric distribution is used: 
\begin{equation} \label{ldelay}
P\left (t_{c}=t\right )=\frac{p_{c} \left(1-p_{c}\right)^{t}}{1 -(1-p_{c})^{t_{a}+\delta}} 
\end{equation}
where $t=\{0, 1,2, \ldots, t_a+\delta \}$ are the number of time steps and $t_a$ is the active period duration of corresponding active copy. Truncation ensures that links are created within $t_{l}+\delta$, i.e. before the active copy expires. In contrast, the link duration $t_{d}$, the stay time of neighbour at the interacted location, does not have a specific upper bound and is generated for each link upon creation through a geometric distribution: 
\begin{equation} \label{ldur}
P\left(t_{d}=t\right)=p_{b} \left(1-p_{b} \right)^{t-1}
%P\left ( t_{d}=t \right )=\rho \left ( 1-\roh \right )^{t-1}
\end{equation}
where $t=\{1,2, \ldots \}$ are the number of time steps. For simplicity, $p_b$ is set to $\rho$ as both probabilities relate to how long nodes stay at a location. Each link with an active copy, thus, has timing characteristics as $t_{s}^{\prime}=t_s+t_c$ and $t_{l}^{\prime}=t_s^{\prime}+t_d$. A link with $t_s^{\prime}\geq t_l$ is an indirect transmission only component. Link can also have indirect component if $t_s^{\prime} < t_l$ and $t_l > t_l^{\prime}$. The above graph generation steps capture the temporal behaviour of SPDT links. The social mixing patterns are integrated by selecting the neighbouring node, as it is described below.

\subsubsection{Social structure}
The underlying social structure of the generated graph depends on the selection of a neighbour node for each link. The simplest way of selecting the neighbour node is to pick a node randomly. However, social network analysis has shown that the neighbour selection for creating a link follows a memory-based process. Thus, the reinforcement process~\cite{karsai2014time} is applied to realistically capture the repeated interactions between individuals. In this process, a neighbour node from the set of already contacted nodes is selected with probability $P(n_{t}+1)=n_{t}/(n_{t}+\eta)$ where $n_{t}$ is the number of nodes the host node already contacted up to this time $t$. On the other hand, a new neighbour node is selected with the probability $1-P(n_t+\eta)$. Here, the size of the contact set, the number of nodes a node contact over the observation period, depends on the $\lambda$. This is because the system forces a host node to select a new neighbour when neighbours from the current contact set are already selected during an active period even if $P(n_t+1)$ is true. On the other hand, a new node is selected as a neighbour with the probability $1-P(n_t+1)$. To maintain local clustering among nodes, $\mu$ proportion new neighbour is selected from the neighbours of neighbour nodes. In addition, when a node $j$ is chosen as a new neighbour by node $i$ in the heterogeneous networks, it is selected with the probability proportional to its $\lambda_j$ as nodes with higher $\lambda$ will be neighbours to the more nodes~\cite{alessandretti2017random}. This ensures nodes with higher potential to create links also have a higher potential to receive links.

%%%%%%%%%%%% SECTION-3 %%%%%%%%%%%%%%%%%%%%%
\section{Materials and Methods}
The performance of the proposed graph model is studied by comparing the network properties of the generated networks with the networks properties of the empirical SPDT graph and analysing its ability to simulate SPDT diffusion. The SPDT model requires appropriate model parameters to generate contact networks. We have used empirical GPS location updates from Shanghai city to estimate the model parameters and location updates from Beijing city for evaluating the model, based on the dataset described below.

\subsection{Data set}

This study applies location update information from users of a social discovery network \emph{Momo}\footnote{https://www.immomo.com}. The Momo App enables users to interact with nearby users by sharing their current locations. Whenever a user launches the Momo app, the current location is forwarded to the Momo server. The server sends back the latest location updates of all users nearby. These location updates have been previously collected by the authors of~\cite{thilakarathna2016deep} using a set of network API communicating with Momo server. The API retrieved location updates every 15 minutes over a period of 71 days (from May to October 2012). The data set contains 356 million location updates from about 6 million Momo users around the world, but primarily in China. Each database entry includes GPS coordinates of the location, time of update and user ID. For this study, the updates from Shanghai and Beijing are separated as they are the cities with the highest number of updates for the period of 32 days from 17 September, 2012 to 19 October, 2012. This data contains almost 56 million location updates from 0.6 million users. 

All possible disease transmission links according to the SPDT diffusion model definition are extracted from the location updates of Momo users. To create an SPDT link between a host user (assumed infected with disease) $v$ and a neighbour user $u$ (susceptible user), it is required to find the arrival times ($t_s, t_s^{\prime}$) and departure times ($t_l, t_l^{\prime}$) of two users. As the first step, it is identified that an infected host user $v$ is staying at a location. Consecutive updates, $X=\{(x_{1},t_{1}),(x_{2},t_{2}),\ldots (x_{k},t_{k})\}$ where $x_{i}$ are the co-ordinate values and $t_{i}$ are the update times, from a user $v$ within a radius of 20m (travel distance of airborne infection particles~\cite{han2014risk}) of the initial update's location $x_{1}$ are indicative of the user staying within the same proximity of $x_1$. A threshold is set for time difference of any two updates to 30 minutes to make sure an infected host remain within the same proximity, as longer gaps may indicate a data gap in the user pattern. Then, the central co-ordinate in the update $X$ is searched where the distances from each update to all other updates are added together and the update $x_c$ with the minimum sum is selected as central co-ordinate. For the host user $v$, its visit to the proximity of $x_{c}$ will represent a valid visit if a susceptible user $u$ has location updates starting at $t^{'}_{1}$ while $v$ is present, or within $\delta$ seconds after $v$ leaves the area. The user $u$ should have at least two updates within 20m of $x_{c}$ to be valid to ensure that it is in fact staying at the same proximity, and therefore can be exposed to the infected particles, rather than simply passing by. The stay period of host user $v$ at the proximity of $x_c$ is ($t_{l}=t_{k}, t_s=t_{1}$), where $t_k$ represents the end of the current stay period. If $u$'s last update within 20m around $x_{c}$ is $(x^{'}_{j},t^{'}_{j})$, the created SPDT link has a link duration ($t_l^{\prime}=t^{\prime}_{j},t_s^{\prime}=t^{'}_{1}$) due to active visit ($t_{l}=t_{k}, t_s=t_{1}$). All links to other users for this active visit ($t_{l}=t_{k}, t_s=t_{1}$) are computed. Similarly, all visits made by $v$ are searched over the updates of 32 days and SPDT links are extracted. This process is executed for all users present in the data set to find all possible SPDT disease transmission links and provide a real contact network with SPDT links among users. An SPDT link is noted as $e_{vu}=\left(v(t_s,t_l),t_s^{\prime},t_l^{\prime}\right)$ which means that a user $v$ visit a location during $(t_s,t_l)$ where another user $u$ is present for the duration $(t_s^{\prime},t_l^{\prime})$ where $t_s^{\prime} \geq t_s$. Each link between the two same users are distinguished by the time intervals $(t_s,t_l)$ and $(t_s^{\prime},t_l^{\prime}
)$ as there can be multiple links to a neighbour for the same stay interval $(t_s,t_l)$ of the host user. 

The above process is executed for all host users present in the data set to find all possible SPDT disease transmission links and provide a SPDT contact network of 338K users. This network includes possible direct and indirect transmission links due to direct and indirect co-location interactions among users. However, users appear in the system for on average 3-4 days and then disappear for the remainder of the simulation period. Thus, the link density in the network is sparse and the network is called Sparse SPDT network (SDT). In this sparse SPDT network, infected individuals cannot apply their full potential to infect other individuals due to absence from the networks and thus underestimate diffusion dynamics. Thus, we reconstruct a dense SPDT network (DDT) from this network repeating the links from the available days of a user to the missing days for that user~\cite{stehle2011simulation,toth2015role}. If a user has links for multiple days, a day will be randomly chosen and will be copied to a randomly chosen  day where that user has no links. The resulting network is used as the empirical contact network that produces realistic diffusion dynamics. The DDT network is used for simulating SPDT diffusion.

\subsection{Model parameters estimation}
 In the SPDT graph model, contact networks are generated based on the five co-located interaction parameters (CIP). We collect the CIP parameters from real contact networks which are created using locations updates of Momo users. Then, maximum likelihood estimation methods are applied to find the model parameters. 

\subsubsection{CIP data selection}
We select the extracted links of seven days from Shanghai city to estimate model parameters. The seven-day data contains 2.53 million active periods from 126K users. The distribution of active period durations $t_a$ exhibits heterogeneous behaviour and is captured by long tail distributions. Then, the waiting periods $t_w$ are collected. However, the waiting period collection is not straight forward like $t_a$. This is because Momo users do not use the app regularly and many potential visits are not captured. If we assume that individuals in the system are activated at least once a day and one actives at the end of the current day after his last activation at the beginning of previous days, the maximum waiting period will be up to 2 days. A new co-location interaction parameter called activation frequency $h$ is defined which measures the rate of activation of nodes to avoid the irregularity in the waiting period. This is measured with the number of active periods created by a Momo user in a day. The activation frequencies $h$ are collected for all users over each day during the selected seven days. This provides 0.5M activation frequencies for the 126K users. The activation frequencies with active period durations model the inactive periods. As the activation frequencies are daily behaviours of Momo users, modelling active period and waiting period based on $h$ avoids the impact of missing updates. Next, the activation degrees are collected based on the number of other Momo users visiting a location where the host user has been for the corresponding active period. Here, the users are required to visit a location within $\delta=3$ hours as it is found that particles can remain infectious for up to the four hours. The activation degree distribution follows the long-tailed distributions which are found in the many degree distributions of real systems. Then, the link creation delays $t_c$ and link duration $t_d$ are collected for all links created over the seven days. These also capture all possible lengths of $t_c$ and $t_d$ and can be described by the long tail distributions.

\subsubsection{Goodness of fit}
\begin{figure}[h!]
\centering
\includegraphics[width=0.45\linewidth, height=5.0cm]{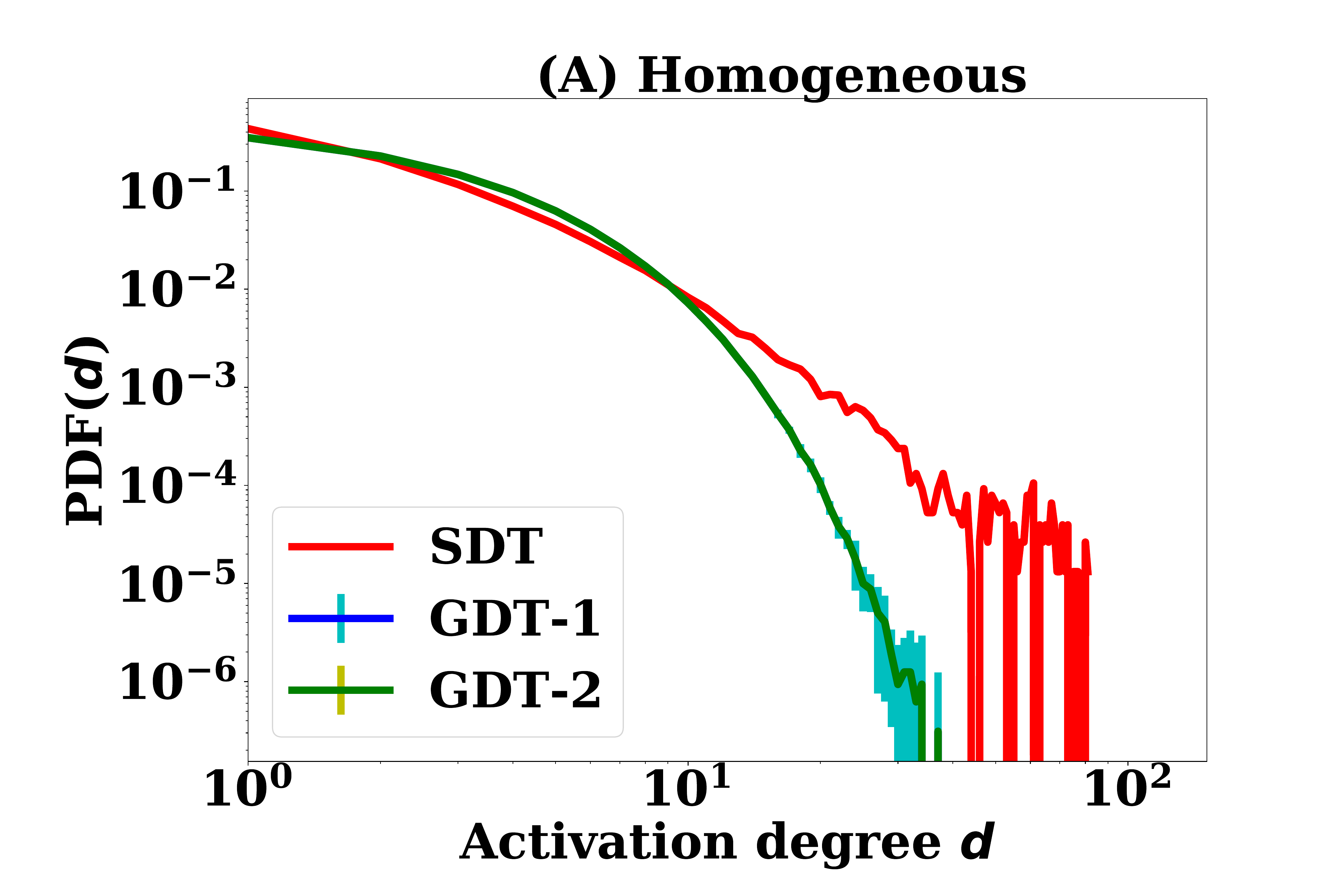}~
\includegraphics[width=0.45\linewidth, height=5.0cm]{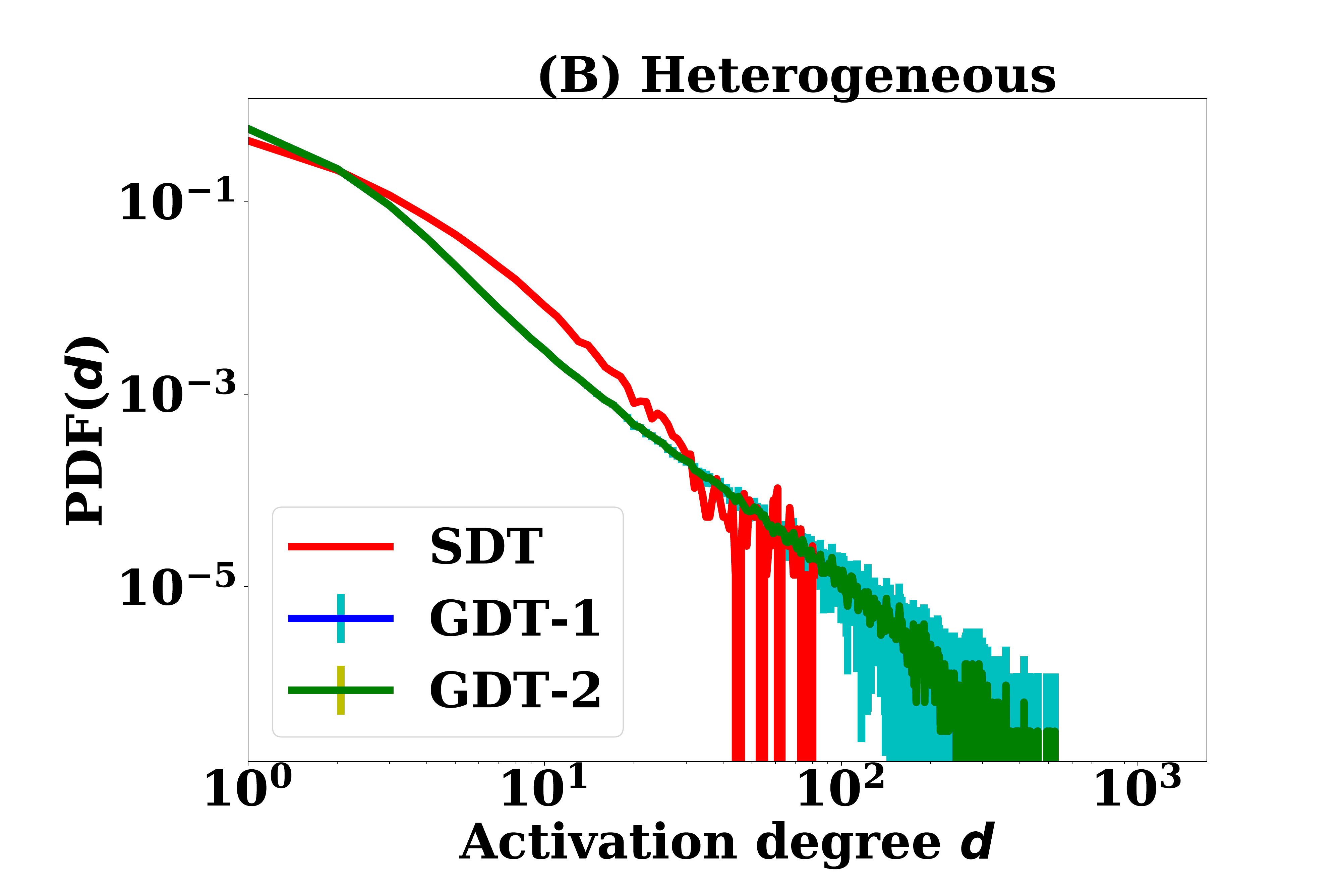}
\caption{Fitting activation degree CIP of generated networks GDT-1 and GDT-2 with DDT network: (A) nodes are homogeneous - having same contact potential and (B) nodes are heterogeneous - having different contact potential}
\label{fig:fitd}
\end{figure}

Using the developed graph model, a synthetic SPDT contact network (GDT-1, generated SPDT network) is created with 126K nodes for 7 days applying the estimated model parameters. Then, the CIP parameters of GDT-1 are compared with the parameters of real SPDT network (SDT) constructed by 7 days updates of 126K users from Beijing city. To understand the model's response across network sizes, another synthetic SPDT contact network (GDT-2) of 0.5M nodes for 7 days is generated. Results of 100 runs for each network are presented in Figures~\ref{fig:fita},~\ref{fig:fitd} and ~\ref{fig:fitl} where periods are based on time steps of $\Delta t =5$ minutes. The root squared error (RSE) is calculated between the generated and real data distributions of CIP parameters as:
\begin{equation}\label{eq:error}
RSE=\sqrt{\sum_{i=1}^{m}(x_i-y_i)^{2}}
\end{equation}
where observed values are grouped in $m$ bins as they are discrete, $x_i$ is the proportion of observations for the i$^{th}$ bin, $y_i$ is the proportion of empirical dataset values in the i$^{th}$ bin. As the RSE is computed from the proportion values, bins are naturally weighted so that bins representing larger proportions of events have higher contributions to the error calculation. Note that the bins up to the last bin of $x_i$ are considered assuming that the discarded values of $y_i$ provide very small error as the selected distributions are long-tailed. The distributions of network CIP parameters are presented in log scale with RSE error bars for GDT-1, GDT-2 and SDT in Figures~\ref{fig:fita},~\ref{fig:fitd} and~\ref{fig:fitl}.

\subsubsection{Fitting activation parameters}

\begin{figure}[h!]
\centering
\includegraphics[width=0.33\linewidth, height=4.0cm]{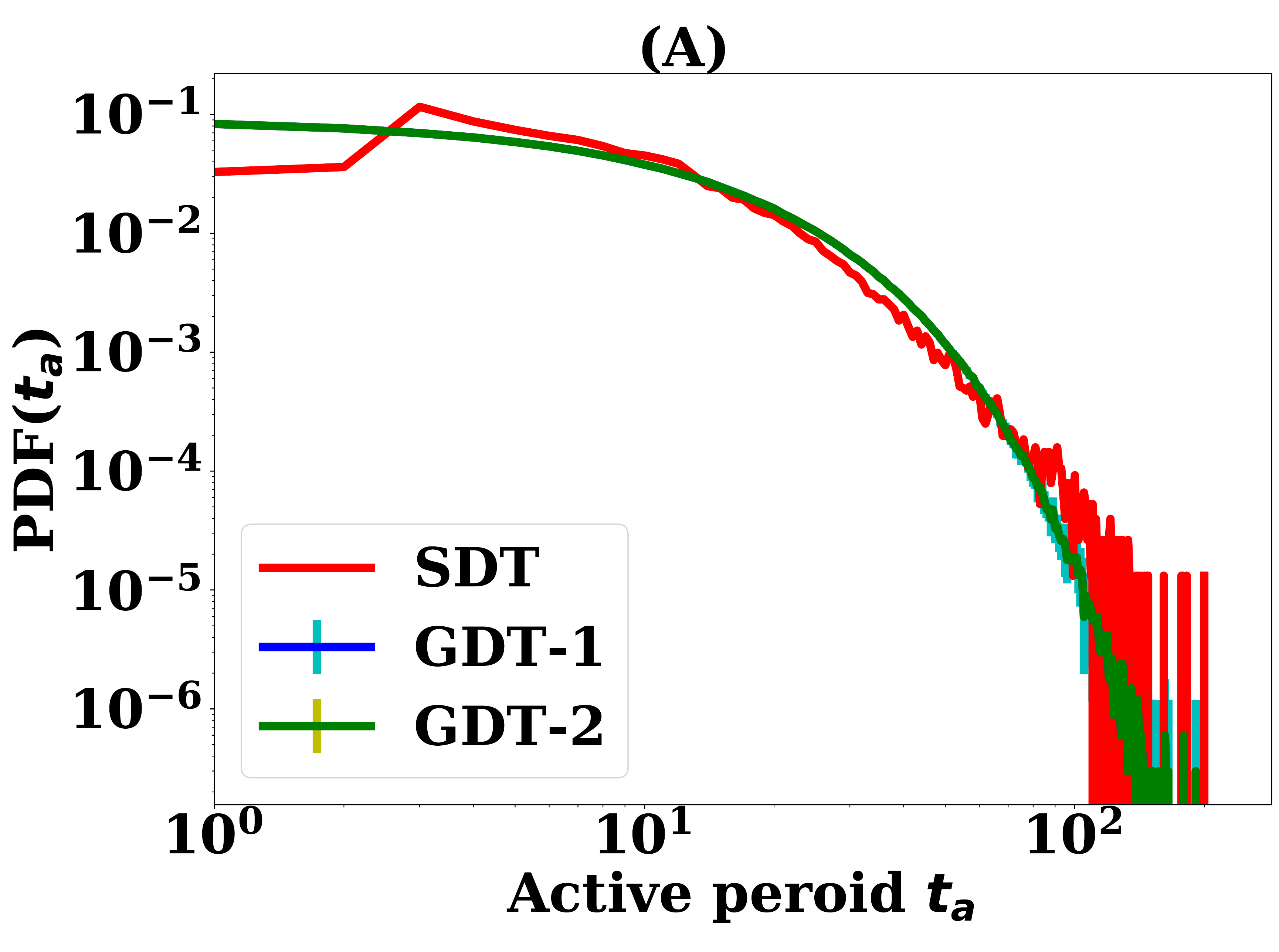}~
\includegraphics[width=0.32\linewidth, height=4.0cm]{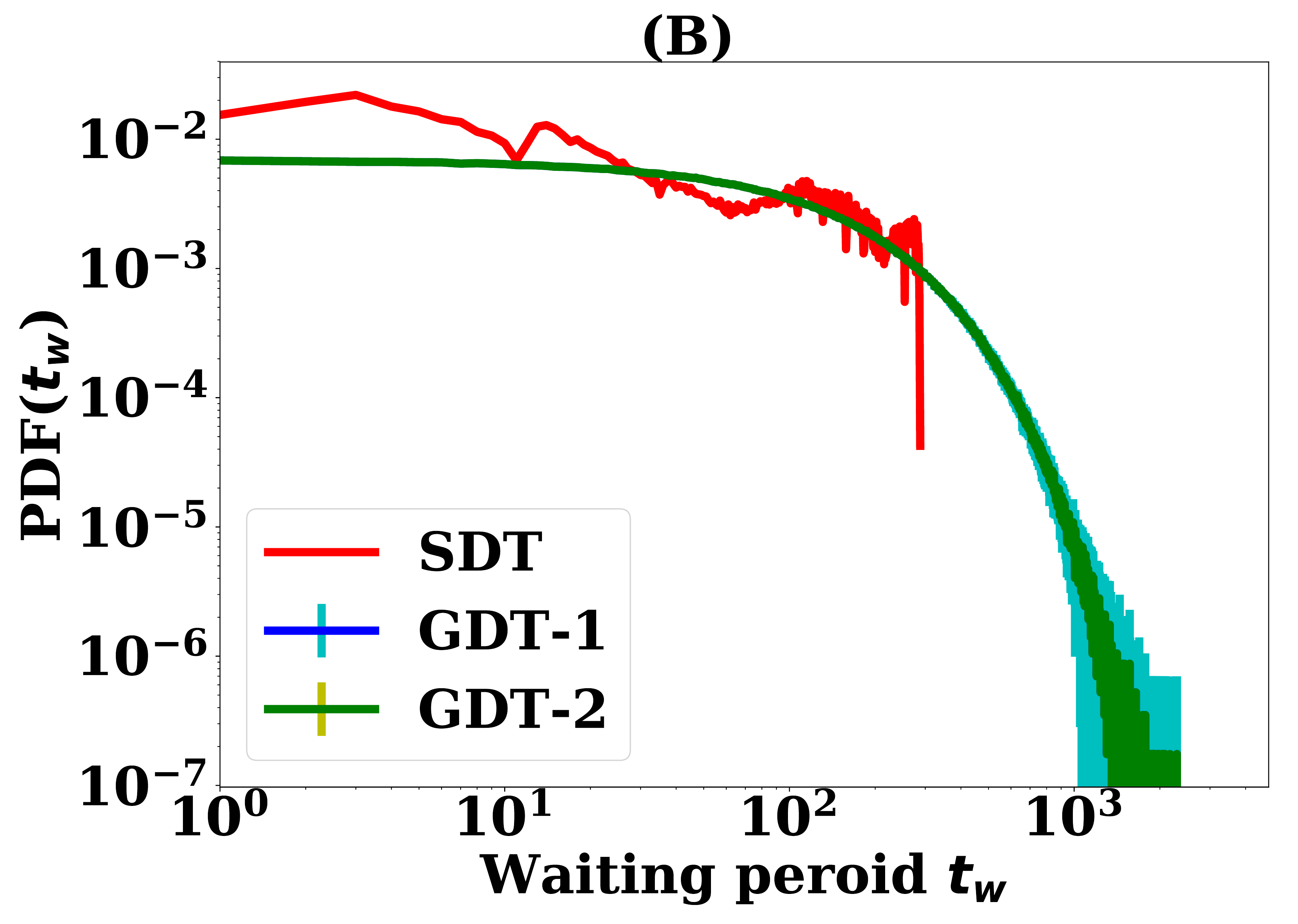}
\vspace{1em}
\includegraphics[width=0.33\linewidth, height=4.0cm]{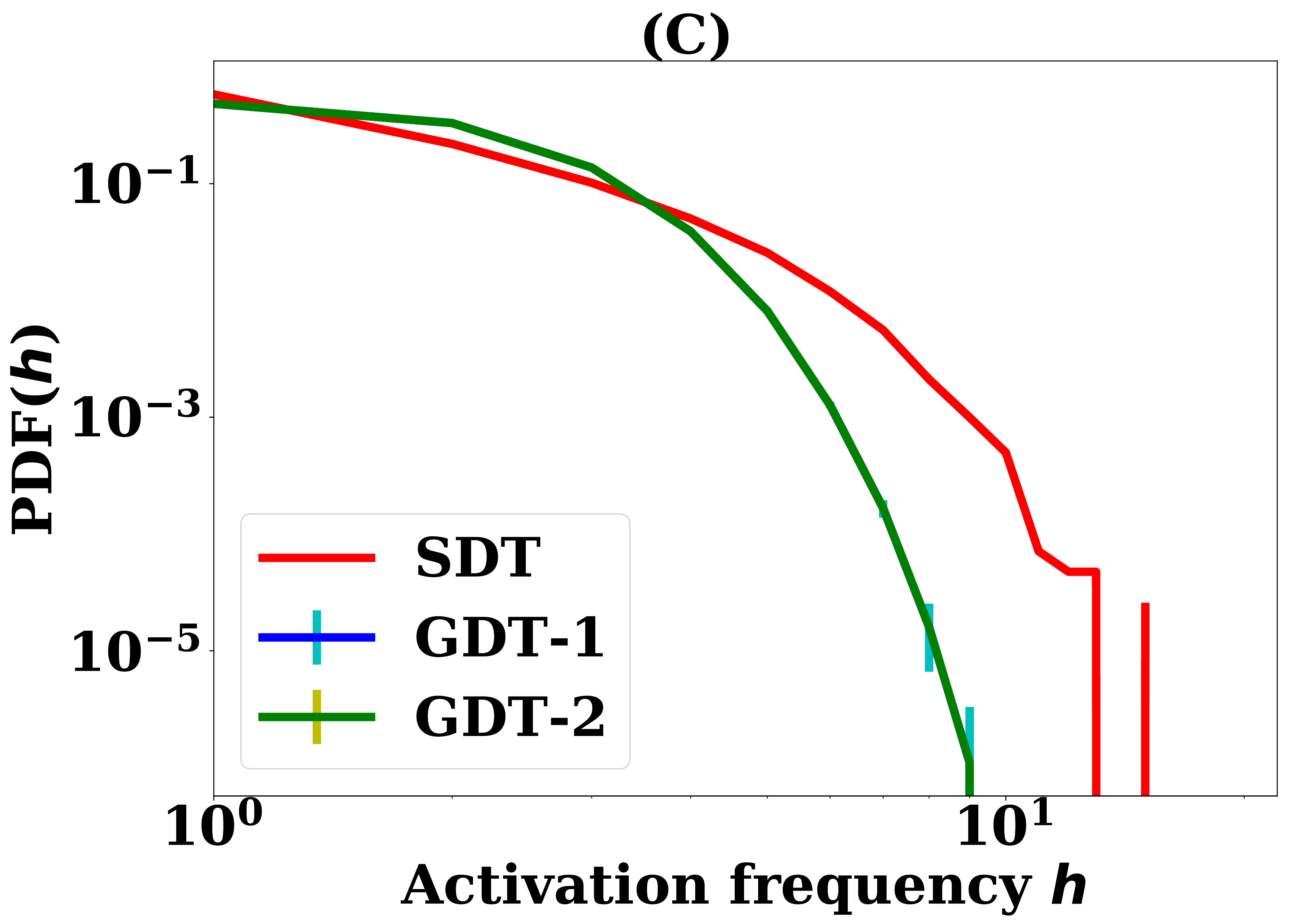}
\caption{Fitting activation CIP parameters. The CIP parameters of generated networks GDT-1, having same number of nodes of DDT network, and GDT-2, having 0.5M nodes, are compared with the CIP of DDT network: (A) active period $t_a$, (B) waiting period $t_w$ and (C) activation frequency $h$. The period is the number of time steps and $h$ is the number of activation in a day}
\label{fig:fita}
\end{figure}

In this model, the active periods $t_a$ are drawn from a geometric distribution with the scaling parameter $\rho$. The value of $\rho$ can be estimated using a sample data set of $t_a$ to the following MLE condition of geometric distribution:
\begin{equation}\label{geomle}
\hat \rho =\frac{m}{\sum_{l=0}^{m}s_{l}}
\end{equation}
where $m$ is the size of the selected sample set. About $m=2.53 M$ real activation periods made by the $126K$ momo users over the 7 days are used here. This large set of active periods contains the active durations from workdays and weekends. The above MLE Equation~\ref{geomle} estimates $\hat \rho=0.085$. The distributions of generated $t_a$ for both networks GDT-1 and GDT-2 are shown in Figure~\ref{fig:fita} A. The RSE error for GDT-1 is 0.0876 with standard deviation (STD) of 0.00041 while RSE for GDT-2 is very similar giving 0.08758 with STD of 0.00018. The model with fitted parameters consistently generate the active periods $t_a$ for nodes of both network sizes. The distribution of active periods is also supported by the findings of other studies where they found the contact duration is broadly distributed~\cite{scherrer2008description,hui2005pocket}. In these studies, contact duration means staying at a location.

The activation periods $t_a$ have similar patterns for all individuals. However, the inactivation periods, $t_w$ waiting time between two active periods, depends on how frequently individuals visit public places. Thus, the waiting periods $t_w$ are fitted based on the fitting of activation frequencies $h$. The number of transition events from inactive to active states or active to inactive states will be the activation frequency during an observation periods. According to Equation~\ref{son}, the probability of transition event $0 \rightarrow 1$ at a time step is: 
\[p_{01}=\frac{\rho q}{q+\rho}\]
Thus, the number of transition events $0 \rightarrow 1$ during $z$ time steps (here number of time steps in a day) represents the number of activation events $h$. The probability of $h$ activation for a node is given by the binomial distribution as: 
\begin{equation*}
Pr(h\mid q)=\begin{pmatrix}
z\\ 
h
\end{pmatrix}\left(\frac{\rho q}{q+\rho}\right)^{h}
\end{equation*}
The term $\frac{\rho q}{q+\rho}$ becomes small as $\rho=0.083$, as $z$ is usually large and $q<1$. Thus, the above equation can be approximated by a Poisson distribution as:
\begin{equation}
Pr(h\mid q)=\frac{\left(\frac{z \rho q}{q+\rho}\right)^ {h}e^{-\frac{z \rho q}{q+\rho}}}{h!}
\end{equation}
The MLE condition for the Equation~\ref{actf} is derived in the Appendix as 
\begin{equation}\label{actf}
\frac{qz\rho}{q+\rho}=\frac{1}{l}\sum_{i=1}^{l}h_i
\end{equation}

In the selected Momo data set, the users are not present every day. Thus, it is assumed that the number of activations made during a day is the representation of the activation frequency. As the focus is to model the waiting time with a geometric distribution, it makes sense to activate the node with variable frequency over the observation days. The day activation frequencies of $126K$ users are collected over the seven days. Applying the activation sample set $h=\{h_1,h_2,\ldots h_r\}$ of size r=300K to MLE Equation~\ref{actf} gives $\hat q =0.0048$. The generated activation frequencies for GDT-1 and GDT-2 are presented in Figure~\ref{fig:fita} B, with RSE of 0.100 and STD of 0.005 for both GDT-1 and GDT-2 compared to real activation frequencies. The plotted waiting periods in Figure~\ref{fig:fita} C also follows the distribution of real $t_w$ with RSE around 0.102 in both networks. The obtained STD of error is 0.01. The $t_w$ is characterized by the irregularity of using Momo Apps by users.

\subsubsection{Fitting activation degree}
The activation degree is drawn from the geometric distribution with the scaling parameter $\lambda$. If we assume that nodes in the network are homogeneous, $\lambda$ can be estimated using the following equation 
\begin{equation}\label{geomleac}
\hat \rho =\frac{m}{\sum_{l=0}^{m}d_{l}}
\end{equation}
The activation degree sample set $d=\{d_{1},\ldots,d_{m}\}$ of size m=518K are used to the MLE equation. The estimated value for $\hat \lambda $ is 0.32. The generated activation degree distributions for 
GDT-1 and GDT-2 and comparison with real one are presented in Figure~\ref{fig:fitd} A. The RSE error is 0.017 with STD of 0.004. The fluctuating error at the tail is due to data sparsity and has little effect on the RSE. 

When nodes in the network has heterogeneous propensity to contact the other individuals, the distribution of $d$ in the network will be given for any $\lambda$ as:
\begin{equation*}
Pr(d)=\frac{\beta }{\xi^{\beta}-1}\int_{\xi}^{1}( \lambda^{d-\beta-2} -\lambda^{d-\beta-1}) d\lambda
\end{equation*}
\begin{equation}\label{eq:dmle}
=\frac{\beta}{\xi^{-\beta} - 1}\left(\frac{1-\xi^{d-\beta-1}}{d-\beta-1}-\frac{1-\xi^{d-\beta}}{d-\beta}\right)
\end{equation}
The MLE condition for the Equation ~\ref{eq:dmle} is derived in the Appendix which provide the following two equations. The MLE condition for $\beta$ assuming $\psi \approx 1 $ is given as 
\begin{equation}\label{eq:dp}
0=\frac{m}{\beta}-\frac{m \xi^\beta \ln\xi}{\xi^\beta-1}+\sum_{k=1}^{n}\frac{\frac{\xi^{d_k-\beta -1} \ln\xi}{d_k-\beta-1}-\frac{1-\xi^{d_k-\beta -1}}{(d_k-\beta -1)^2}-\frac{\xi^{d_k-\beta} \ln\xi}{d_k-\beta}-\frac{1-\xi^{d_k-\beta}}{(d_k-\beta)^2}}{\frac{1-\xi^{d_k-\beta-1}}{d_k-\beta-1}-\frac{1-\xi^{d_k-\beta}}{d_k-\beta}}
\end{equation}
where $m$ is the length of the data set $d=\{d_{1},\ldots,d_{m}\}$. Then the lower limit of power law distribution is estimated with the following MLE equation.
\begin{equation}\label{eq:dl}
0=\frac{m \beta\xi^{-\beta -1}}{\xi^{-\beta}-1}+\sum_{k=1}^{m}\frac{\frac{(d_k-\beta)\xi^{d_k-\beta -1}}{d_k-\beta}-\frac{(d_k-\beta-1)\xi^{d_k-\beta -2}}{d_k-\beta-1}}{\frac{1-\xi^{d_k-\beta-1}}{d_k-\beta-1}-\frac{1-\xi^{d_k-\beta}}{d_k-\beta}}
\end{equation}
As the MLE equations do not have closed form solutions, the python module called fsolve is used to estimate the best value of $\xi$ and $\beta$. We assume that $\beta=2.5$  since it is known that it is in the range 2-3 in the real world network. We then estimate the value of $\xi=0.26$ using the MLE Equation~\ref{eq:dl}. Using $\xi=0.26$ in Equation~\ref{eq:dp}, the value of $\beta$ is estimated as 2.963. The generated activation degree distributions with the heterogeneous potential are presented in the Figure~\ref{fig:fitd}. The RSE error is 0.0152 with STD of 0.004. There are few values with high degree and that can happen in the real networks.

\begin{figure}[h!]
\centering
\includegraphics[width=0.45\linewidth, height=5.0cm]{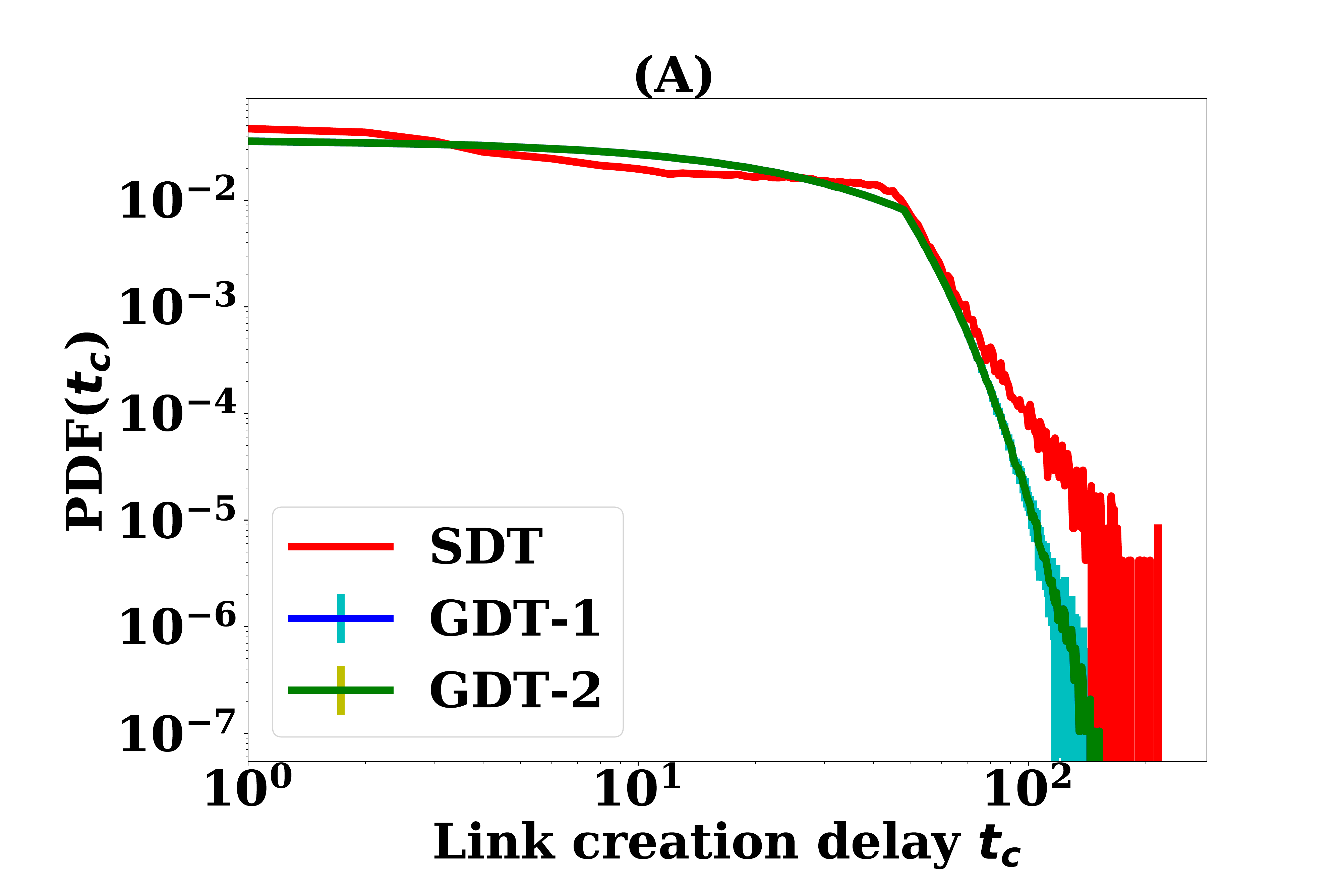}~
\includegraphics[width=0.45\linewidth, height=5.0cm]{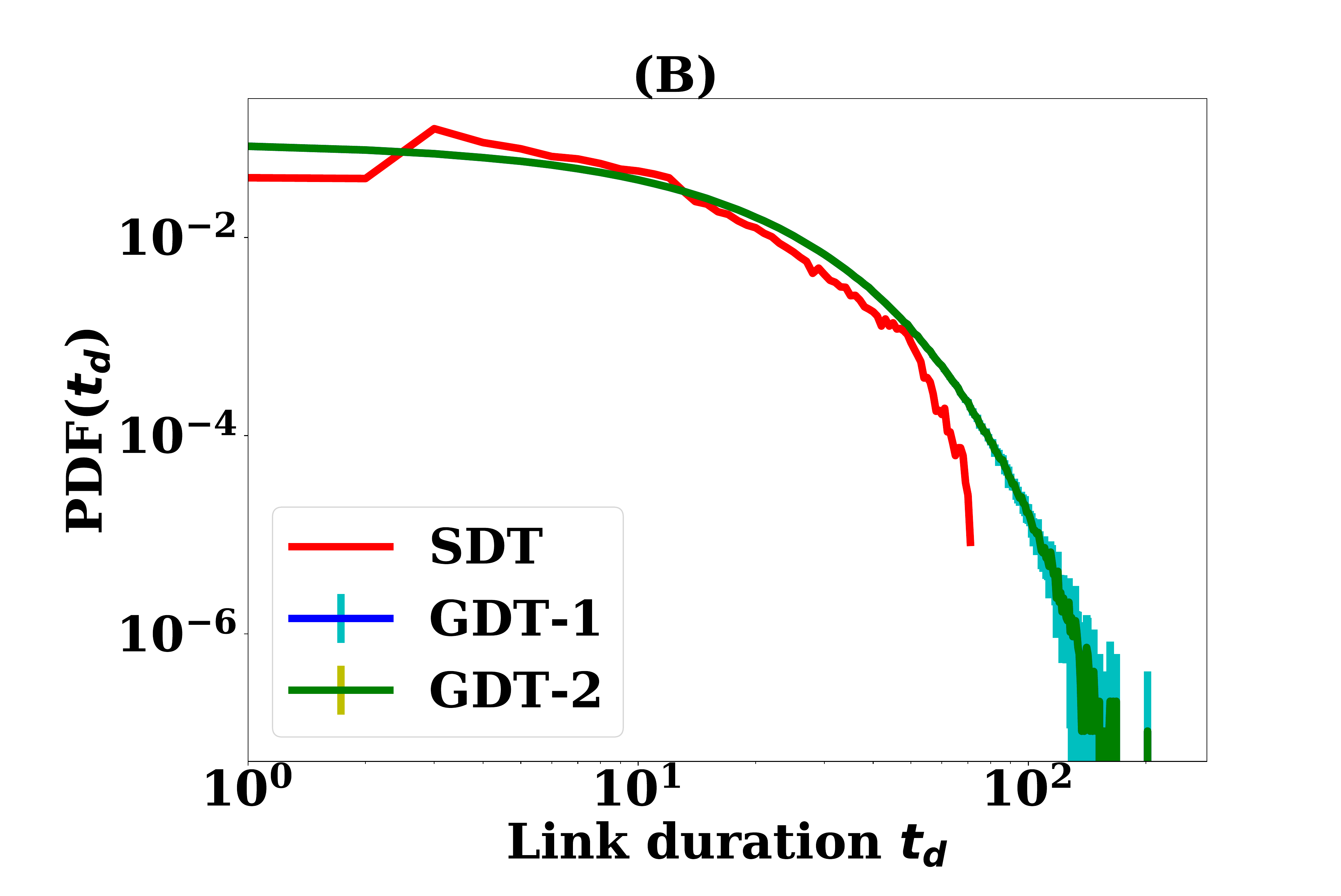}
\caption{Fitting link creation CIP of DDT network with the generated networks GDT-1 and GDT-2: (A) link creation delay $t_c$ and (B) link duration $t_d$. The values of $t_c$ and $t_d$ are given with the number of time steps}
\label{fig:fitl}
\end{figure}

\subsubsection{Fitting link creation parameters}
The link creation delay $t_c$ is drawn with the truncated geometric equation. The MLE condition is not a straight forward as the MLE of geometric equation. The MLE for the truncated equation is derived as follows:
\begin{equation*}
0=p_c \sum_{k=1}^{n} \frac{\sum_{l=1}^{m}\frac{(t_{c}^{k}-1)(1-(1-p_c)^{t_a^{l}+\delta})+(t_{a}^{l}+\delta)(1-p_c)^{t_a^l +\delta}}{(1-p_c)(1-(1-p_c)^{t^{l}_{t_a}+\delta})^2}}{\sum_{l=1}^{m} ((1-(1-p_{c})^{t^{l}_{a}+\delta})^{-1}}
\end{equation*}
where $t^{1}_{c},t^{2}_{c},\ldots,t^{n}_{c}$ are sample set of size $n=1.2M$ and $t_a=\{t_a^{1},\ldots,t_a^{m}\}$ with $m=518$K. The estimated value of $\hat p_c$ is 0.02. The generated link creation delays are presented in the Figure~\ref{fig:fitl}A, where the generated $t_c$ have RSE of 0.009 in comparison with the real distribution. The errors was consistent in both GDT-1 and GDT-2. The value of $\rho=0.084$, which is fitted with the active periods, is used to generate the link duration $t_d$. Figure~\ref{fig:fitl}B presents the comparison of generated link duration with real duration which has RSE error of 0.022 with STD 0.004. The six SPDT network parameters of the generated network are a good fit to the real network parameters built by Momo users. The variations for generated networks are low and consistent in both GDT-1 and GDT-2 networks. With the estimated model parameters, synthetic contact networks are generated and their properties are explored to understand the performance of the model.

%%%%%%%% SECTION-4 %%%%%%%%%%%%%%%%%%%%%%%%%%%%%%%
\section{Results and discussion}
We now validate the SPDT graph model by analyzing the generated network properties and the capability of simulating SPDT diffusion process on it. We first  analyse the network properties  to understand the capability of the graph model to generate properties of the real graph. Then, the diffusion dynamics on the generated networks are analyzed to assess its ability to replicate the diffusion dynamics of the real contact networks. The previous section tuned the model's parameters  using the location updates from Shanghai. This section uses the location updates from Beijing to compare the network properties and diffusion dynamics. The SPDT graph properties and diffusion dynamics are studied generating networks with both homogeneous and heterogeneous degree distributions. 

\subsection{Network Properties}
The previous section has focused on estimating co-location interaction parameters (CIP) to fit the SPDT graph which is based on the activity-driven network modelling to empirical data. This section explores the fitted model's ability to reproduce the static and temporal properties of empirical networks. 
\begin{figure}[h!]
\centering
\includegraphics[width=0.8\linewidth, height=3.2cm]{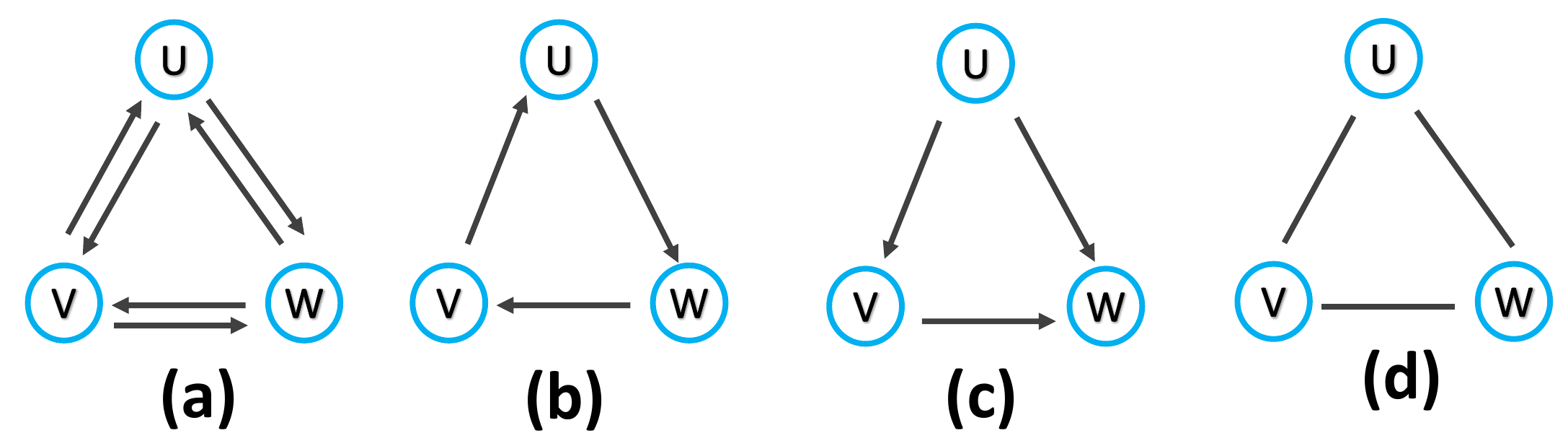}
 \caption{Triangles for various SPDT link configuration among nodes}
      \label{fig:triangle}
\end{figure}

\subsubsection{Static Properties}
The SPDT contact networks generated by the developed graph model are converted to static contact networks where a directed edge between two nodes is created if they have at least an SPDT link from the host node to neighbouring node at any time over the observation period. Then, two static network metrics, degree centrality and clustering coefficients, are studied. The degree centrality quantifies the extent of a node's connectedness to other nodes~\cite{freeman1978centrality}, i.e. how many nodes a host node contacts during the observation period. In a disease spread context, nodes with higher degree centrality get infected quickly as well as infect a higher number of other nodes~\cite{de2014role}. While degree centrality highlights the node connectivity, the clustering coefficient represents how a node is locally connected with the neighbour, i.e. the local social contact structure of nodes in the network~\cite{laurent2015calls}. The local clustering coefficient is defined as the ratio of the number of triangles present among the neighbours and the possible maximum triangles among neighbours. If a node is already infected in a cluster, it is not infected further by other nodes of the same cluster. Thus, the spreading varies based on the clustering coefficient and the realistic clustering coefficient is required in the generated network. To compute the clustering coefficient, the directions of links are neglected. However, the triangles still preserve the social structure formed by friends of friends as shown in Figure~\ref{fig:triangle}. All the triangles formed with directional links are the same as the triangle made by Figure~\ref{fig:triangle}. Thus, the generated directed network is converted into an undirected network for computing clustering coefficients.

\begin{figure}[h!]
\includegraphics[width=0.47\linewidth, height=4.8cm]{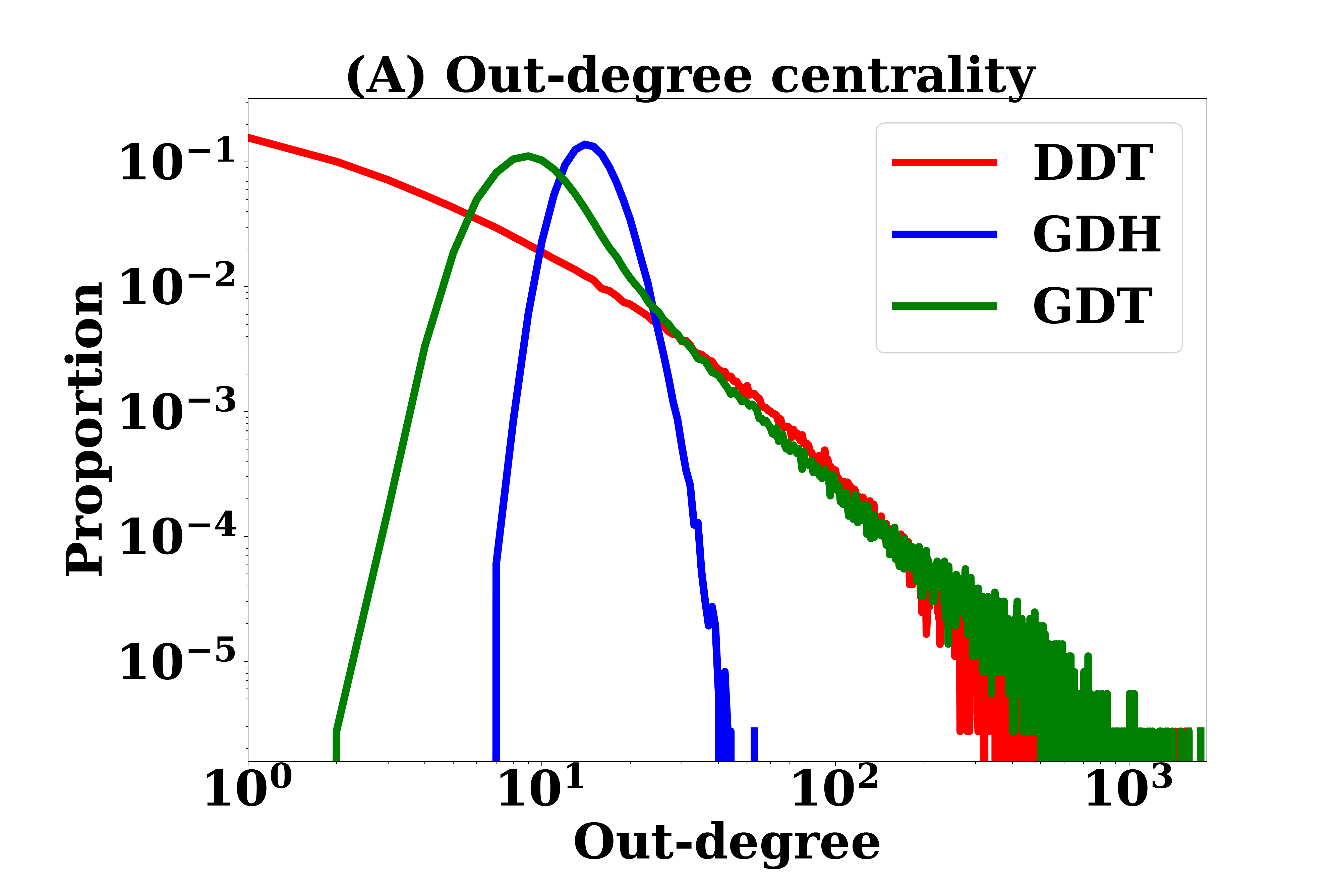}~
\includegraphics[width=0.47\linewidth, height=4.8cm]{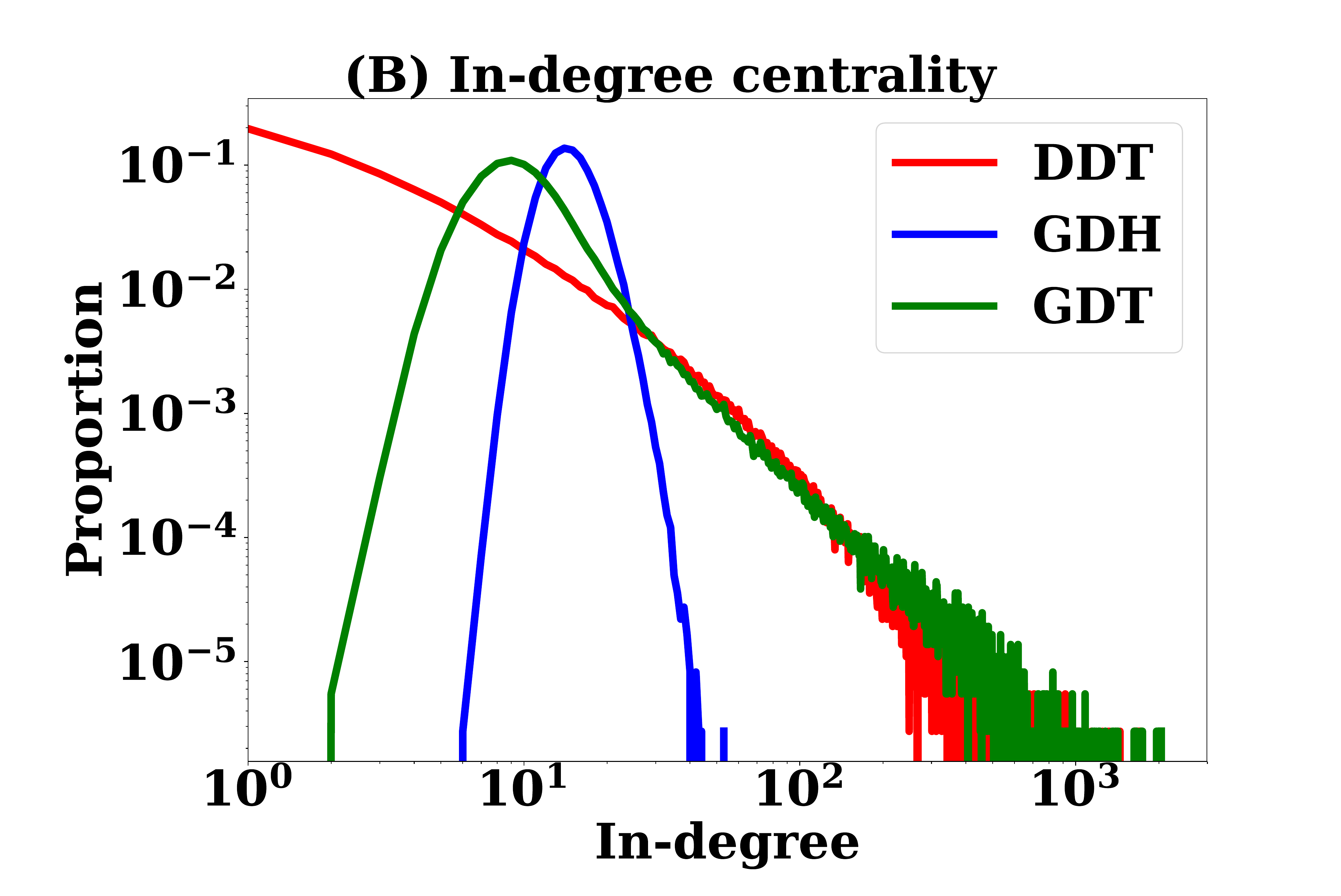}\\

\includegraphics[width=0.47\linewidth, height=4.8cm]{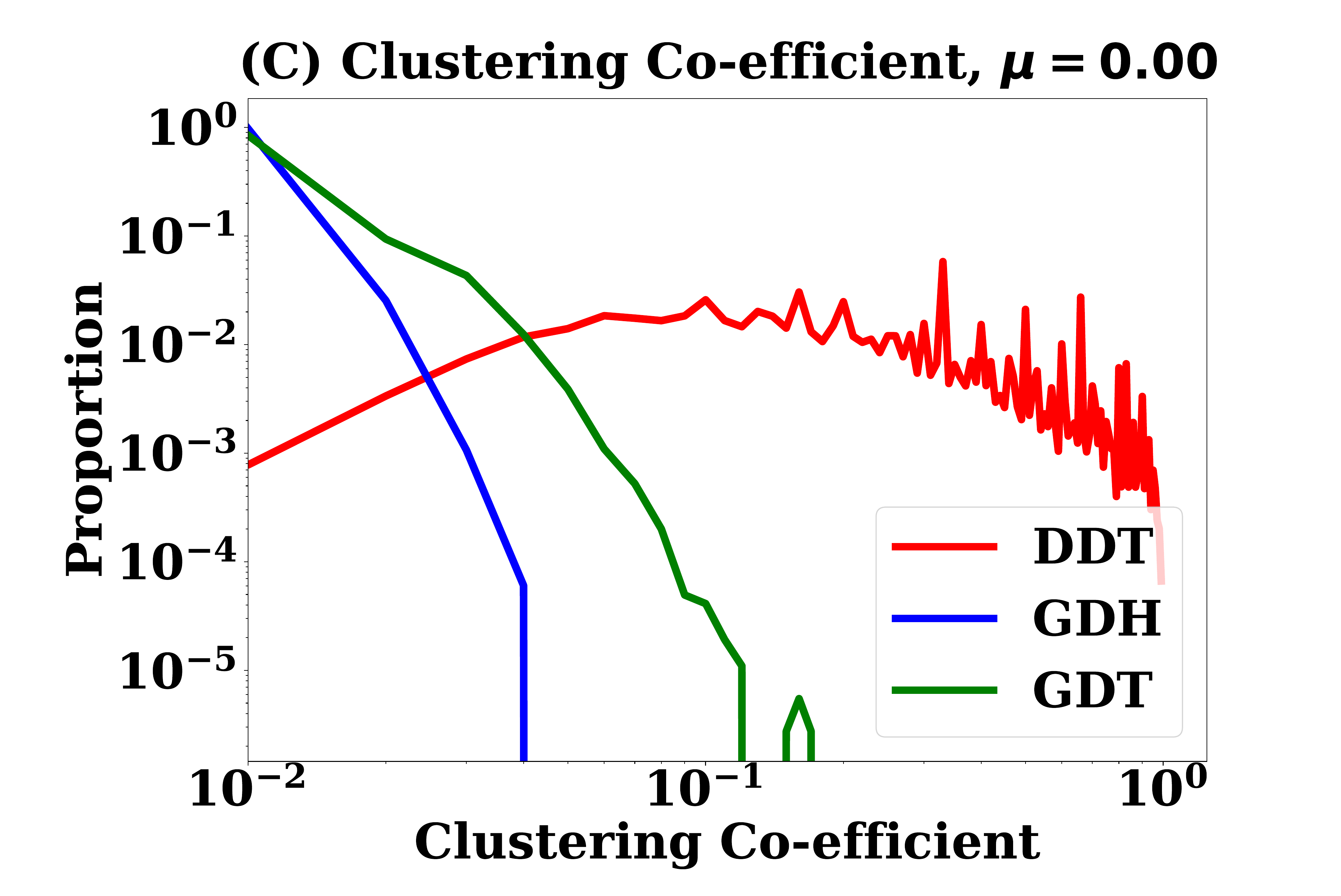}~
\includegraphics[width=0.47\linewidth, height=4.8cm]{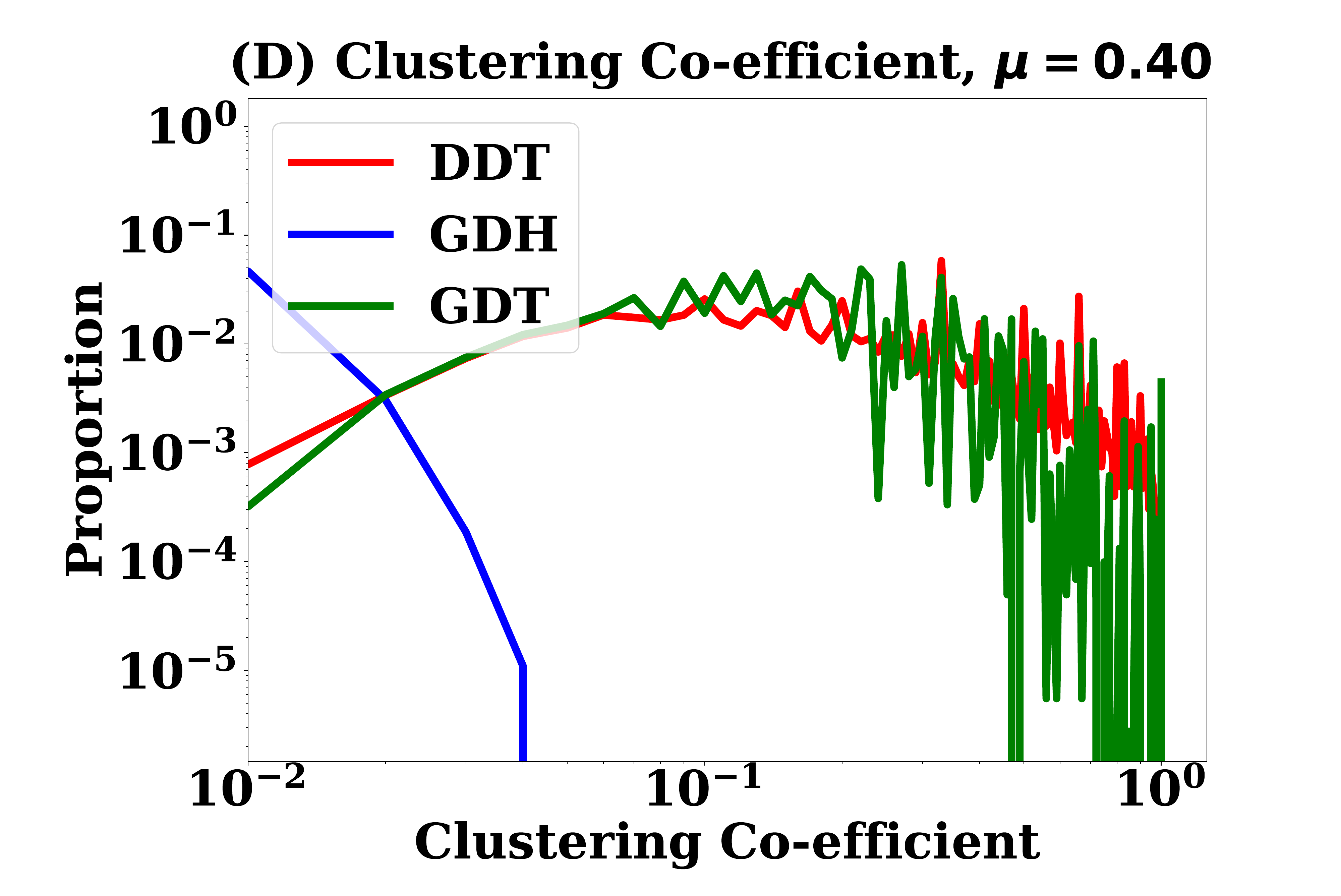}\\

\includegraphics[width=0.47\linewidth, height=4.8cm]{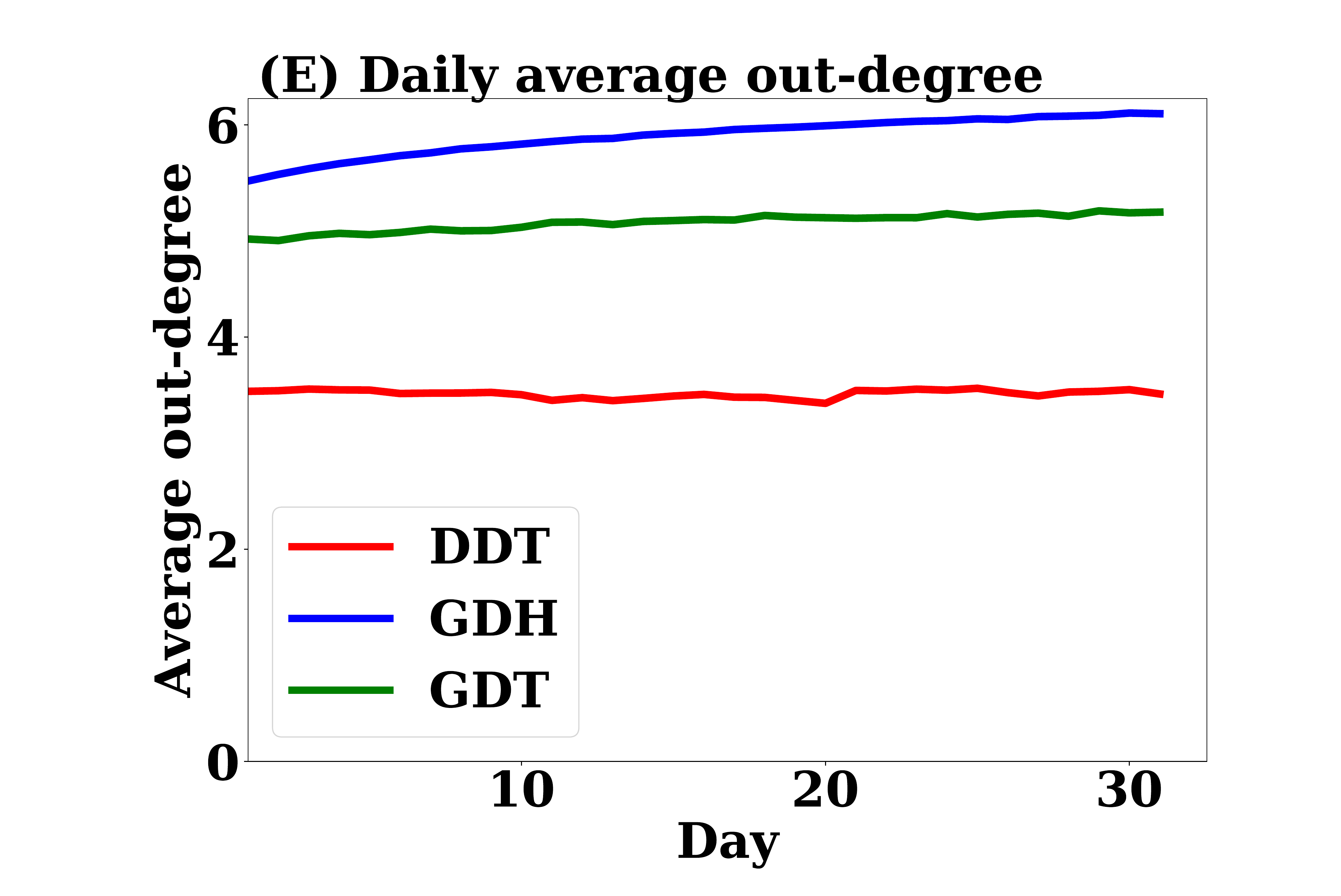}~
\includegraphics[width=0.47\linewidth, height=4.8cm]{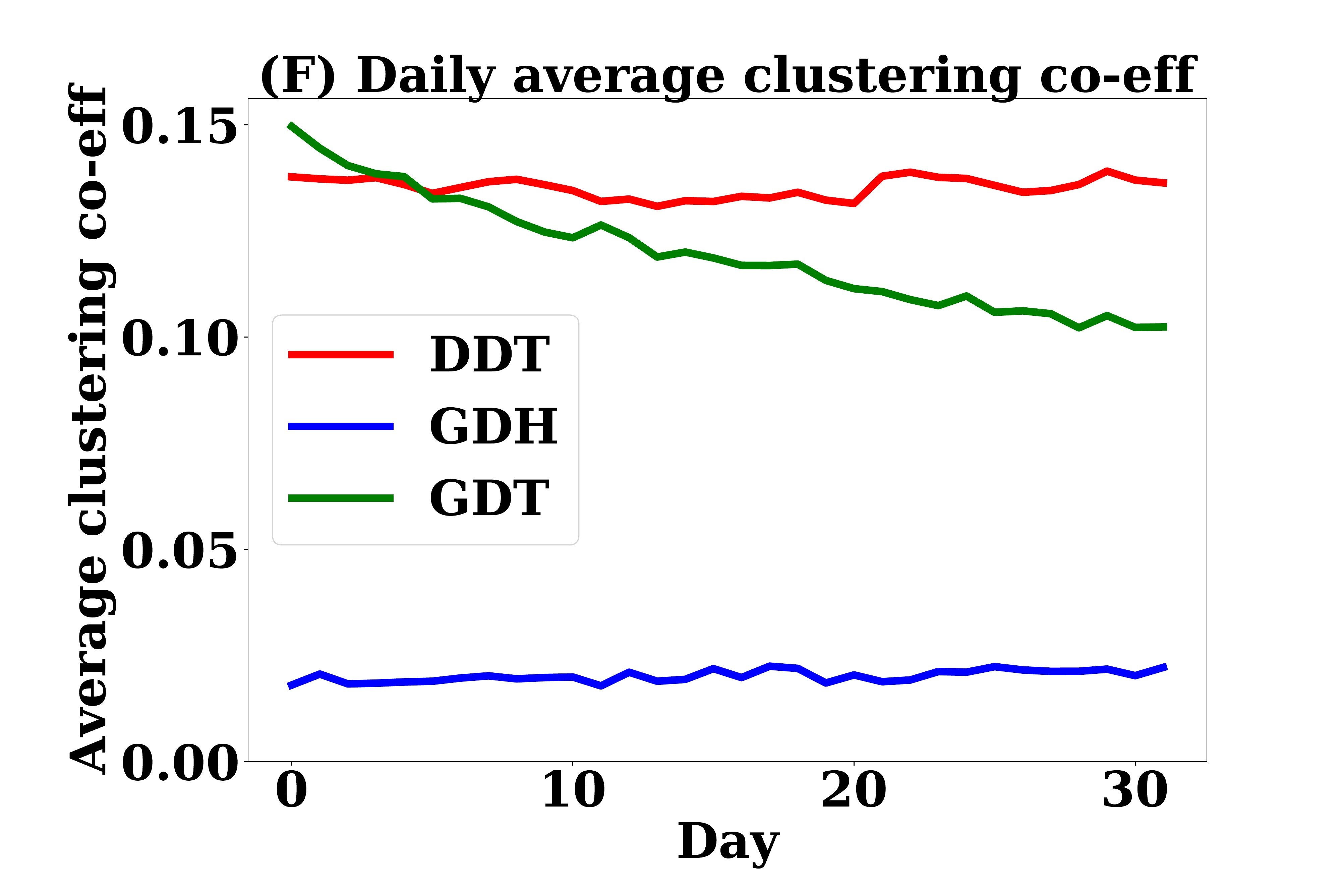}

 \caption{Static network properties in the real contact network DDT, homogeneous contact network GDH and heterogeneous contact network GDT: A) out-degree centrality, B) In-degreee centrality, C) Clustering co-efficient with $\mu=0$, D) clustering co-efficient with $\mu=0.40$, E) Daily average out-degree, and F) Daily average clustering co-efficient}
      \label{fig:staticpro}
\end{figure}
In the developed model, the growth of the contact set of a node is determined by $\lambda$ and the neighbour selection process defined by $P(n_t+1)=n_t/(n_t+\eta)$. The value of $\eta$ controls the degree to which nodes expand their contact set size. In this study, $\eta$ is empirically set to 0.1, and the selection of an optimal value of $\eta$ is beyond the scope of this paper. The growing of contact set size depends on $\lambda$ which is the same for all nodes in the homogeneous network (GDH) and varies across nodes in the heterogeneous network (GDT). However, setting $\eta=0.1$ creates opportunities to receive new neighbour nodes. The computed degree centrality and clustering coefficients for both homogeneous and heterogeneous networks are presented in Figure~\ref{fig:staticpro}. In degree centrality, another desirable feature is that nodes which have more directed links to other nodes also receive more links. Thus, the degree centrality is analysed for both out-degree and in-degree and which have the same trends (Figure~\ref{fig:staticpro}A and Figure~\ref{fig:staticpro}B). 

The contact set size distribution in the DDT network follows the long-tailed distribution. The developed GDT network are also long-tailed, but the distribution of lower degree nodes deviates from that of the DDT network. This is more practical compared to the DDT network where many individuals have contact set sizes of one. In reality one individual will have contact with more than one individuals during 32 days period. Thus, the developed model overcomes the limitation of the DDT network for not growing contact set sizes. As $\eta$ increases, the contact set sizes can be larger. We have also generated synthetic network (GDH) with homogeneous degree distribution. The degree distribution in the GDH network follows the Poisson distribution as the activation degree in the GDH network is generated with a geometric distribution and with the small parameter value. Therefore, the homogeneous intensity to involve in interaction with others can not capture the real contact set size distribution. 

In the developed graph, local clustering is formed with selecting $\mu$ proportion neighbours from the neighbours of neighbouring nodes. The clustering coefficient for two values $\mu=0$ and $\mu=0.4$ are studied in both heterogeneous and homogeneous networks. The results are presented in Figure~\ref{fig:staticpro}C and Figure~\ref{fig:staticpro}D where nodes are binned based on their clustering coefficient with a step of 0.01. If $\mu$ is set to zero, the clustering coefficient is low. The clustering coefficient in the GDH network is 0.005 and 353K nodes are in the first bin, i.e 353k nodes having clustering coefficient below 0.01 while 93K nodes have clustering coefficient below 0.01. The clustering coefficient improves a bit in the GDT network with long-tailed activation degree distribution where 307k nodes having a clustering coefficient below 0.01 and average clustering coefficient in the GDT with $\mu=0$ is 0.008. Then, $\mu$ is set to 0.4 which is found analysing Momo data set. With this value, the GDT network improves significantly to form realistic clustering coefficient and the number of nodes in the first bin is 45K. However, the clustering coefficient is still not realistic in the GDH networks and have similar trends of $\mu=0$. The heterogeneous degree distribution contributes to making the local clustering as many lower degree nodes can create local cluster through the higher degree nodes connect.

This experiment also studies the average daily degree and clustering coefficient of nodes. This is measured converting the everyday dynamic graph as a static graph where a pair of nodes is linked if once they have contacted. This can indicate how many other nodes a node contact during each day and what is their clustering coefficient. The results are presented in Figure~\ref{fig:staticpro}E and Figure~\ref{fig:staticpro}F. It is observed that nodes in DDT network contact on average 4 nodes which are around 5 nodes in the GDT network. As the contact set size grows in the GDT network, it has a high average contact degree over the day. The GDH network has the highest average daily degree. This is because nodes in the GDH network have moderate activation degree and do not have many nodes with the low degree as it is in the DDT and GDT networks. The daily average clustering coefficient is low in the GDH network. The average daily clustering coefficient in the GDT network is similar to DDT network at the early days. However, it decreases as simulation time goes. This is because the contact set sizes grow in the GDT network and nodes have more options to select neighbour nodes. Therefore, common neighbours among nodes decrease and hence clustering coefficients are low.

\subsubsection{Temporal Properties}
The proposed graph model generates the temporal activities of nodes through SPDT link creation. The temporal properties should match those of real networks. Two commonly used temporal network central properties to characterise the network: closeness and betweenness are studied. The node closeness measures the geodesic distance between all pairs of nodes in the network. The node betweenness is measured as the fraction of geodesics distances, which is connected between all node pairs in the network, passes through a node~\cite{holme2015modern}. The generated network is converted to the temporal network $G_{T}=(Z,L_{T})$ where $Z$ is the set of nodes and set $L^{T}\subseteq V\times Z\times [0,T]$ of time-stamped links $(u,v;t)\in L_{T}$. In airborne disease propagation, susceptible individuals who get infected do not start infecting immediately and go through an incubation period before they can infect others. In this study, the incubation period is set to one day, i.e. individuals that become infected on a simulation day can start spreading the disease the next day after infection. Thus, all links of a day between a pair of nodes are aggregated over the day and the timestamp $t_j$ represents the interaction day $j$. If a disease is transmitted from node $u$ to node $v$ through a link $(u,v;t_{j})$, the disease will be transmitted from node $v$ to another node $w$ with the link $(v,w;t_{j+k})$ where $1\leq k\leq 5$ as an node is considered to be infectious for the next five days. The temporal properties are measured based on the time-respecting paths among nodes. A time-respecting path between an infected node $u$ to a susceptible node $v$ may create through a sequence of time-stamped links as follows:
\[(w_{0},w_{1};t_{1}),(w_{1},w_{2};t_{2}),\ldots,(w_{l-1},w_{l};t_{l})\]
where $w_{0}=u,w_{l}=v$ and the sequence of time-stamped links should be $t_{1}\leq t_{2}\leq \ldots \leq t_{l}$. The time difference between two consecutive links of a time-respecting path can be at most 5.

\begin{figure}[h!]
\includegraphics[width=0.45\linewidth, height=4.8cm]{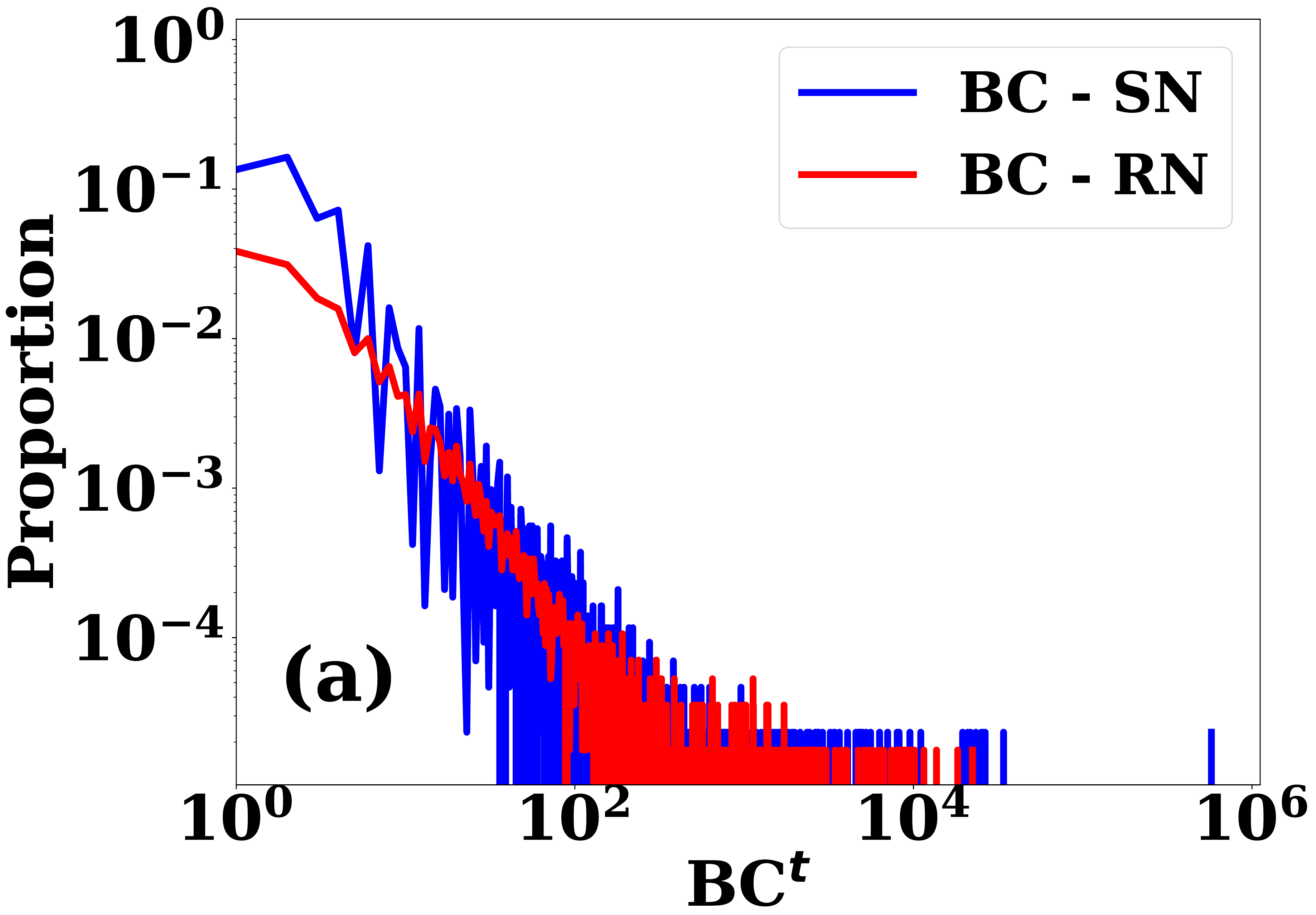}~  \includegraphics[width=0.45\linewidth, height=4.8cm]{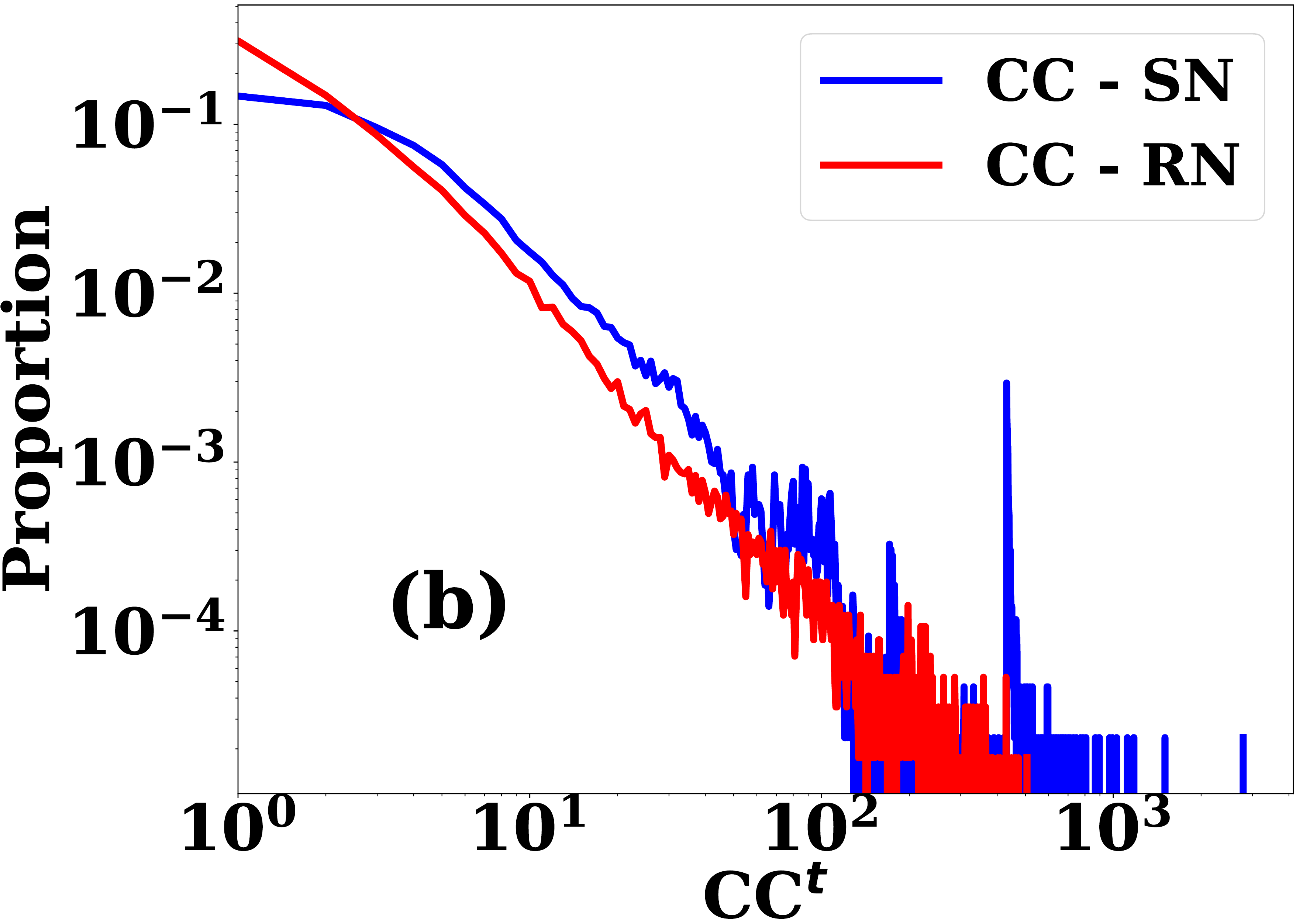} 
\caption{Temporal network properties in the generated contact network and comparison with that of real contact network : A) Betweenness centralities and B) Closeness centralities}
      \label{fig:tempro}
\end{figure}

In the temporal network, the betweenness of a node is defined as the number of shortest time-respecting paths pass through it. However, the temporal betweenness depends on the starting time $t_{1}$ of time-respecting paths as the path length between the same pair of nodes may change if the starting time is shifted. Thus, shortest paths between a pair of nodes are computed at all starting times and the shortest paths regardless of starting time are counted as the set to measure betweenness. The temporal betweenness centrality $BC^{t}$ of an node $v$ is measured as:
\[BC^{t}(v)=\sum_{u\not= v \not= u}^{}\mid Q^{t}\left(u,w; v\right)\mid\]
where $Q^{t}$ is the set of shortest time-respecting paths between $u$ and $w$ passing through $v$ starting any time during observation period~\cite{scholtes2016higher}. In a similar way, the temporal closeness centrality $CC^{t}$ based on the shortest time respecting paths is defined as: 
\[CC^{t}(v)=\sum_{u\not= v}^{}\frac{1}{dist^{t}\left(u,v\right)}\]
where $dist^{t}$ is the distance of the shortest time-respecting path from $u$ to $v$ starting at any time during the observation period. The temporal network analysis methods proposed by the authors of~\cite{scholtes2016higher} are used in this experiment. The generated networks of 32 days are used and the results obtained are presented in Figure~\ref{fig:tempro} for both the real network (RN) and synthetic network (SN). The betweenness distributions of $BC^{t}$ and $CC^{t}$ have similar trends. However, the synthetic network has a higher number of nodes with a lower $BC^{t}$. This is because the synthetic network is more random than the real network. In the synthetic network, some nodes may have higher activation degree as it depends on the power law distributed $\lambda_{i}$ and hence have higher $CC^{t}$. The synthetic network overall produces similar network properties of the real network.

%%%%%%%%%%  SECTION - 5 %%%%%%%%%%%%%%
\subsection{Diffusion dynamics}
We now focus on  the capability of simulating the SPDT process by the generated networks. Accordingly, airborne disease spreading is simulated on the generated synthetic contact networks and the real contact networks. We first study the similarities in  diffusion dynamics on various networks and then evaluate the model's parameter sensitivity. The experimental setup and diffusion analysis are presented below.

\subsubsection{Experimental setup}
For the real contact networks, the DDT contact networks constructed in Section 3 are selected since this network has shown realistic diffusion dynamics due to dense link density~\cite{shahzamaROS}. In addition, the developed model is expected to generate realistic diffusion dynamics. Homogeneous (GDH) and heterogeneous (GDT) synthetic contacts networks among 364K nodes are generated for 32 days as the DDT network contains 364K nodes and 32 days period. The simulations are also conducted on another synthetic graph constructed according to the basic activity driven networks (ADN) model~\cite{perra2012activity} to understand how well the proposed model captures diffusion dynamics compared to the current graph models that are only based on direct transmission links. In this graph, nodes activate at each time step $\Delta t$ with a probability $\phi$ and generate $m$ links to others nodes. The probabilities $\vartheta$ of activation of nodes are assigned with a power law degree distribution. The number of links $m$ created during activation can be constant for all nodes or heterogeneous. In this study, both types of $m$ are used and homogeneous (BDH) and heterogeneous (BDT) basic ADN networks with homogeneous and heterogeneous degree distribution are generated. For fitting activation potential $\vartheta$, the daily activation frequencies $h$ of Momo users are used. The analysis of Momo users shows that their average stay periods $\Delta t$ are 50 minutes. Therefore, the number of time steps in one day is 28.8 and the potential is $\vartheta=28.8/h$. Thus, the activation frequencies are converted to activation potential $\vartheta$ and fitted with a power law degree distribution with a lower limit of $\vartheta$ is 0.02 and an upper limit of 0.18 while power law exponent is 2.95. For the activation degree, it is found that the average activation degree is $m=3$. For heterogeneous degree distribution, the activation degree distribution from the developed SPDT graph model are being used. With these fitted parameters BDH and BDT networks are also generated for 364K nodes over 32 days.

For propagating disease on the selected contact networks, we consider a generic Susceptible-Infected-Recovered (SIR) epidemic model. In this model, nodes remain in one of the three compartments, namely, Susceptible (S), Infectious (I) and Recovered (R). If a node in the susceptible compartment receives a SPDT link from a node in the infectious compartment, the former is subject to exposure $E_l$ of infectious pathogens for both direct and indirect transmission links according to the following equation
\begin{equation}
%\begin{split}
%E_l & =\frac{gp}{Vr^2}\left[r\left(t_i-t_s^{\prime}\right)+ e^{rt_{l}}\left(e^{-rt_i}-e^{-rt_l^{\prime}} \right)\right]\\
%& +\frac{gp}{Vr^2}\left(e^{-rt_l^{\prime}}-e^{-rt_s^{\prime}} \right)e^{rt_{s}}
%\end{split}
E_l =\frac{gp}{Vr^2}\left[r\left(t_i-t_s^{\prime}\right)+ e^{rt_{l}}\left(e^{-rt_i}-e^{-rt_l^{\prime}} \right)\right]
+\frac{gp}{Vr^2}\left(e^{-rt_l^{\prime}}-e^{-rt_s^{\prime}} \right)e^{rt_{s}}
%\end{split}
\end{equation}
where g is the particle generation rate of infected individual, p is the pulmonary rate of susceptible individual, V is the volume of the interaction area, r is the particles removal rates from the interaction area, $t_s$ is the arrival time of infected individual, $t_l$ is the leaving time of infected individual, $t_s^{\prime}$ is the arrival time of susceptible individuals and $t_l^{\prime}$ is the leaving time of susceptible from the interaction location and $t_i$ is given as follows: $t_i=t_l^{\prime}$ when SPDT link has only direct component, $t_i=t_l$ if SPDT link has both direct and indirect components, and otherwise $t_i=t_s^{\prime}$. If $t_s<t_s^{\prime}$, $t_s$ is set to $t_s^{\prime}$ for calculating appropriate exposure~\cite{shahzamaROS}. If a susceptible individual receives $m$ SPDT links from infected individuals during an observation period, the total exposure $E$ is 
\begin{equation}\label{eq:expo}
E=\sum_{k=0}^{m}E_{l}^{k}
\end{equation}
where $E_{l}^{k}$ is the received exposure for k$^{th}$ link. The probability of infection for causing disease can be determined by the dose-response relationship defined as 
\begin{equation}\label{eq:prob}
P_I=1-e^{-\sigma E}
\end{equation}
where $\sigma$ is the infectiousness of the virus to cause infection~\cite{fernstrom2013aerobiology}. This value depends on the disease types and even virus types. If the susceptible individual move to the infected compartment with the probability $P_I$, it continues to produce infectious particles over its infectious period $\tau$ days until they enter the recovered state, where $1/\tau$ is the rate of recovering from the disease. In this model, no event of births, deaths or entry of new individual are considered.

The simulations are step forwarded in our experiments with one day interval~\cite{stehle2011simulation,toth2015role}. We chose an initial set of 500 seed nodes randomly in each experiment to start simulations assuming that it will be capable of showing the full epidemic curve in the studied simulation duration. During each day of disease simulation, the received SPDT links for each susceptible individual from infected individuals are separated and infection causing probabilities are calculated by Eqn.~\ref{eq:prob}. The volume $V$ of proximity in Eqn.~\ref{eq:expo} is fixed to 2512 m$^{3}$ assuming that the distance, within which a susceptible individual can inhale the infectious particles from an infected individual, is 20m and the particles will be available up to the height of 2m~\cite{han2014risk,fernstrom2013aerobiology}. The other parameters are assigned as follows: particle generation rate $g=0.304$ PFU (plaque-forming unit)/s and pulmonary rate $q=7.5$ liter/min ~\cite{yan2018infectious,lindsley2015viable,han2014risk}. Infectious particles may require 7.5 min to 300 min to be removed from interaction areas after their generation. We assign $r=\frac{1}{60 b}$ to Eqn~\ref{eq:expo} where $b$ is particle removal time randomly chosen from [7.5-300] min given a median particle removal time $r_t$. The parameter $\sigma$ is set to 0.33 as the median value of required exposures for influenza to induce disease in 50$\%$ susceptible individuals is 2.1 PFU~\cite{alford1966human}. Susceptible individuals stochastically switch to the infected states in the next day of simulation according to the Bernoulli process with the infection probability $P_I$ (Eqn~\ref{eq:prob}). Individual stays infected up to $\tau$ days randomly picked up from 3-5 days maintaining $\bar{\tau}=3$ days (except when other ranges are mentioned explicitly)~\cite{huang2016insights}. The epidemic parameters: disease prevalence $I_p$, number of infected nodes in the network, and cumulative infection $I_a$, number of the infected individuals up to a day are calculated. To characterize the diffusion reproduction ability, the absolute percentage variation (APV) for infection events are calculated for each network with the reference of real contact networks as: 
\[ APV=100\times \frac{I_r-I_o}{I_r}\]
where $I_r$ is the number of infection events in the real network and $I_o$ is the infection event in the corresponding observed graph. For the both $I_p$ and $I_a$ are measured for each network.

\begin{table} \label{difsum}
\caption{Summary of diffusion dynamics on contact networks}
\vspace{1em}
\centering
\begin{tabular}{|c|c|c|c|c|c|}
\hline
Networks & $r_t$ & $I_p$(max) & $I_a$(max) & average DV (\%) & average AV (\%)  \\ \hline
\multirow{3}*{DDT} & 60 & 1731  & 11221 & 0 & 0 \\ 
\cline{2-6}
 & 40 & 1332  & 8671 & 0 & 0 \\  \cline{2-6}
 & 20 & 966  & 6460 & 0 & 0 \\ \hline
 
\multirow{3}*{GDT} & 60 & 2094  & 11929 & 15 & 10 \\ 
\cline{2-6}
 & 40 & 1531  & 9584 & 18 & 12 \\ \cline{2-6}
 & 20 & 987  & 6853 & 22 & 13 \\ \hline
 
\multirow{3}*{BDT} & 60 & 1007 & 7809  & 29 & 37 \\ 
\cline{2-6}
 & 40 & 673  & 5666 & 30 & 38 \\ \cline{2-6}
 & 20 & 335  & 3170 & 45 & 44 \\ \hline
 
\multirow{3}*{GDH} & 60 & 698  & 3600 & 61 & 50 \\ 
\cline{2-6}
 & 40 & 626  & 2110 & 69 & 53 \\ \cline{2-6}
 & 20 & 586  & 1592 & 70 & 51 \\ \hline
 
 \multirow{3}*{BDH} & 60 & 600  & 1892 & 73 & 62 \\ 
\cline{2-6}
 & 40 & 581  & 1645 & 73 & 59 \\ \cline{2-6}
 & 20 & 562  & 1386 & 72 & 55 \\ \hline

\end{tabular}
\end{table}

\begin{figure}[h!]
\begin{tikzpicture}
    \begin{customlegend}[legend columns=5,legend style={at={(0.12,1.02)},draw=none,column sep=2ex }, legend entries={DDT, GDT, GDH, BDT, BDH}]
    \addlegendimage{mark=triangle,solid,line legend, color=blue}
    \addlegendimage{mark=x,solid, color=red}   
    \addlegendimage{mark=square,solid, color=green}
    \addlegendimage{mark=o, color=cyan}
    \addlegendimage{mark=star, color=yellow}
    \end{customlegend}
 \end{tikzpicture}
\vspace{3ex}
\centering
\includegraphics[width=0.45\linewidth, height=5cm]{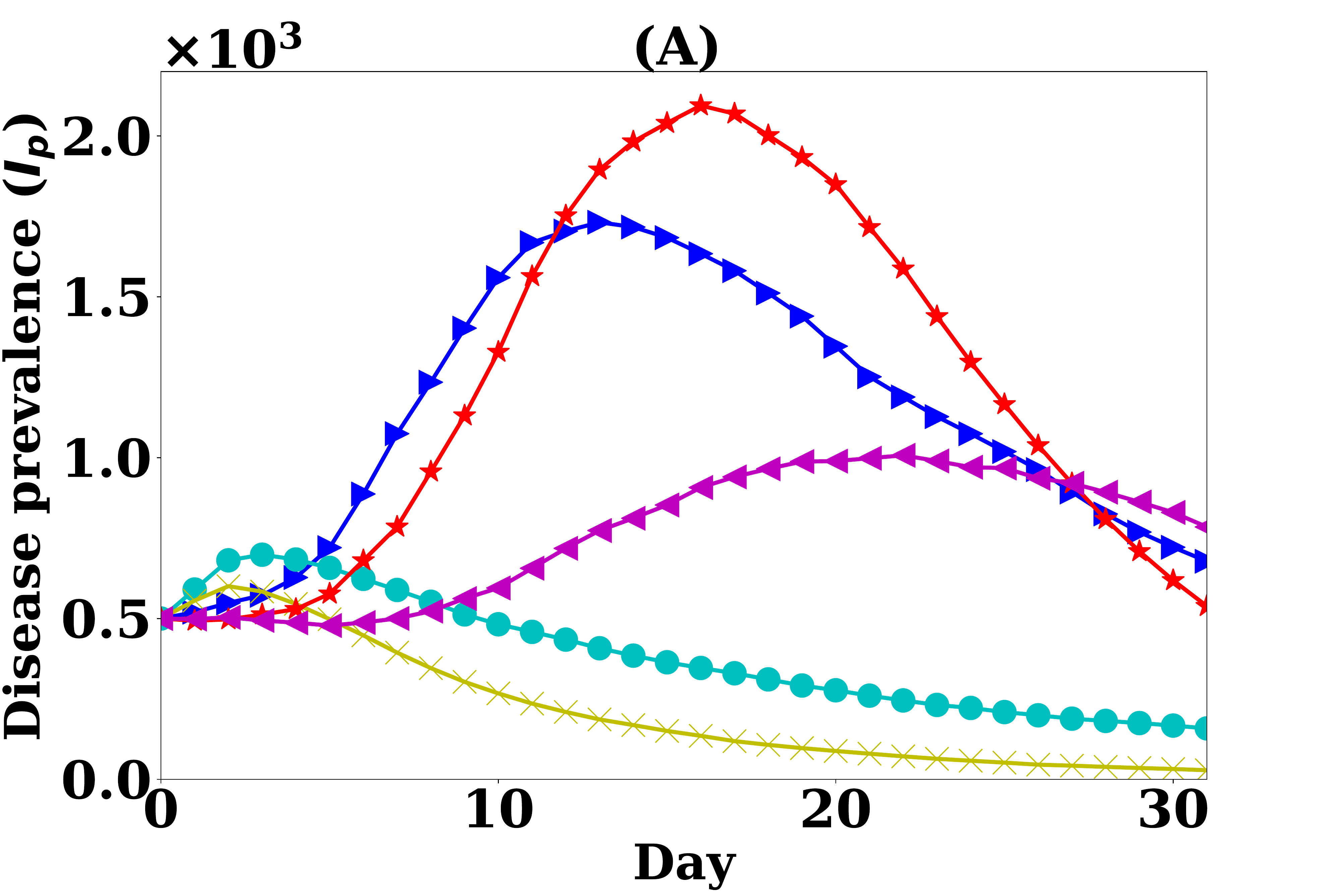}~
\includegraphics[width=0.45\linewidth, height=5cm]{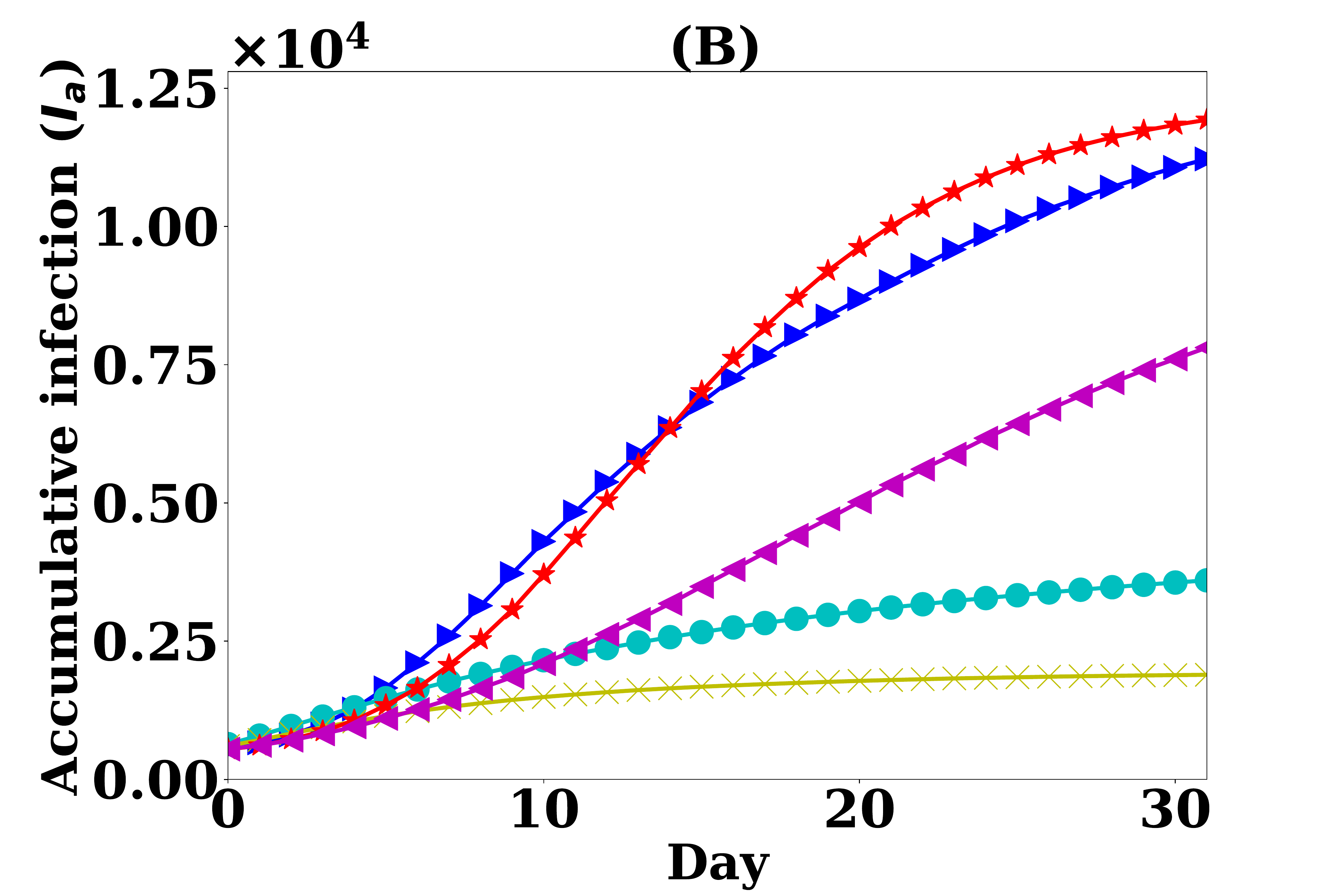}\\
% \vspace{1em}
\includegraphics[width=0.45\linewidth, height=5cm]{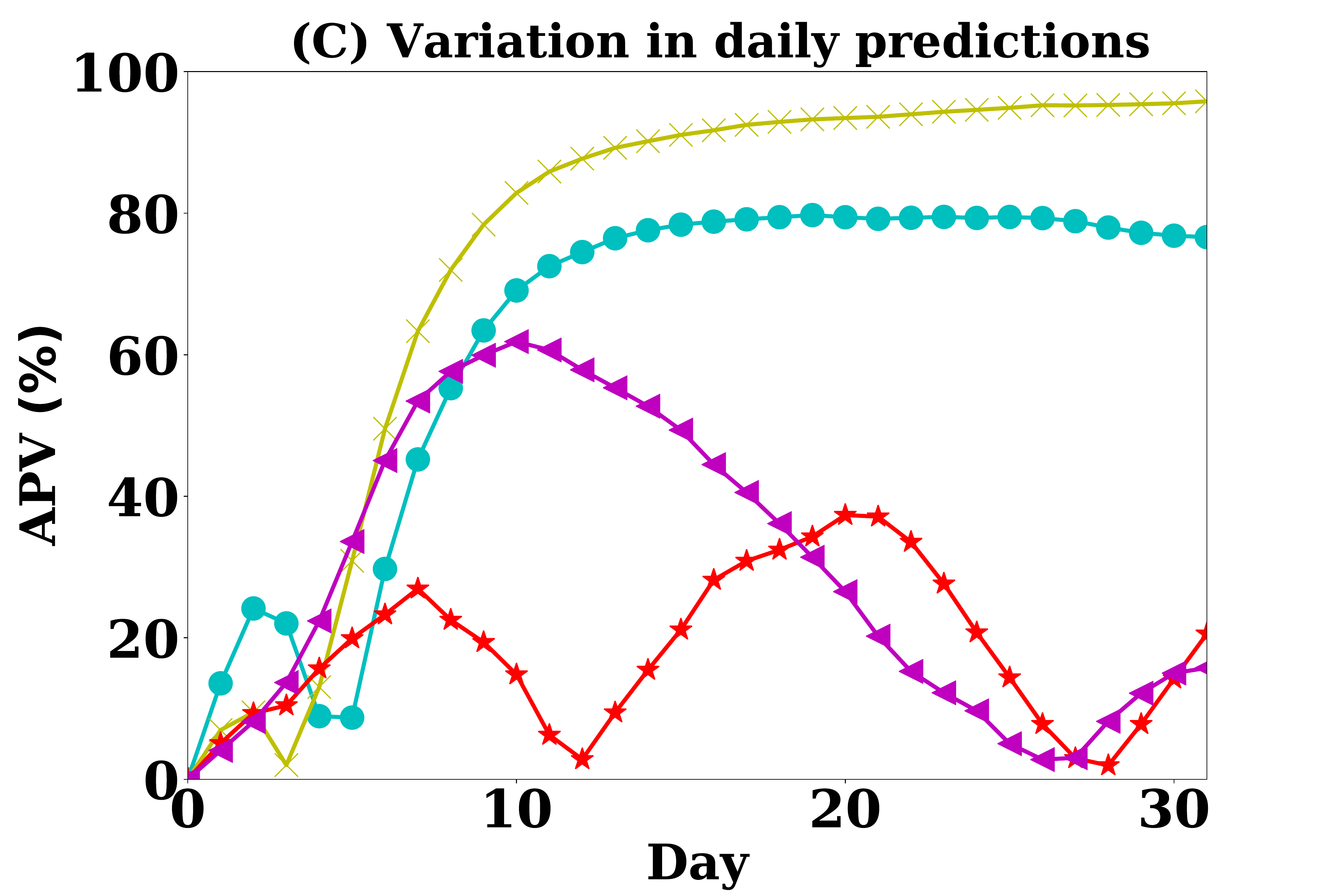}~
\includegraphics[width=0.45\linewidth, height=5cm]{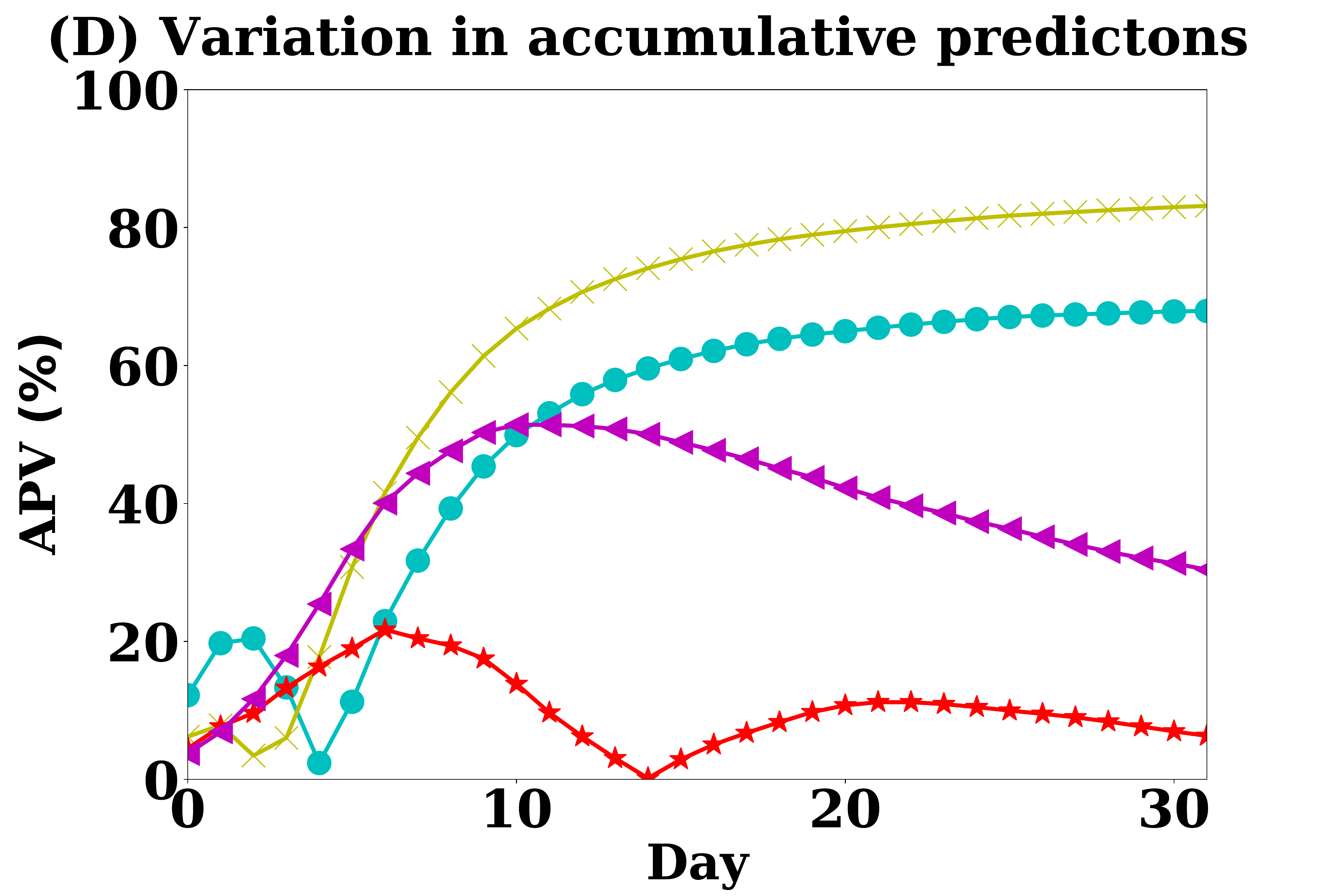}
\caption{Diffusion dynamics on various contact networks: A) disease prevalence dynamics, B) accumulative infection over simulation days, C) variations in the daily prediction comparing to real contact network and D) variation in the accumulative prediction.}
\label{fig:nethmdf}
\end{figure}

\subsubsection{Diffusion analysis}
The selected four networks (BDH, BDT, GDH and GDT) are generated with 364K nodes for 32 days and simulations are run with the selected disease parameters. The infection probability $P_I$ in Equation~\ref{eq:expo} is calculated with particle decay rates $r_t=60$ min. The average results for 1,000 realisations are shown in Figure~\ref{fig:nethmdf}. The proposed graph model with heterogeneous degree distributions (GDT) has diffusion dynamics that are close to real contact networks (DDT). The GDT network has higher disease prevalence $I_p$ than the DDT network (Fig.~\ref{fig:nethmdf}A). This is because nodes in the GDT network grow their contact sizes substantially compared to DDT network where links between two nodes are repeated. Thus, infected nodes in the DDT network repeatedly forward SPDT links to the same susceptible nodes and disease propagates faster in the DDT network at the early days of simulations. On the other hand, infected nodes in the GDT networks forward links to more susceptible nodes and disease transmission happen slowly in the GDT network. Thus, the initial spreading rates in the DDT network is faster than the GDT network. For the same reason, the GDT network carries the growth of $I_p$ for a longer time and to a higher value. The total infection in GDT network is also close to the total of the real network (Figure~\ref{fig:nethmdf}B). On the other hand, the BDT networks show increasing of $I_p$. But, the diffusion dynamics are underestimated significantly and total infection is comparatively low. This is because the BDT network does not have indirect transmission links. Thus, the growth of $I_p$ is not carried out in this network as it is in the DDT and GDT networks. On the other hand, homogeneous networks cannot replicate the diffusion dynamics of real contact networks. Initially, the disease prevalence $I_p$ increases for a few days in the GDH network and declines. A similar trend is found for the BDH network. This is because the homogeneous networks do not have higher degree nodes that have strong disease spreading potential and forward disease to different points of the network. In the homogeneous networks, the neighbour nodes quickly get infected and infection resistance force grows due to reducing susceptible neighbour nodes. The prediction variations in APV for all networks are shown in the Figure~\ref{fig:nethmdf} C and D. The GDT network shows the lowest variation in daily predictions among all networks compared to the DDT network. This has on average 15\% variation while the BDT network shows on average 29\% variation. However, the variations are high for the BDT network with the maximum variation being 60\%. The accumulative prediction variation in the GDT network has an average of 10\%, while this is very high for the BDT networks at 37\%. For the other networks, the GDH network has on average 61\% daily variation and BDH network has 70\% variation. Though it has a higher variation in daily prediction, the proposed model still maintains a cumulative prediction variation around 10\%. The proposed graph model is up to 27\% better than current models for predicting disease spread dynamics.

\begin{figure}[h!]
\begin{tikzpicture}
    \begin{customlegend}[legend columns=5,legend style={at={(0.12,1.02)},draw=none,column sep=2ex }, legend entries={DDT, GDH, BDH, GDT, BDT, $r_t=60$, $r_t=40$, $r_t=20$}]
    \addlegendimage{solid,line legend, color=blue}
    \addlegendimage{solid, color=red}   
    \addlegendimage{solid, color=cyan}
    \addlegendimage{color=yellow}
    \addlegendimage{color=black}
    \addlegendimage{mark=x}
    \addlegendimage{mark=o}
    \addlegendimage{mark= triangle}
    \end{customlegend}
 \end{tikzpicture}
\vspace{3ex}
\centering
\includegraphics[width=0.45\linewidth, height=5cm]{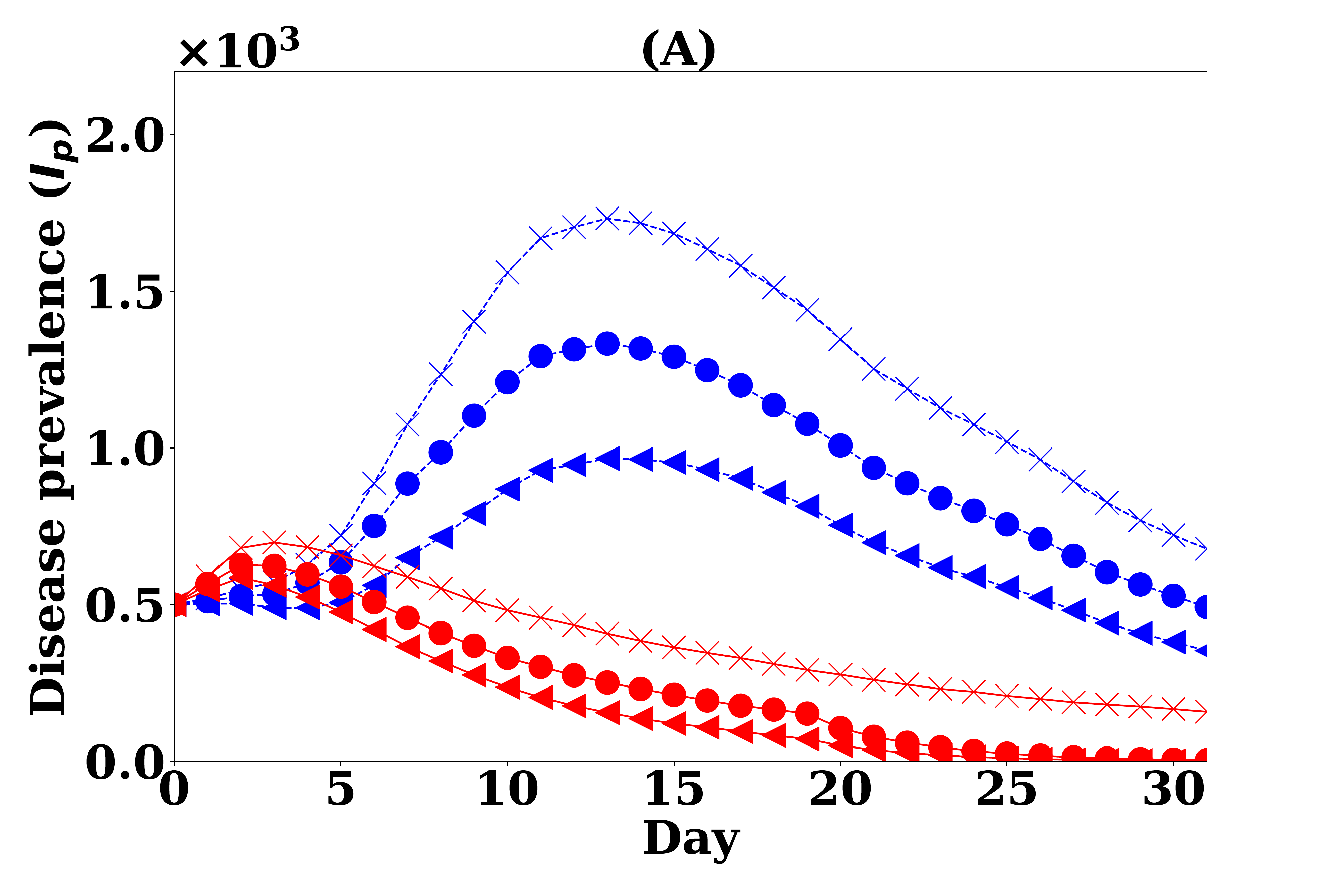}~
\includegraphics[width=0.45\linewidth, height=5cm]{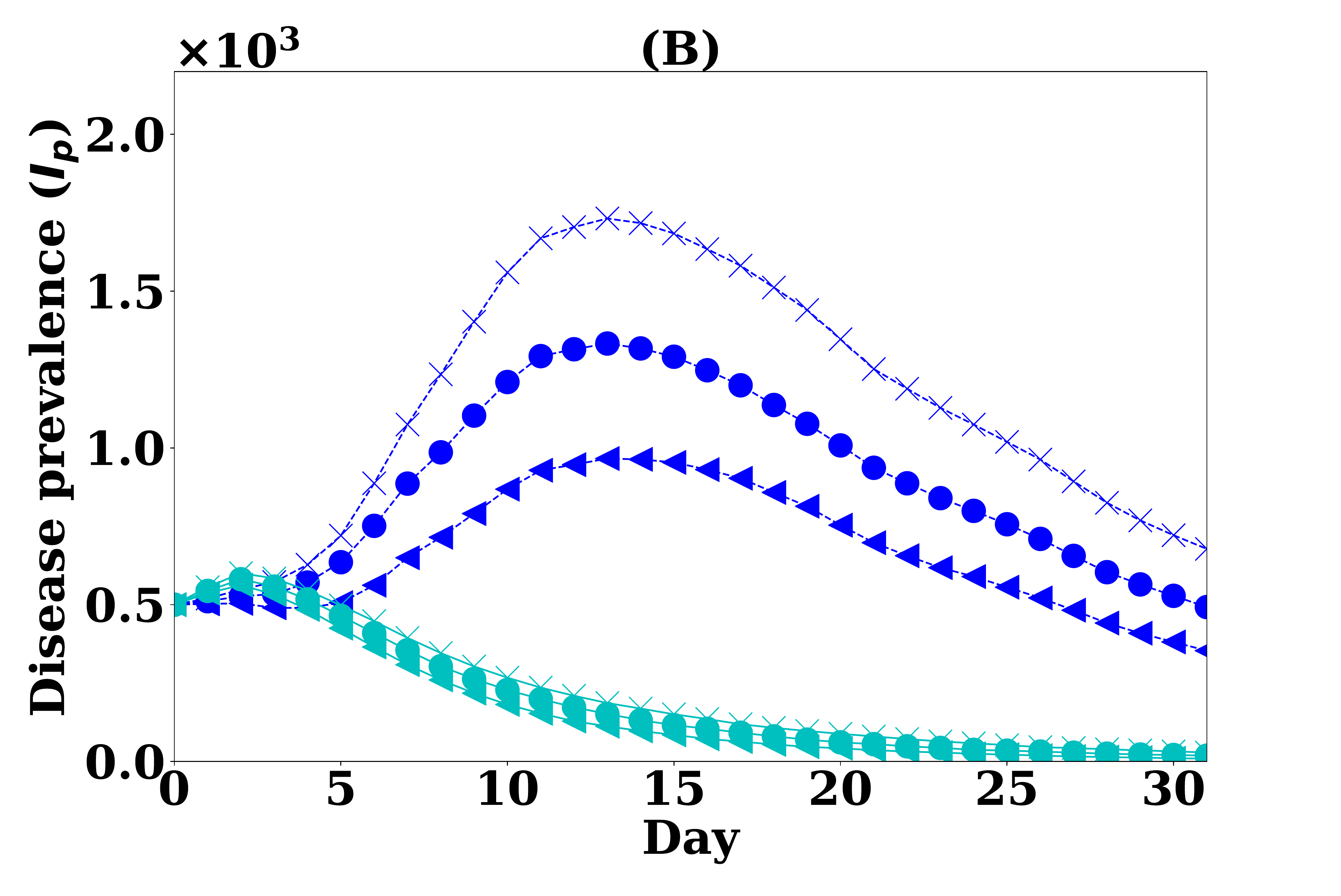}\\
% \vspace{0.5em}
\includegraphics[width=0.45\linewidth, height=5cm]{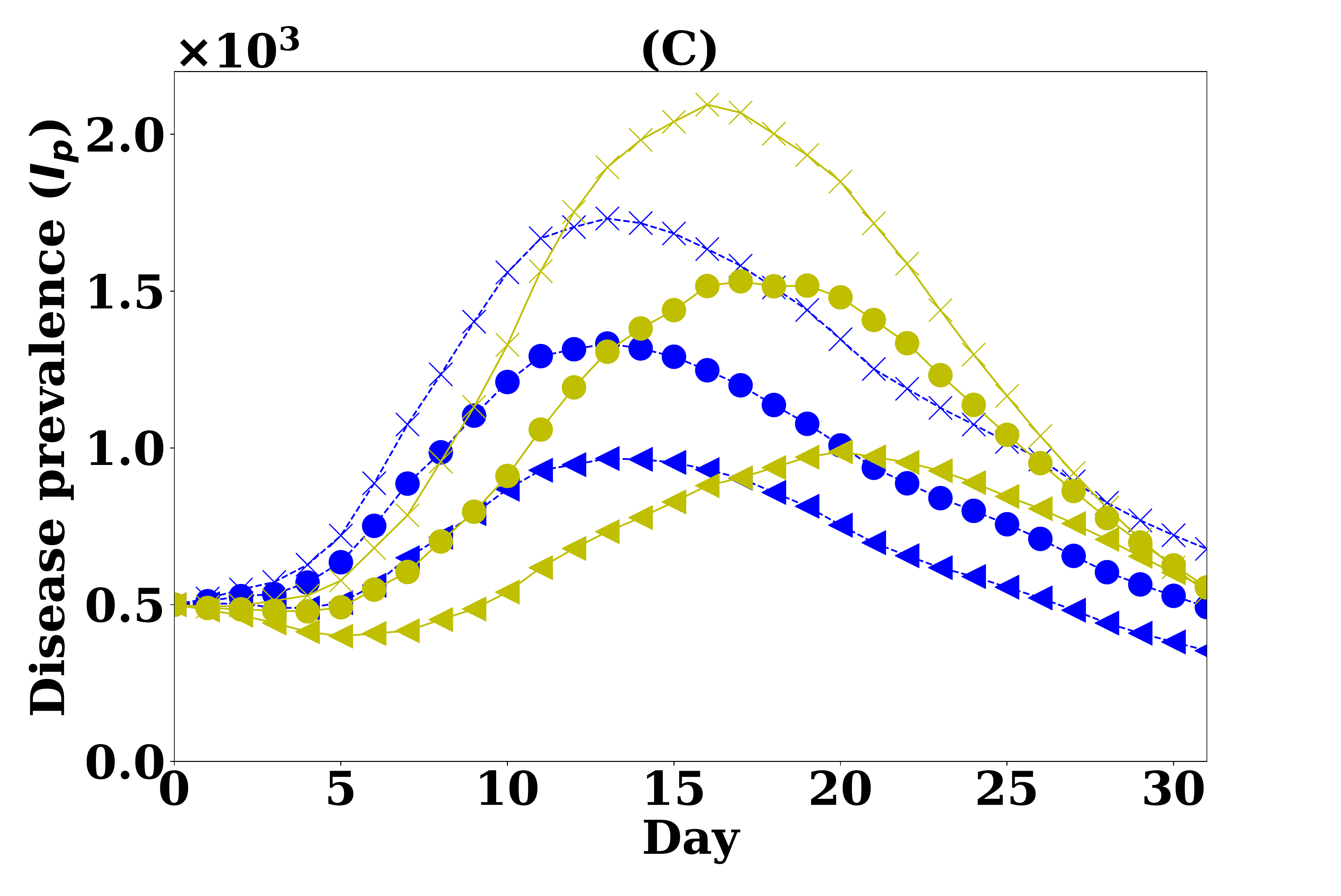}~
\includegraphics[width=0.45\linewidth, height=5cm]{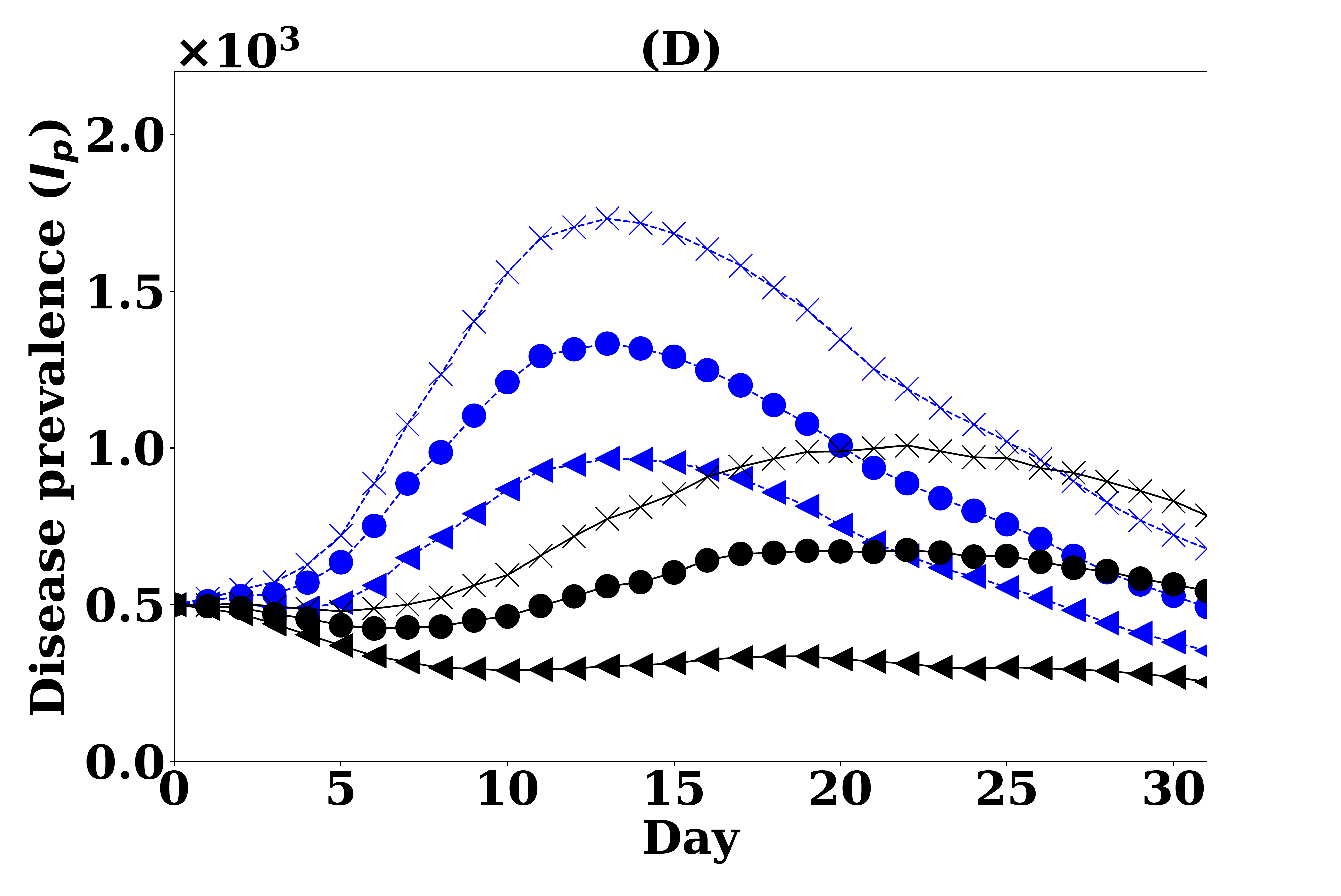}
\caption{Sensitivity analysis of the contact networks with various particle decay rates $r_t$. The diffusion dynamics of DDT network at various $r_t$ are compared with that of the generated networks: A) GDH network, B) BDH network, C) GDT network and D) BDT network}
\label{fig:netsens}
\end{figure}

\begin{figure}[h!]
\begin{tikzpicture}
    \begin{customlegend}[legend columns=5,legend style={at={(0.12,1.02)},draw=none,column sep=2ex }, legend entries={DDT, GDH, BDH, GDT, BDT, $\sigma=0.33$, $\sigma=0.40$, $\tau=4$ days}]
    \addlegendimage{solid,line legend, color=blue}
    \addlegendimage{solid, color=red}   
    \addlegendimage{solid, color=cyan}
    \addlegendimage{color=yellow}
    \addlegendimage{color=black}
    \addlegendimage{mark=triangle}
    \addlegendimage{mark=o}
    \addlegendimage{mark= x }
    \end{customlegend}
 \end{tikzpicture}
\vspace{3ex}
\centering
\includegraphics[width=0.45\linewidth, height=5cm]{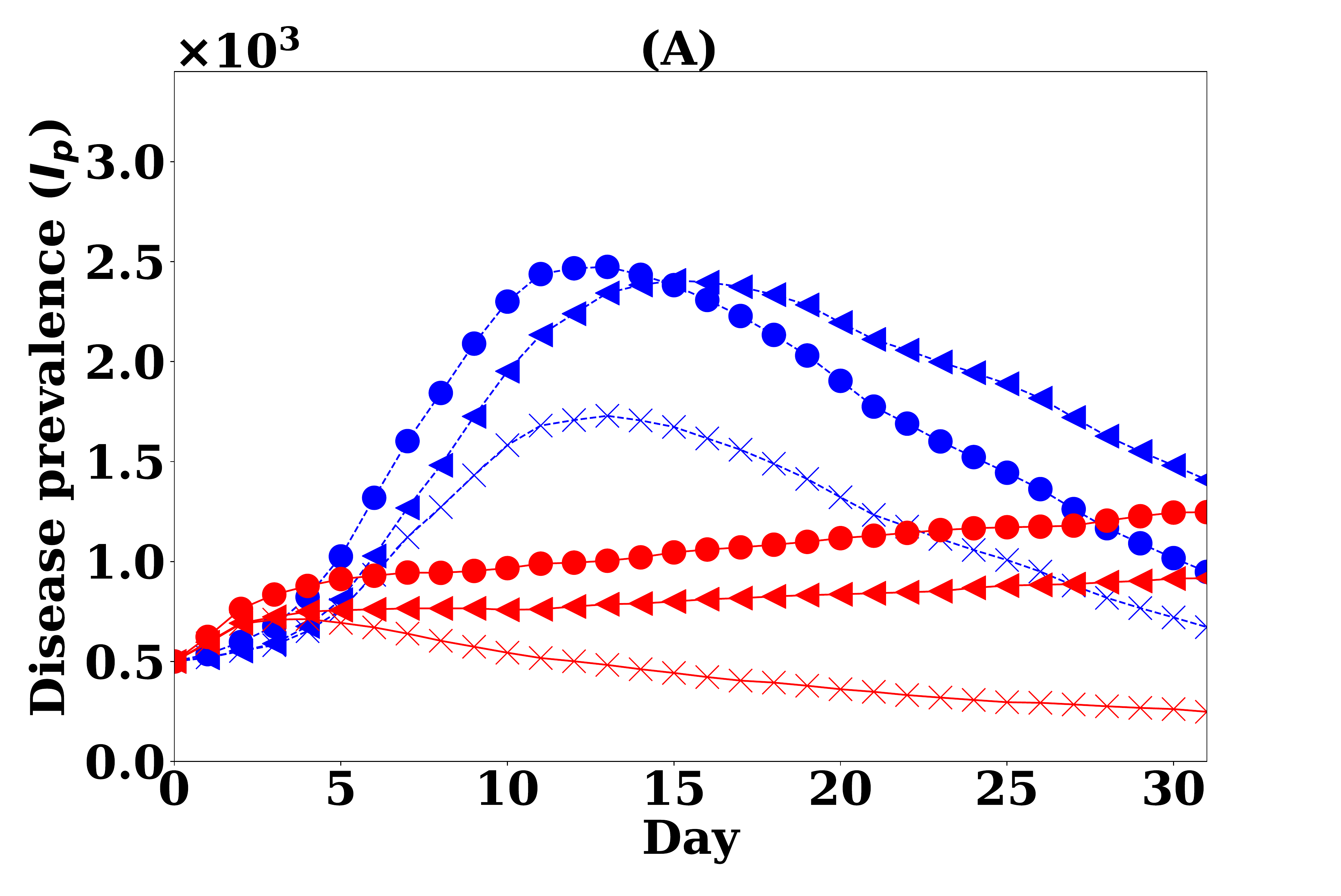}~
\includegraphics[width=0.45\linewidth, height=5cm]{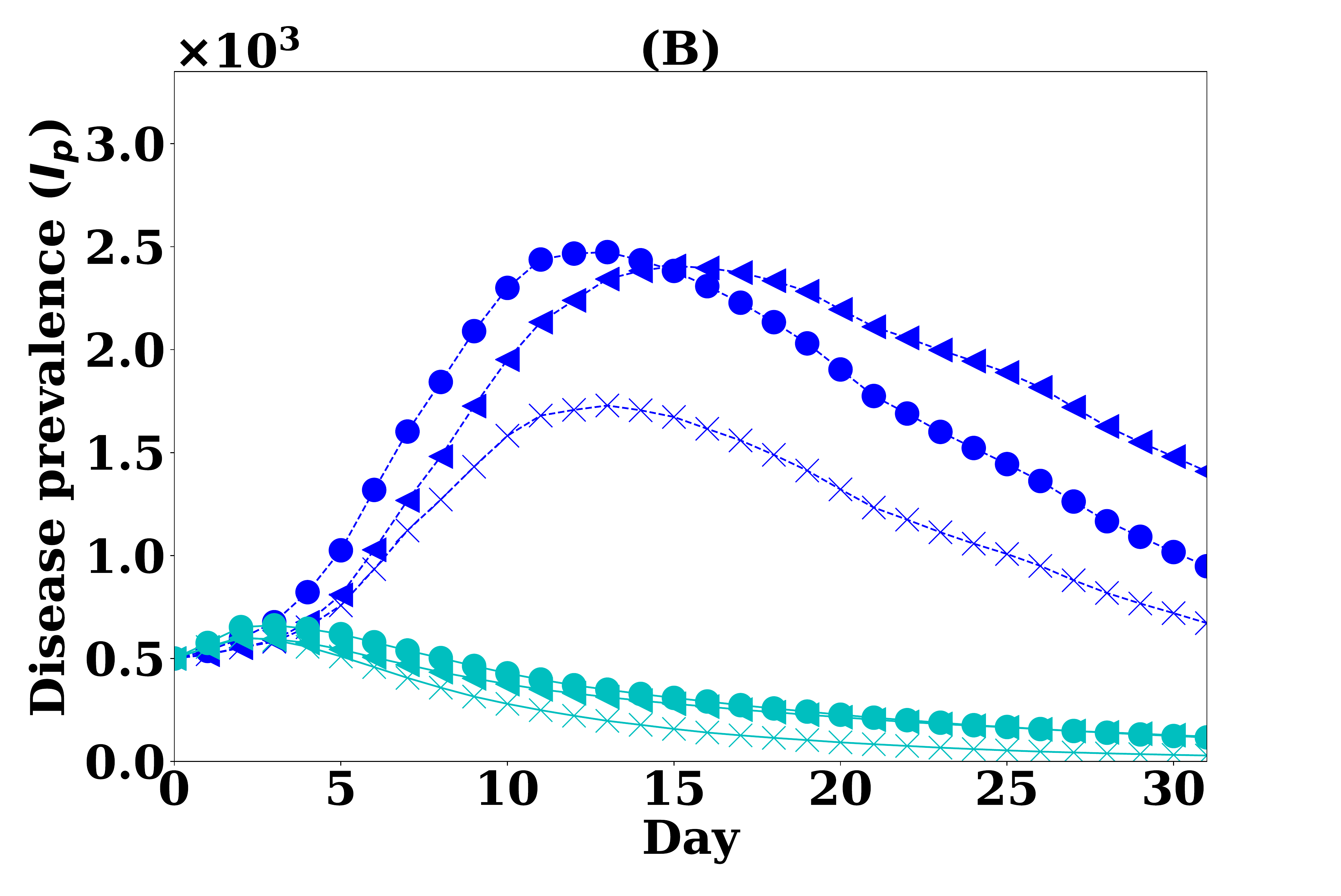}\\
% \vspace{0.5em}
\includegraphics[width=0.45\linewidth, height=5cm]{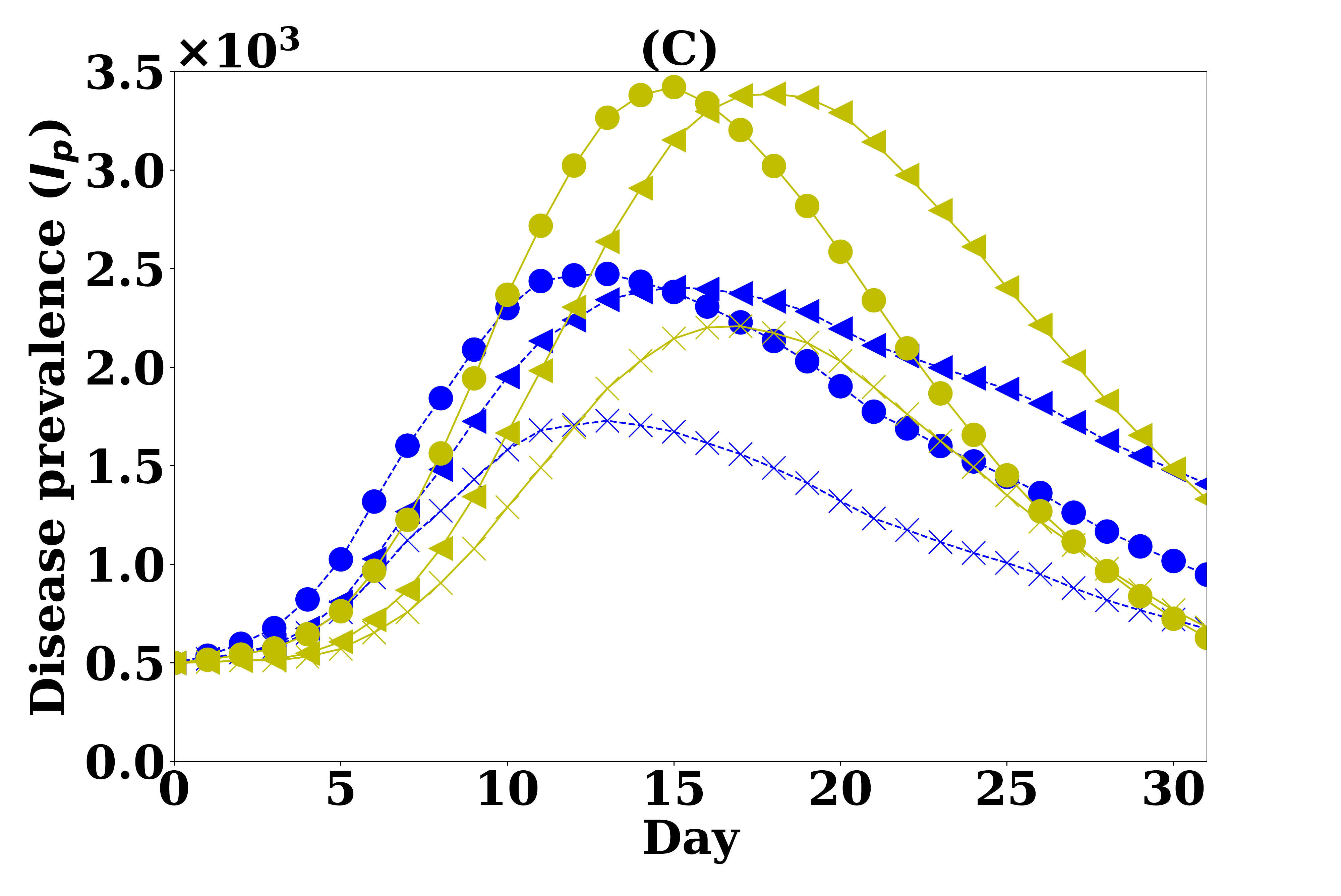}~
\includegraphics[width=0.45\linewidth, height=5cm]{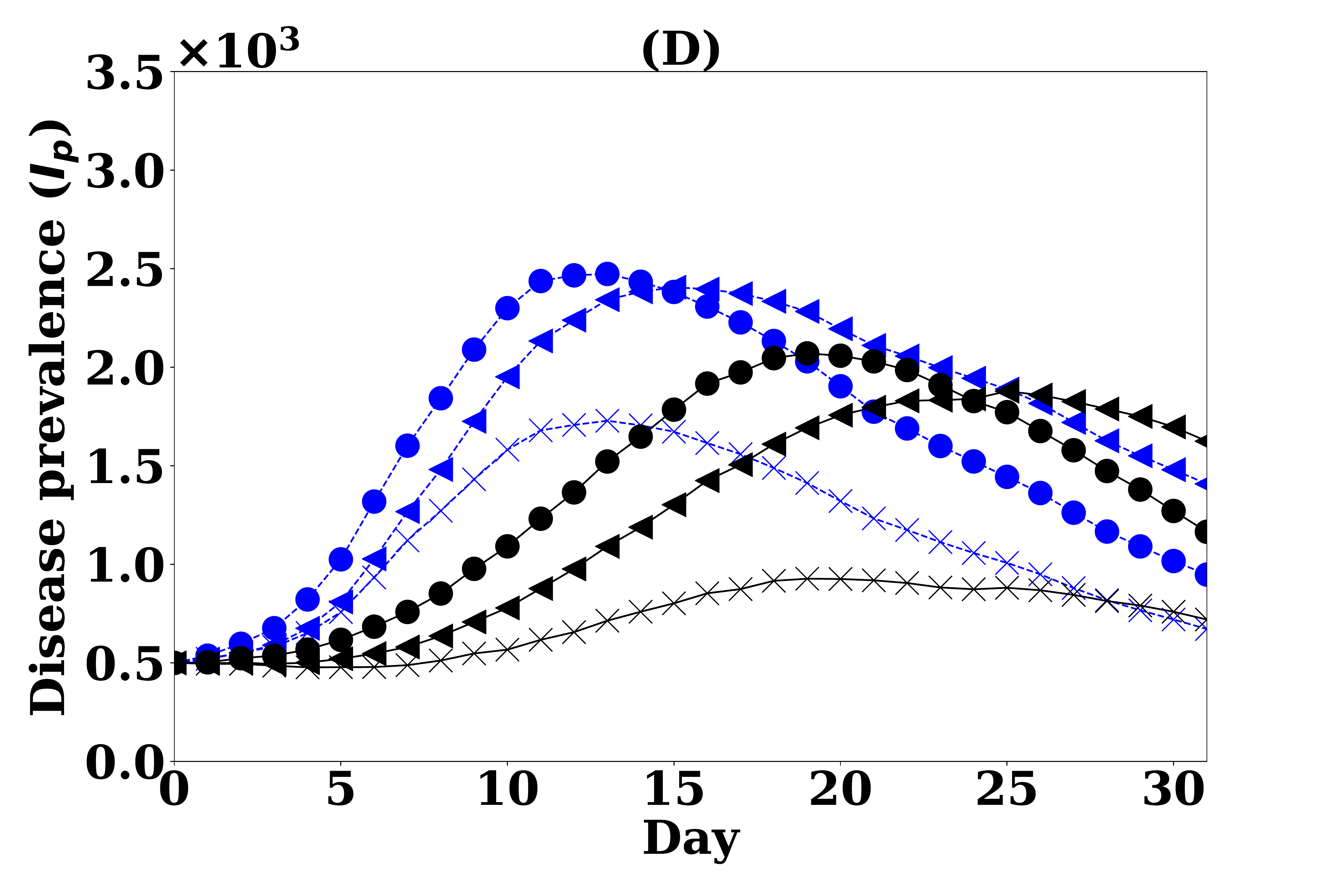}
\caption{Sensitivity analysis of the contact networks with disease parameters infectiousness $\sigma$ and infectious period $\tau$. The diffusion dynamics of DDT network are compared with that of the generated networks: A) GDH network, B) BDH network, C) GDT network and D) BDT network}
\label{fig:netdis}
\end{figure}

Now, the model's sensitivity to the various diffusion parameters is studied. The SPDT model is highly sensitive to the particle decay rates $r_t$. Thus, disease spreading is simulated for three different values of $r_t=\{20,40,60\}$ min for all networks and results are presented in Figure~\ref{fig:netsens}. The homogeneous network GDH can model the variation in diffusion dynamics with $r_t$. However, it underestimates the diffusion at all $r_t$. The prediction variations are increased with $r_t$ where the average prediction variation reaches to 70\% with the average accumulative variation of 51\%. The other homogeneous network BDH cannot vary the diffusion with $r_t$, as this network has only direct transmission links. The BDH network always has more than 72\% variation in daily prediction for any value of $r_t$. The proposed heterogeneous network GDT shows increasing daily prediction variation. However, it is still 23\% better than the BDT networks at particle decay rates $r_t=20$ min. The model also maintains the same 32\% improvement in the cumulative prediction variations at $r_t=20$ min compared to the GDT model. Although the prediction variation increases with the raising of particle decay rates, the model still predicts well compared to other models.

The diffusion dynamics are also influenced by the properties of contagious items. Therefore, the model should respond well to varying the properties in the contagious items as well. Now, we vary the infectiousness $\sigma$ and infectious period $\tau$ are varied keeping $r_t=60$ min constant to analyse how the generated network responds. The results are presented in Figure~\ref{fig:netdis}. The disease prevalence gets stronger in all networks for increasing infectiousness from $\sigma=0.33$ to from $\sigma=0.40$. However, the increase in $I_p$ are different in different networks. The variation in the GDT network follow the same trends of the DDT network while the remaining networks deviate significantly. In the GDT network, the average daily variation in $I_p$ compared to that of DDT network is 13\%. For the other networks, the average daily variations in $I_p$ increases for increasing $\sigma=0.40$ and the maximum is 70\% in the GDH networks. Therefore, the GDT network responds similarly to DDT network in changing $\sigma$. The similar trends are found for the increase in infectious period $\tau=4$ days. The average daily variation in $I_p$ for GDT network is 12\% while all others networks show higher average daily variation in $I_p$ with a maximum in BDH network. The overall responses of GDT network for changing in the properties of contagious items follow the trends of the DDT network.

\begin{figure}[h!]
\centering
\includegraphics[width=0.45\linewidth, height=5cm]{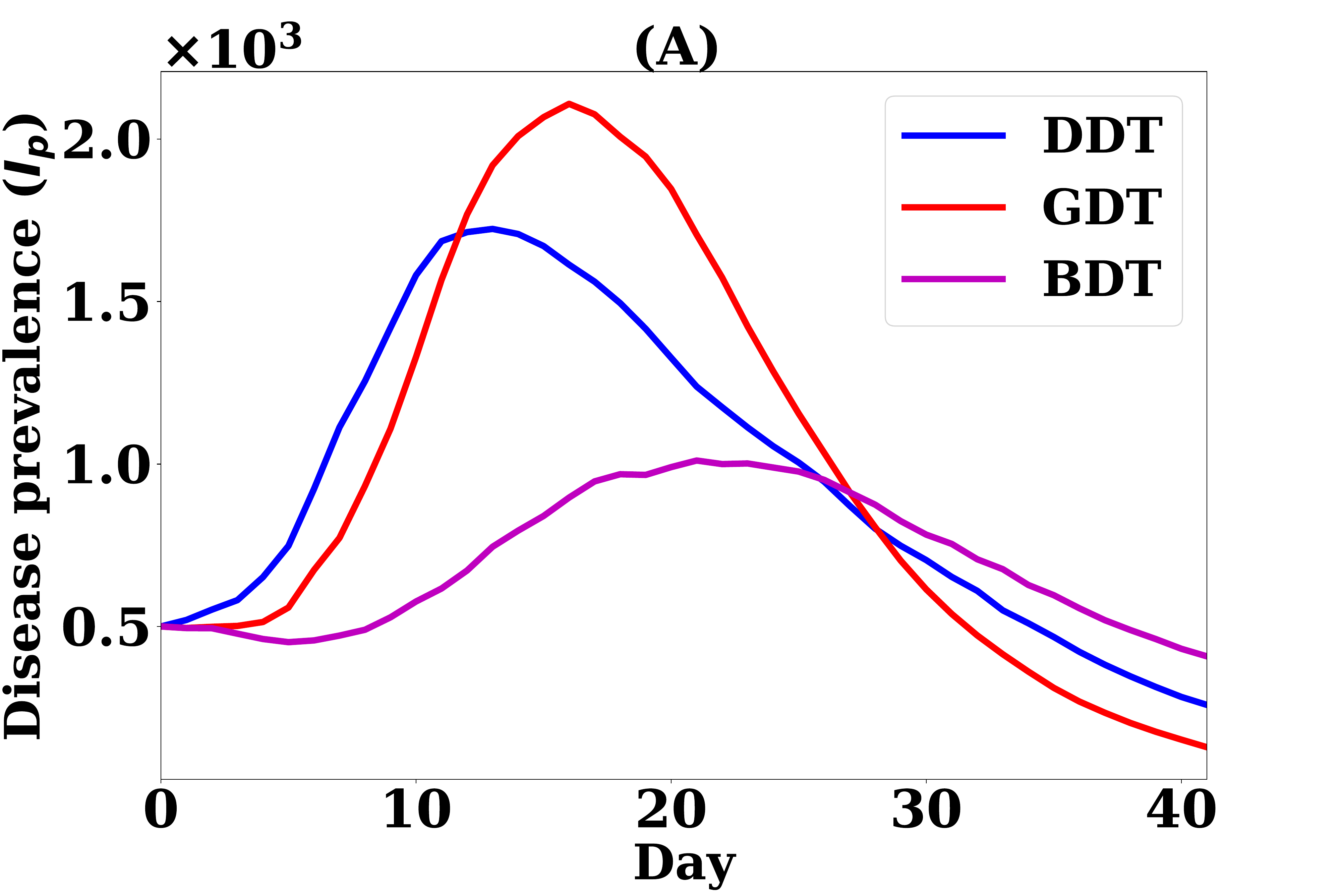}~
\includegraphics[width=0.45\linewidth, height=5cm]{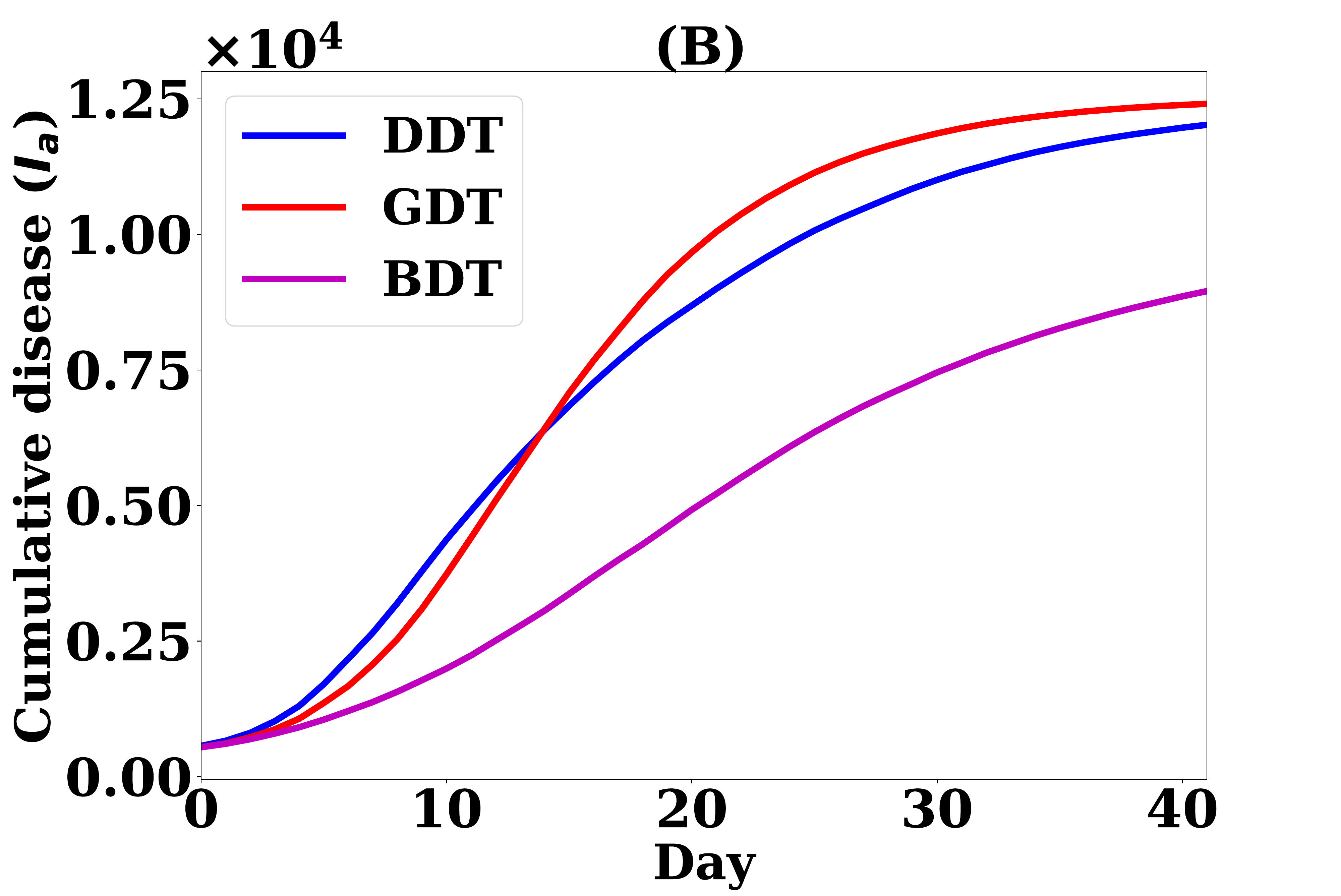}
\caption{Sensitivity of the model with observation period. The diffusion dynamics of DDT network are compared with that of the generated networks: A) disease prevalence dynamics and B) accumulative infection over simulation days}
\label{fig:netsll}
\end{figure}

\begin{figure}[h!]
\begin{tikzpicture}
    \begin{customlegend}[legend columns=5,legend style={at={(0.12,1.02)},draw=none,column sep=2ex }, legend entries={GDT, BDT, 0.364M, 0.50M, 1.0M}]
    \addlegendimage{dash dot,line legend, color=black}
    \addlegendimage{solid, color=black}   
    \addlegendimage{solid, color=red}
    \addlegendimage{solid, color=blue}
    \addlegendimage{solid, color=green}
    \end{customlegend}
 \end{tikzpicture}
\vspace{4ex}
\centering
\includegraphics[width=0.45\linewidth, height=5cm]{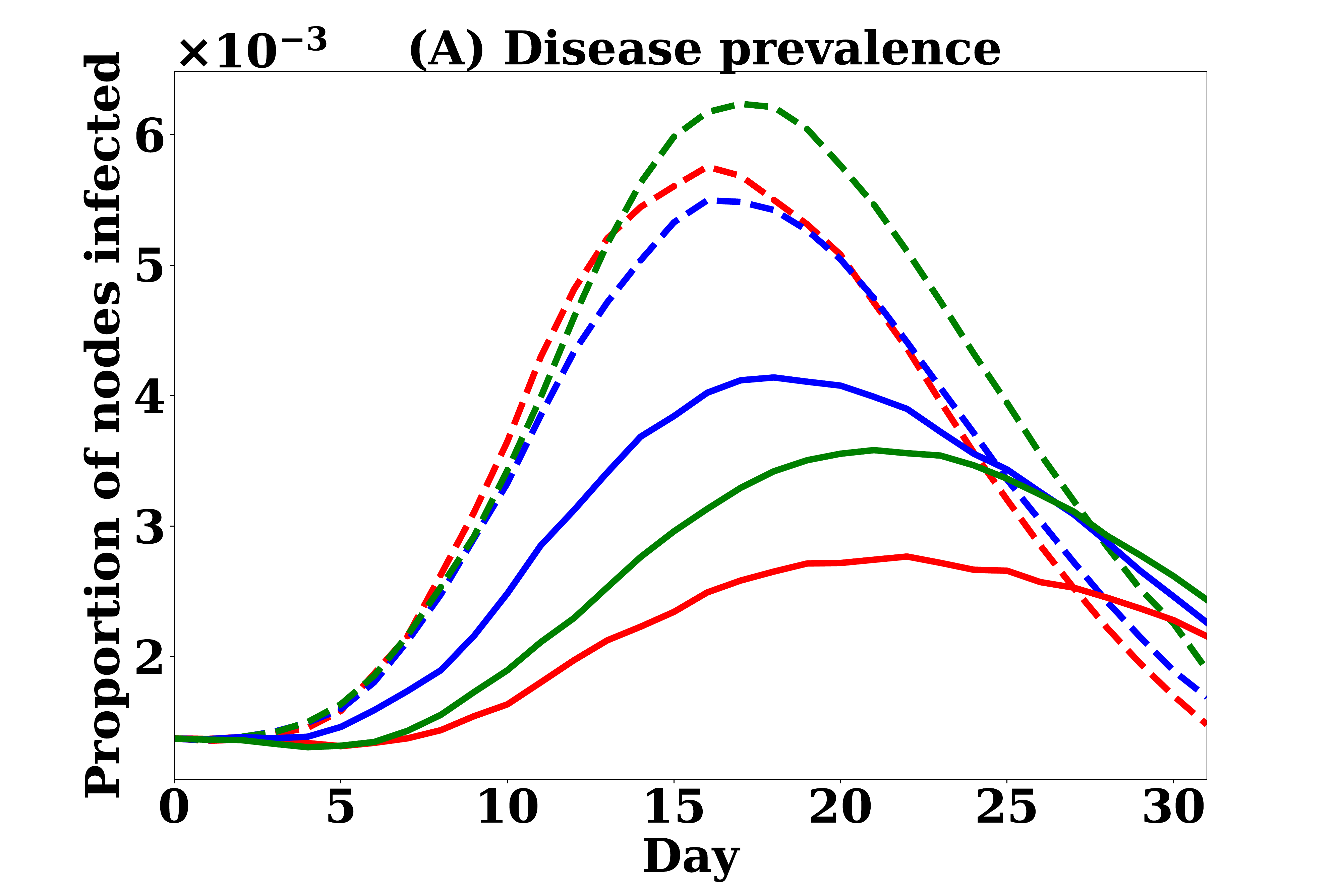}~
\includegraphics[width=0.45\linewidth, height=5cm]{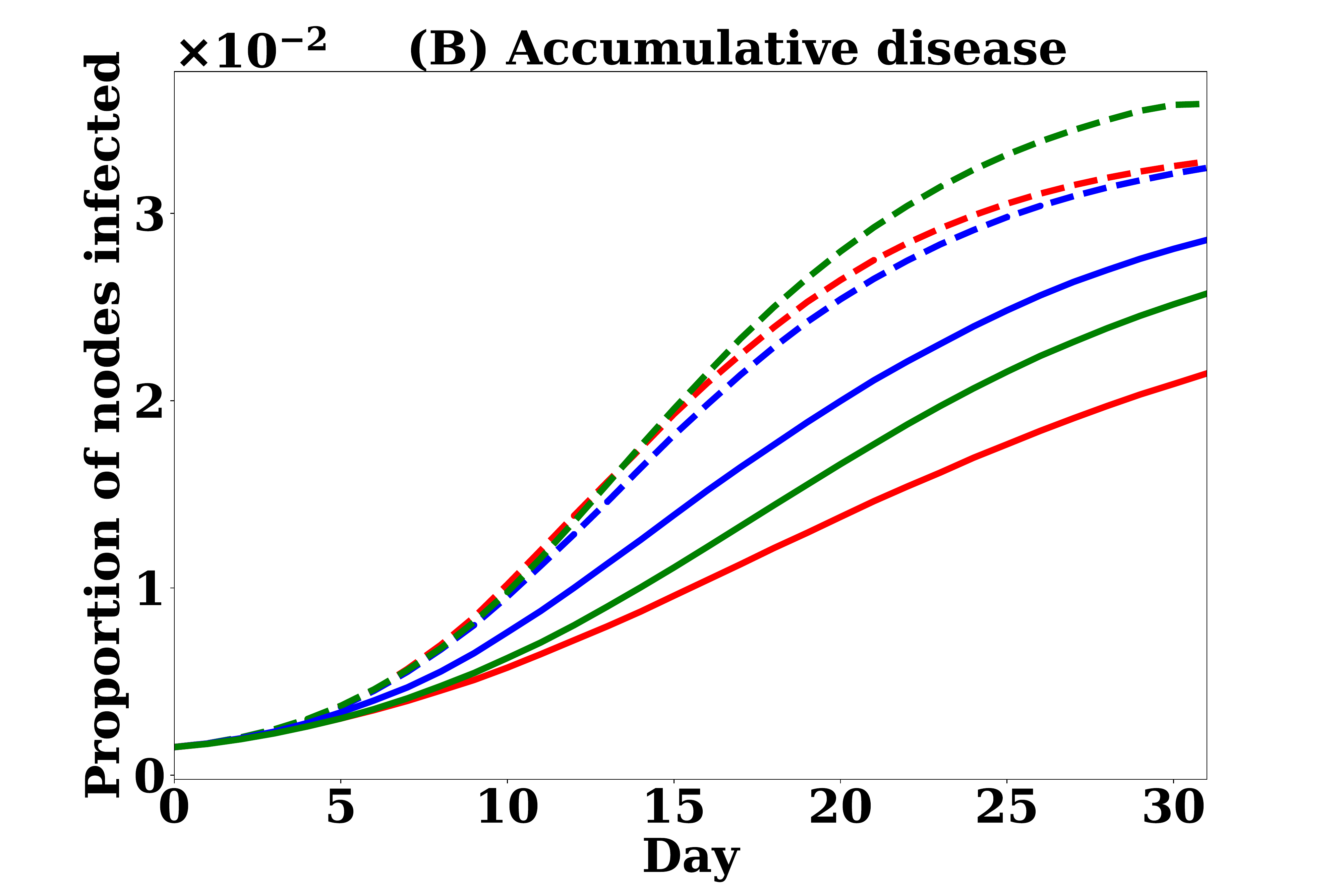}
\caption{Sensitivity of the graph model with network sizes. The diffusion dynamics of DDT network are compared with that of the generated networks: A) disease prevalence dynamics and B) accumulative infection over simulation days}
\label{fig:netszl}
\end{figure}

To understand the response of the graph model for larger scale simulation with large network sizes, the simulations are also run on various sizes of the networks. Two dimensions are considered here: increasing the length of the simulation period and number of nodes. In this experiment, only heterogeneous networks are considered as they assume more realistic diffusion dynamics. For this study, a GDT network and a BDT network with 364K nodes are generated for 42 days period. Then, the real contact network DDT is extended to a network of 42 days repeating all links of a day randomly picked up from 32 days and placing to a random day within days 33 to 42. The results are presented in Figure~\ref{fig:netsll}. The diffusion dynamic is consistent in the GDT network. We have found the similar amount of variation in $I_p$ to the previous experiments. The total infection in the GDT network becomes almost the same at the end of the simulation period. However, the BDT network keep continuing underestimation of diffusion dynamic and total infection as that was in the previous experiments. Thus, our GDT network perform better even with longer simulation period. The model response is also analyzed generating the network with a large number of nodes. For these experiments, the GDT and BDT networks are selected and generated with 0.364M, 0.5M and 1M nodes for 32 days. The simulations are run with the same proportion of seed nodes and results are presented in the Figure~\ref{fig:netszl}. The results show that the GDT network behaves consistently for diffusion dynamic and accumulative diffusion while BDT network shows some variations for increasing node numbers. The generated contact network GDT by our model can capture the characteristics of DDT network efficiently than the other graph models.  

%%%%%%%%%%%%%%%% CONCLUSION %%%%%%%%%%%%%%%5

\section{Conclusion}
In this paper, a novel generative graph model is developed and analysed to support the study of SPDT diffusion dynamics. The developed SPDT graph can capture both direct and indirect interactions for modelling the spread of contagious items. The synthetic dynamic contact networks generated by this graph model are capable of capturing SPDT dynamics, applying reinforcement for repetitive interactions and heterogeneous propensity to engage in interactions. This paper has demonstrated how the model can be generated based on empirical data from a large social networking application and reproduce both network properties and diffusion dynamics of empirical contact networks. The developed model shows 30\% higher efficiency to generate the diffusion dynamics of the real SPDT contact networks. Analysis of the sensitivity to the SPDT diffusion model parameters has demonstrated that the SPDT graph model responds similarly to that of the real SPDT graph. The model is also consistent for a longer simulation period and large network size. 

The previous studies show that contact degree distribution follows a power law. In the SPDT graph model, there are two types of contact degree distribution: activation degree and overall contact set size over the observation period. While both degree distributions follow the power law, we find that using the  homogeneous degree distribution leads to an underestimation of the spread as diffusion is not driven by the higher degree individuals. This is because the disease spreading force is accelerated gradually by the higher degree nodes, which is not reflected in homogeneous networks. ADN networks with power law degree distribution have also not shown similar diffusion dynamics as they can not cover the distribution of indirect links. The integration of active copies of nodes in our model facilitates the inclusion of indirect links and can better mimic the real network. 

The clustering co-efficient is also changed in the network by adding indirect links. The neighbour selection mechanism controls the growth of clustering co-efficient. Survival of the active copy for indirect transmission period can create links and maintain the local clustering co-efficient. The stay time of an individual in a location is found to be long tailed distribution and the developed model for stay time captured the distribution of stay time very closely with simple modelling parameters of transition probability. 

The developed SPDT graph model can be extended for other applications as well. For modelling user activities across multiple social networks, the concept of active copies can be exploited to generate underlying interaction networks. We can create an active copy of a node when they live in an online social network (OSN) and the active copies expire when a user leaves the OSN. Moreover, the concurrent existence of active copies can represent concurrent activities of a user in different OSN~\cite{kong2013inferring,zhang2017identifying}. The graph model can also be applied to study queen message dissemination in social ant-colonies where ants spray a chemical encoded with the queen message and the message is transferred to the ants that encountered the chemical~\cite{richardson2015beyond}. Vector borne disease (transmitting via mosquitoes, flies and mites etc.) spreading can be studied using the SPDT model as well. For example, if a virus is transferred from an infected individual to mosquitoes, the virus can be transferred to the susceptible individuals when the infected individual moves to the other locations~\cite{adams2009man}. Similarly, propagation of mobile malware can be explored by the SPDT graph model where a malicious mobile phone worm can be transferred to the Wi-Fi router and this can be transferred to the susceptible phones when the infected phone is not present~\cite{tang2011exploiting}.  

There are several potential and interesting future research directions on the SPDT graph modelling. In the current SPDT graph, we have only considered that the indirect transmission links can be created for the three hour period as the significant infectious particles can persist up to three hours. However, the indirect links can be created for a longer period for other applications. It is interesting to explore how to create indirect links for any period  of times and how it interacts with the decay rates of infectious items. We have studied the network properties such as degree distributions, clustering co-efficient and temporal properties through simulations. It would be interesting to find the network properties analytically as well as the analytic diffusion dynamics. The community structure and sub-graphs also play vital roles in the modelling of diffusion processes on the contact networks. This could be another interesting research direction where we can investigate how the community structure and sub-graphs are changed for including the indirect links and what is their impact on diffusion dynamics.

\section*{Acknowledgment}

This study was supported by the Australian Research Council Grant (ARC - DP170102794) and Commonwealth Scientific and Industrial Research Organisation (CSIRO). Md.S. was partially supported by the CSIRO and ARC- DP170102794. B.M. was partially supported by the ARC-DP170102794. R.J. and F.d.H. were supported by the CSIRO. The authors gratefully acknowledge the Distributed Sensing System Group, Data61, CSIRO for providing research facilities for this research. 


\bibliographystyle{comnet}
\bibliography{references}

\begin{thebibliography}{00}

\bibitem{adams2009man}
Adams, B. {\&} Kapan, D.~D. (2009)  Man bites mosquito: understanding the
  contribution of human movement to vector-borne disease dynamics. {\em PloS
  one}, \textbf{4}(8), e6763.

\bibitem{alessandretti2017random}
Alessandretti, L., S., K., B., A. {\&} P., N. (2017)  Random walks on
  activity-driven networks with attractiveness. {\em Physical Review E},
  \textbf{95}(5), 052318.

\bibitem{alford1966human}
Alford~RH, Kasel~JA, G.~P. {\&} V, K. (1966)  Human influenza resulting from
  aerosol inhalation. {\em Proceedings of the Society for Experimental Biology
  and Medicine}, \textbf{122}(3), 800--804.

\bibitem{chattopadhyay2014fitting}
C., S. {\&} Murthy, C.and~P., S.~K. (2014)  Fitting truncated geometric
  distributions in large scale real world networks. {\em Theoretical Computer
  Science}, \textbf{551}.

\bibitem{de2014role}
De~Arruda, G.~F., B., A.~L., Rodr{\'\i}guez, P.~M., R., F.~A., M., Y. {\&}
  da~F.~C., L. (2014)  Role of centrality for the identification of influential
  spreaders in complex networks. {\em Physical Review E}, \textbf{90}(3),
  032812.

\bibitem{fernstrom2013aerobiology}
Fernstrom, A. {\&} Goldblatt, M. (2013)  Aerobiology and its role in the
  transmission of infectious diseases. {\em Journal of pathogens},
  \textbf{2013}.

\bibitem{freeman1978centrality}
Freeman, L.~C. (1978)  Centrality in social networks conceptual clarification.
  {\em Social networks}, \textbf{1}(3), 215--239.

\bibitem{han2014risk}
Han, Z., Weng, W., Huang, Q. {\&} Zhong, S. (2014)  A Risk Estimation Method
  for Airborne Infectious Diseases Based on Aerosol Transmission in Indoor
  Environment. In {\em Proceedings of the World Congress on Engineering},
  volume~2.

\bibitem{holme2015modern}
Holme, P. (2015)  Modern temporal network theory: a colloquium. {\em The
  European Physical Journal B}, \textbf{88}(9), 234.

\bibitem{huang2016insights}
Huang, C., Liu, X., Sun, S., Li, S.~C., Deng, M., He, G., Zhang, H., Wang, C.,
  Zhou, Y., Zhao, Y.  et~al. (2016)  Insights into the transmission of
  respiratory infectious diseases through empirical human contact networks.
  {\em Scientific Reports}, \textbf{6}.

\bibitem{hui2005pocket}
Hui, P., Chaintreau, A., Scott, J., Gass, R., Crowcroft, J. {\&} Diot, C.
  (2005)  Pocket switched networks and human mobility in conference
  environments. In {\em Proceedings of the 2005 ACM SIGCOMM workshop on
  Delay-tolerant networking}, pages 244--251. ACM.

\bibitem{jacquet2010information}
Jacquet, P., Mans, B. {\&} Rodolakis, G. (2010)  Information propagation speed
  in mobile and delay tolerant networks. {\em IEEE Transactions on Information
  Theory}, \textbf{56}(10), 5001--5015.

\bibitem{karsai2014time}
Karsai, M., Perra, N. {\&} Vespignani, A. (2014)  Time varying networks and the
  weakness of strong ties. {\em Scientific reports}, \textbf{4}, srep04001.

\bibitem{keeling2008modeling}
Keeling, M.~J. {\&} Rohani, P. (2008) {\em Modeling infectious diseases in
  humans and animals}.
Princeton University Press.

\bibitem{kong2013inferring}
Kong, X., Zhang, J. {\&} Yu, P.~S. (2013)  Inferring anchor links across
  multiple heterogeneous social networks. In {\em Proceedings of the 22nd ACM
  international conference on Information \& Knowledge Management}, pages
  179--188. ACM.

\bibitem{kovacs2019network}
Kov{\'a}cs, I.~A., Luck, K., Spirohn, K., Wang, Y., Pollis, C., Schlabach, S.,
  Bian, W., Kim, D.-K., Kishore, N., Hao, T.  et~al. (2019)  Network-based
  prediction of protein interactions. {\em Nature communications},
  \textbf{10}(1), 1240.

\bibitem{laurent2015calls}
Laurent, G., S.{\"a}ki, J. {\&} Karsai, M. (2015)  From calls to communities: a
  model for time-varying social networks. {\em The European Physical Journal
  B}, \textbf{88}.

\bibitem{lindsley2015viable}
Lindsley, W.~G., Noti, J.~D., Blachere, F.~M., Thewlis, R.~E., Martin, S.~B.,
  Othumpangat, S., Noorbakhsh, B., Goldsmith, W.~T., Vishnu, A., Palmer, J.~E.
  et~al. (2015)  Viable influenza A virus in airborne particles from human
  coughs. {\em Journal of occupational and environmental hygiene},
  \textbf{12}(2), 107--113.

\bibitem{moon2019spatio}
Moon, S.~A., Cohnstaedt, L.~W., McVey, D.~S. {\&} Scoglio, C.~M. (2019)  A
  spatio-temporal individual-based network framework for West Nile virus in the
  USA: Spreading pattern of West Nile virus. {\em PLoS computational biology},
  \textbf{15}(3), e1006875.

\bibitem{nagatani2019epidemic}
Nagatani, T., Ichinose, G. {\&} Tainaka, K.-i. (2019)  Epidemic spreading of
  random walkers in metapopulation model on an alternating graph. {\em Physica
  A: Statistical Mechanics and its Applications}, \textbf{520}, 350--360.

\bibitem{ogura2019optimal}
Ogura, M., Preciado, V.~M. {\&} Masuda, N. (2019)  Optimal Containment of
  Epidemics over Temporal Activity-Driven Networks. {\em SIAM Journal on
  Applied Mathematics}, \textbf{79}(3), 986--1006.

\bibitem{perra2012activity}
Perra, N., Gon{\c{c}}alves, B., Pastor-Satorras, R. {\&} Vespignani, A. (2012)
  Activity driven modeling of time varying networks. {\em Scientific reports},
  \textbf{2}.

\bibitem{porter2016network}
Porter, M.~A. {\&} Bianconi, G. (2016)  Network analysis and modelling: Special
  issue of European Journal of Applied Mathematics. {\em European Journal of
  Applied Mathematics}, \textbf{27}(6), 807--811.

\bibitem{richardson2015beyond}
Richardson, T.~O. {\&} Gorochowski, T.~E. (2015)  Beyond contact-based
  transmission networks: the role of spatial coincidence. {\em Journal of The
  Royal Society Interface}, \textbf{12}(111), 20150705.

\bibitem{rushton2019transmission}
Rushton, S.~P., Sanderson, R.~A., Reid, W.~D., Shirley, M.~D., Harris, J.~P.,
  Hunter, P.~R. {\&} O'Brien, S.~J. (2019)  Transmission routes of rare
  seasonal diseases: the case of norovirus infections. {\em Philosophical
  Transactions of the Royal Society B}, \textbf{374}(1776), 20180267.

\bibitem{scherrer2008description}
Scherrer, A., Borgnat, P., Fleury, E., Guillaume, J.-L. {\&} Robardet, C.
  (2008)  Description and simulation of dynamic mobility networks. {\em
  Computer Networks}, \textbf{52}(15), 2842--2858.

\bibitem{scholtes2016higher}
Scholtes, I., W., N. {\&} G., A. (2016)  Higher-order aggregate networks in the
  analysis of temporal networks: path structures and centralities. {\em The
  European Physical Journal B}, \textbf{89}(3), 61.

\bibitem{sekamatte2019individual}
Sekamatte, M., Riad, M.~H., Tekleghiorghis, T., Linthicum, K.~J., Britch,
  S.~C., Richt, J.~A., Gonzalez, J. {\&} Scoglio, C.~M. (2019)
  Individual-based network model for Rift Valley fever in Kabale District,
  Uganda. {\em PloS one}, \textbf{14}(3), e0202721.

\bibitem{shahzamal2017airborne}
Shahzamal, M., Jurdak, R., Arablouei, R., Kim, M., Thilakarathna, K. {\&} Mans,
  B. (2017)  Airborne Disease Propagation on Large Scale Social Contact
  Networks. In {\em Proceedings of the 2nd Int. Workshop on Social Sensing},
  pages 35--40. ACM.

\bibitem{shahzamaROS}
Shahzamal, M., Jurdak, R., Mans, B. {\&} de~Hoog, F. (2009)  Indirect
  Interactions Influence Contact Network Structure and Diffusion Dynamics. {\em
  Royal Society Open Science}, \textbf{6}(8), 190845.

\bibitem{shahzamal2018impact}
Shahzamal, M., Jurdak, R., Mans, B., El~Shoghri, A. {\&} De~Hoog, F. (2018)
  Impact of Indirect Contacts in Emerging Infectious Disease on Social
  Networks. In {\em Pacific-Asia Conference on Knowledge Discovery and Data
  Mining}, pages 53--65. Springer.

\bibitem{shahzamal2016smartphones}
Shahzamal, M. {\&} Pervez, M.~F. (2016)  Smartphones based warning messaging
  system for marine fisheries and its characteristics. In {\em 2016 5th
  International Conference on Informatics, Electronics and Vision (ICIEV)},
  pages 111--116. IEEE.

\bibitem{starnini2013modeling}
Starnini, M., Baronchelli, A. {\&} Pastor-Satorras, R. (2013)  Modeling human
  dynamics of face-to-face interaction networks. {\em Physical review letters},
  \textbf{110}(16), 168701.

\bibitem{stehle2011simulation}
Stehl{\'e}, J., Voirin, N., Barrat, A., Cattuto, C., Colizza, V., Isella, L.,
  R{\'e}gis, C., Pinton, J.-F., Khanafer, N., Van~den Broeck, W.  et~al. (2011)
   Simulation of an SEIR infectious disease model on the dynamic contact
  network of conference attendees. {\em BMC medicine}, \textbf{9}(1), 87.

\bibitem{sun2015contrasting}
Sun, K., Baronchelli, A. {\&} Perra, N. (2015)  Contrasting effects of strong
  ties on SIR and SIS processes in temporal networks. {\em The European
  Physical Journal B}, \textbf{88}(12), 326.

\bibitem{sze2010review}
Sze~To, G. {\&} Chao, C. (2010)  Review and comparison between the Wells--Riley
  and dose-response approaches to risk assessment of infectious respiratory
  diseases. {\em Indoor Air}, \textbf{20}(1), 2--16.

\bibitem{tang2011exploiting}
Tang, J., Mascolo, C., Musolesi, M. {\&} Latora, V. (2011)  Exploiting temporal
  complex network metrics in mobile malware containment. In {\em 2011 IEEE
  International Symposium on a World of Wireless, Mobile and Multimedia
  Networks}, pages 1--9. IEEE.

\bibitem{thilakarathna2016deep}
Thilakarathna, K., Seneviratne, S., Gupta, K., Kaafar, M.~A. {\&} Seneviratne,
  A. (2016)  A deep dive into location-based communities in social discovery
  networks. {\em Computer Communications}.

\bibitem{toth2015role}
Toth, D.~J., Leecaster, M., Pettey, W.~B., Gundlapalli, A.~V., Gao, H., Rainey,
  J.~J., Uzicanin, A. {\&} Samore, M.~H. (2015)  The role of heterogeneity in
  contact timing and duration in network models of influenza spread in schools.
  {\em Journal of The Royal Society Interface}, \textbf{12}(108), 20150279.

\bibitem{xu2017synthetic}
Xu, Z., Glass, K., Lau, C.~L., Geard, N., Graves, P. {\&} Clements, A. (2017)
  A synthetic population for modelling the dynamics of infectious disease
  transmission in American Samoa. {\em Scientific reports}, \textbf{7}(1),
  16725.

\bibitem{yan2018infectious}
Yan, J., Grantham, M., Pantelic, J., de~Mesquita, P. J.~B., Albert, B., Liu,
  F., Ehrman, S., Milton, D.~K., Adamson, W., Beato-Arribas, B.  et~al. (2018)
  Infectious virus in exhaled breath of symptomatic seasonal influenza cases
  from a college community. {\em Proceedings of the National Academy of
  Sciences}, page 201716561.

\bibitem{zhang2017identifying}
Zhang, P., Lu, T., Gu, H. {\&} Gu, N. (2017)  Identifying user identity across
  social network sites based on overlapping relationship and social
  interaction. In {\em Proceedings of the 12th Chinese Conference on Computer
  Supported Cooperative Work and Social Computing}, pages 25--32. ACM.

\bibitem{zhang2016modelling}
Zhang, Y., C., J., Z., S.-M., Z., Q. {\&} L., X. (2016)  Modelling temporal
  networks of human face-to-face contacts with public activity and individual
  reachability. {\em The European Physical Journal B}, \textbf{89}(2), 26.

\end{thebibliography}

\begin{appendices}
\section{Appendix}
\subsection{Formulating Maximum Likelihood Estimation for Node Activation}
The distribution function for the frequency of node activation given $z$ number of time step in the system, $\rho$ scaling parameter for active periods $t_a$ and the waiting time parameter $q$, we can write
\begin{equation*}
Pr(h\mid q)=\frac{\left(\frac{z \rho q}{q+\rho}\right)^ {h}e^{-\frac{z \rho q}{q+\rho}}}{h!}
\end{equation*}

Thus, the Likelihood function for a given set $h=\{h_1,h_2\ldots,h_m\}$ is 
\begin{equation*}
L(q\mid h)=\prod_{i=1}^{m}\frac{\left(\frac{z \rho q}{q+\rho}\right)^ {h_1}e^{-\frac{z \rho q}{q+\rho}}}{h_1!}
\end{equation*}

Taking log on the both sides as it increases monotonically, we obtain
\begin{equation*}
\ln\left( L\left(q\mid h\right)\right)=\ln \left(\frac{zq\rho}{\rho +q}\right)\sum_{i=1}^{m}h_i-\frac{mzq\rho}{q+\rho}-\sum_{i=1}^{m}\ln h_i
\end{equation*}

Differentiating partial derivatives with respect to $q$, we get
\begin{equation*}
\frac{\partial \ln \left(L\right)}{\partial q}=\frac{\rho}{q(\rho+q)}\sum_{i=1}^{m}h_i-\frac{mz\rho^2}{(q+\rho)^2}
\end{equation*}

At the maximum, $\frac{\partial \ln \left(L\right)}{\partial q}=0$ and we can write

\begin{equation*}
0=\frac{1}{q}\sum_{i=1}^{m}h_i-\frac{mz\rho}{q+\rho}
\end{equation*}

Thus, the maximum likelihood estimation condition will be
\begin{equation}
\frac{qz\rho}{q+\rho}=\frac{1}{m}\sum_{i=1}^{m}h_i
\end{equation}

\subsection{Link creation delay distribution}

The probability function for link creation delay $t$ is given by the following equation
\begin{equation*}
f\left ( t_{c}\mid p_{c} ,t_{a}  \right )=\frac{p_{c} (1-p_{c})^{t_{c}-1} }{1-(1-p_{c})^{t_{a}+\delta}}
\end{equation*}
The probability of $t_c$ given $p_c$ can be written according to law of total probability as
\begin{equation*}
f\left(t_c\mid p_c\right)=\int f\left(t_c\mid p_c,t_a\right) f\left(t_a\right) dt_a
\end{equation*}
Approximating the above equation for m values of $t_a$ applying quadrature, we get
\begin{equation*}
f\left(t_c\mid p_c\right)\approx \frac{1}{m} \sum_{l=1}^{m} f\left(t_c\mid p_c,t^{l}_{a}\right)
\end{equation*}
\begin{equation*}
\approx \frac{1}{m} \sum_{l=1}^{m} \frac{p_{c} (1-p_{c})^{t_{c}-1} }{1-(1-p_{c})^{t^{l}_{a}+\delta}}
\end{equation*}

Now we define a likelihood function $L(p_c\mid t_c)$ for the probability distribution function $f(t_c\mid p_c)$. Function $L(p_c\mid t_c)$ will be maximizes for the observations $t^{1}_{c},t^{2}_{c},\ldots,t^{n}_{c}$. The approximated likelihood function can be written as
\begin{equation*}
L(p_c\mid t_c)\approx \prod_{k=1}^{n}\frac{1}{m} \sum_{l=1}^{m} \frac{p_{c} (1-p_{c})^{t^{k}_{c}-1} }{1-(1-p_{c})^{t^{l}_{a}+\delta}}
\end{equation*}
Since the log function is monotonically increasing, we can maximize $\ln L(p_c\mid t_c)$ instead $L(p_c\mid t_c)$. Therefore, we obtain
\begin{equation*}
\ln L(p_c\mid t_c)= n\ln(p_c)-n\ln(m) +\sum_{k=1}^{n} \ln \left(\sum_{l=1}^{m} \frac{(1-p_{c})^{t^{k}_{c}-1} }{1-(1-p_{c})^{t^{l}_{a}+\delta}}\right)
\end{equation*}
Differentiating the above equation, we get
\begin{equation*}
\frac{\partial}{\partial p_c}\left(\ln L(p_c\mid t_c)\right)=\frac{n}{p_c}-\sum_{k=1}^{n} \frac{\sum_{l=1}^{m}\frac{(t_{c}^{k}-1)(1-p_c)^{t_{c}^{k}-2}(1-(1-p_c)^{t_a^{l}+\delta})+(t_{a}^{l}+\delta)(1-p_c)^{t_c^k +t_a^l +\delta -2}}{(1-(1-p_c)^{t^{l}_{t_a}+\delta})^2}}{\sum_{l=1}^{m}\frac{(1-p_{c})^{t^{k}_{c}-1} }{1-(1-p_{c})^{t^{l}_{a}+\delta}}}
\end{equation*}
For maximizing, $\hat p_c$ we set $\frac{\partial}{\partial p_c}\left(\ln L(p_c\mid t_c)\right)=0$ and get
\begin{equation}
0=\frac{n}{p_c}-\sum_{k=1}^{n} \frac{\sum_{l=1}^{m}\frac{(t_{c}^{k}-1)(1-(1-p_c)^{t_a^{l}+\delta})+(t_{a}^{l}+\delta)(1-p_c)^{t_a^l +\delta}}{(1-p_c)(1-(1-p_c)^{t^{l}_{t_a}+\delta})^2}}{\sum_{l=1}^{m} ((1-(1-p_{c})^{t^{l}_{a}+\delta})^{-1}}
\end{equation}
We can solve this equation numerically to find $\hat p_c$.

\subsection{Activation degree distribution}

The probability density function for activation degree $d$ given $\lambda$ is,
\begin{equation*}
f(d\mid \lambda)=(1-\lambda)\lambda^{d-1}
\end{equation*}
and $\lambda$ is drawn from the following power law distribution
\begin{equation*}
f\left( \lambda \right)= \frac{\beta \lambda ^{-(\beta+1)}}{\xi^{-\beta}-\psi^{-\beta}}
\end{equation*}
Therefore, the probability distribution of $d$ for any $\lambda$ can be found according to Law of Total Probability as follows
\begin{equation*}
f(d)=\frac{\beta}{\xi^{-\beta} - \psi^{-\beta}}\int_{\xi}^{\psi}(1-\lambda)\lambda^{d-1}\lambda^{-(\beta+1)} d\lambda=\frac{\beta }{\xi^{\beta}-\psi^{\beta}}\int_{\xi}^{\psi}( \lambda^{d-\beta-2} -\lambda^{d-\beta-1}) d\lambda
\end{equation*}
\begin{equation*}
=\frac{\beta}{\xi^{-\beta} - \psi^{-\beta}}\left(\frac{\psi^{d-\beta-1}-\xi^{d-\beta-1}}{d-\beta-1}-\frac{\psi^{d-\beta}-\xi^{d-\beta}}{d-\beta}\right)
\end{equation*}

Now we would like to derive the maximum likelihood estimator (MLE) conditions to find the parameters $\beta$, $\xi$ and $\psi$ given $d$ samples from real world network. The likelihood function $L$ for the random values of $d=\{d_1,d_2,\ldots d_n\}$ can be written as 
\begin{equation*}
L\left(\beta,\xi,\psi \mid d\right)=\prod_{k=1}^{n}\frac{\beta}{\xi^{-\beta} - \psi^{-\beta}}\left(\frac{\psi^{d_k-\beta-1}-\xi^{d_k-\beta-1}}{d_k-\beta-1}-\frac{\psi^{d_k-\beta}-\xi^{d_k-\beta}}{d_k-\beta}\right)
\end{equation*}
Taking log on both sides, we get
\begin{equation}\label{actd}
\ln L\left(\beta,\xi,\psi \mid d\right)=n\ln\beta-n\ln(\xi^{-\beta} - \psi^{-\beta})+\sum_{k=1}^{n}\ln \left(\frac{\psi^{d_k-\beta-1}-\xi^{d_k-\beta-1}}{d_k-\beta-1}-\frac{\psi^{d_k-\beta}-\xi^{d_k-\beta}}{d_k-\beta}\right)
\end{equation}
To find the maximum of likelihood function $L(\beta,\xi,\psi)$ for the value of $\beta$, we take the partial derivative of \ref{actd} with respect to $\beta$ and set zero to result for the maximum of $\hat \beta$
\begin{equation*}
0=\frac{n}{\beta}-n\frac{\psi^{-\beta}\ln\psi-\xi^{-\beta}\ln\xi}{\psi^{-\beta}-\psi^{-\beta}}+\sum_{k=1}^{n}\frac{\frac{\xi^{d_k-\beta -1} \ln\xi -\psi^{d_k-\beta -1}\ln\psi}{d_k-\beta-1}-\frac{\psi^{d_k-\beta -1}-\xi^{d_k-\beta -1}}{(d_k-\beta -1)^2}-\frac{\xi^{d_k-\beta}\ln\xi-\psi^{d_k-\beta} \ln\psi}{d_k-\beta}-\frac{\psi^{d_k-\beta}-\xi^{d_k-\beta}}{(d_k-\beta)^2}}{\frac{\psi^{d_k-\beta-1}-\xi^{d_k-\beta-1}}{d_k-\beta-1}-\frac{\psi^{d_k-\beta}-\xi^{d_k-\beta}}{d_k-\beta}}
\end{equation*}
Since $\psi \approx 1 $, we assume $\ln \psi = 0$ and $\psi^{x}=1$ for any $x$. We can simplify as 
\begin{equation}
0=\frac{n}{\beta}-\frac{n\xi^\beta \ln\xi}{\xi^\beta-\psi^{-\beta}}+\sum_{k=1}^{n}\frac{\frac{\xi^{d_k-\beta -1} \ln\xi}{d_k-\beta-1}-\frac{1-\xi^{d_k-\beta -1}}{(d_k-\beta -1)^2}-\frac{\xi^{d_k-\beta} \ln\xi}{d_k-\beta}-\frac{1-\xi^{d_k-\beta}}{(d_k-\beta)^2}}{\frac{1-\xi^{d_k-\beta-1}}{d_k-\beta-1}-\frac{1-\xi^{d_k-\beta}}{d_k-\beta}}
\end{equation}\\
If we set $y=$
In the similar way, we can find $\hat \xi$ derivating $L(\beta,\xi,\psi)$ ( Eq.\ref{actd}) with respect to $\xi$ and setting zero for maximal of $\hat \xi$

\begin{equation*}
0=\frac{n \beta\xi^{-\beta -1}}{\xi^{-\beta}-\psi^{-\beta}}+\sum_{k=1}^{n}\frac{\frac{(d_k-\beta)\xi^{d_k-\beta -1}}{d_k-\beta}-\frac{(d_k-\beta-1)\xi^{d_k-\beta -2}}{d_k-\beta-1}}{\frac{\psi^{d_k-\beta-1}-\xi^{d_k-\beta-1}}{d_k-\beta-1}-\frac{\psi^{d_k-\beta}-\xi^{d_k-\beta}}{d_k-\beta}}
\end{equation*}

With simplification we get
\begin{equation}
0=\frac{n(\beta +1)}{\xi^{-\beta}-\psi^{-\beta}}+\sum_{k=1}^{n}\frac{\xi^{d_k}-\xi^{d_k-1}}{(\psi^{d_k-\beta-1}-\xi^{d_k-\beta-1})(d_k-\beta)^{-1}-(d_k-\beta-1)^{-1}(\psi^{d_k-\beta}-\xi^{d_k-\beta})}
\end{equation}

To estimate $\hat \psi$, we derivate $L(\beta,\xi,\psi)$ ( Eq.\ref{actd}) with respect to $\psi$ and setting zero for maximal of $\hat \psi$
\begin{equation*}
0=-\frac{n\beta \psi^{-\beta -1}}{\xi^{-\beta}-\psi^{-\beta}}+\sum_{k=1}^{n}\frac{\frac{(d_k-\beta -1)\psi^{d_k-\beta -2}}{d_k-\beta-1}-\frac{(d_k-\beta)\psi^{d_k-\beta -1}}{d_k-\beta}}{\frac{\psi^{d_k-\beta-1}-\xi^{d_k-\beta-1}}{d_k-\beta-1}-\frac{\psi^{d_k-\beta}-\xi^{d_k-\beta}}{d_k-\beta}}
\end{equation*}
\begin{equation}
0=-\frac{n(\beta +1)}{\xi^{-\beta}-\psi^{-\beta}}+\sum_{k=1}^{n}\frac{\psi^{d_k-1}-\psi^{d_k}}{(\psi^{d_k-\beta-1}-\xi^{d_k-\beta-1})(d_k-\beta)^{-1}-(d_k-\beta-1)^{-1}(\psi^{d_k-\beta}-\xi^{d_k-\beta})}
\end{equation}
\end{appendices}

\end{document}